\pdfoutput=1
\documentclass[11pt]{article}

\usepackage[english]{babel}
\usepackage{xcolor}
\usepackage{graphicx}

\usepackage[colorlinks,allcolors=blue,linktocpage=true,draft=false]{hyperref}

\usepackage{cite}
\usepackage{subcaption}
\usepackage{amsmath}

\usepackage{amssymb}

\usepackage{booktabs}   % for tables

\usepackage[top=15mm,bottom=12mm,left=30mm,right=30mm,foot=12mm,%
includefoot]{geometry}

\allowdisplaybreaks[3]

\numberwithin{equation}{section}

\newcommand{\sect}[1]{section~#1}
\newcommand{\sects}[1]{sections~#1}

\newcommand{\fig}[1]{figure~#1}
\newcommand{\figs}[1]{figures~#1}

\newcommand{\tab}[1]{table~#1}

\newcommand{\eqn}[1]{equation~#1}
\newcommand{\eqs}[1]{equations~#1}

\newcommand{\tev}{\operatorname{TeV}}
\newcommand{\gev}{\operatorname{GeV}}

\newcommand{\ms}{\mskip 1.5mu}
\newcommand{\bs}{\mskip -1.5mu}

\newcommand{\lsim}{\lesssim}

\newcommand{\mvec}[1]{\smash{\vec{\mskip 0.5mu #1}}\mskip 1.5mu}

\newcommand{\jpsi}{J\mskip -2mu/\mskip -0.5mu\Psi}

\newcommand{\chilipdf}{\textsc{ChiliPDF}}

\newcommand{\lum}{\mathcal{L}}

\newcommand{\Rbar}{\overline{R}\ms}

\newcommand{\Rp}[1]{R{}^{\ms \prime}_{#1}}
\newcommand{\Rpbar}[1]{\overline{R}{}^{\ms \prime}_{#1}}

\newcommand{\pr}[2]{{}^{#1}\bs #2}      % with small backspace
\newcommand{\prb}[2]{{}^{#1}\! #2}      % with big backspace
\newcommand{\prn}[2]{{}^{#1} #2}        % without backspace

\newcommand{\conv}[1]{\underset{#1}{\otimes}}

\newcommand{\ycut}{y_{\text{cut}}}

\newcommand{\muy}{\mu^{*\bs}}

\graphicspath{{plots/}}

\begin{document}

\begin{flushright}
DESY-26-023 \\
%\href{https://arxiv.org/abs/}{arXiv: [hep-ph]}
\end{flushright}

\begin{center}
\vspace{4\baselineskip}
\textbf{\Large A quantitative study of two-loop splitting \\
   in double parton distributions
} \\
\vspace{3\baselineskip}
Markus~Diehl and Peter Pl\"o{\ss}l

\vspace{\baselineskip}

Deutsches Elektronen-Synchrotron DESY, Notkestr.~85, 22607 Hamburg, Germany

\vspace{3\baselineskip}

\parbox{0.9\textwidth}{
Double parton distributions at small distances between the two partons are
dominated by a mechanism in which the two observed partons originate from the
splitting of a single parton.  This contribution can be computed in terms of
single-parton distributions and perturbative splitting kernels.  We demonstrate
that two-loop corrections to these kernels can have a substantial quantitative
impact and considerably improve the stability of predictions for double parton
scattering.  We also consider the impact of heavy quark masses in the two-loop
splitting kernels in an approximate manner.

}
\end{center}

\vfill

\newpage

\tableofcontents

\begin{center}
\rule{0.6\textwidth}{0.3pt}
\end{center}

% \newpage

%%%%%%%%%%%%%%%%%%%%%%%%%%%%%%%%%%%%%%%%%%%%%%%%%%%%%%%%%%%%%%%%%%%%%%%%%%%%%%%%

%%%%%%%%%%%%%%%%%%%%%%%%%%%%%%%%%%%%%%%%%%%%%%%%%%%%%%%%%%%%%%%%%%%%%%%%%%%%%%%%

\section{Introduction}
\label{sec:intro}

An intriguing mechanism in proton-proton collisions at high energy is double
parton scattering (DPS), where two partons in each proton enter a
hard-scattering process that produces a part of the observed final state.  This
mechanism is generically suppressed compared with single-parton scattering
(SPS), where the observed particles in the final state originate from a single
hard scatter, but there are various final states and kinematical regions in
which DPS can be important or even dominant.  A prominent example is the
production of like-sign $W$ pairs ($W^+ W^+$ or $W^- W^-$), which was proposed
for studying DPS in \cite{Kulesza:1999zh, Gaunt:2010pi} and experimentally
observed in Run II of the LHC \cite{CMS:2022pio, ATLAS:2025bcb}.

The theoretical description of DPS in QCD rests on pioneering work in the 1980s
\cite{Paver:1982yp, Mekhfi:1983az}, followed by substantial new developments in
the 2010s \cite{Gaunt:2009re, Diehl:2011yj, Gaunt:2011xd, Gaunt:2012dd, Blok:2011bu, Blok:2013bpa, Ryskin:2011kk, Ryskin:2012qx, Manohar:2012jr, Manohar:2012pe, Diehl:2017kgu, Buffing:2017mqm}.  The
inclusion of double (and more generally multiple) parton interactions is also an
integral part of general-purpose event generators for proton-proton collisions
\cite{Sjostrand:2004pf, Bellm:2019icn, Sherpa:2024mfk, Cabouat:2020ssr,
Fedkevych:2025lgp}.  A comprehensive overview of the field up to 2017 is given
in the monograph \cite{Bartalini:2017jkk}, and for recent theoretical or
phenomenological studies we refer to \cite{Jaarsma:2023woo, Andersen:2023hzm,
Peng:2024qpw, Dumitru:2025bib, Blok:2025kvs, Lovato:2025jgh,
Ceccopieri:2025edn}.  Experimentally, double parton scattering has been observed
both at the Tevatron and the LHC, see \cite{Abe:1997xk, Abazov:2015nnn,
Aaij:2016bqq, CMS:2022pio, ALICE:2023lsn, ATLAS:2025bcb} and references therein.
In addition to like-sign $W$ pairs, observed final states in these studies are
$\jpsi$ pairs \cite{D0:2014vql, Aaij:2016bqq, ATLAS:2016ydt, ALICE:2023lsn}, $W
+ \jpsi$ \cite{ATLAS:2019jzd}, four jets \cite{ATLAS:2016rnd, CMS:2021lxi}, $W
+$ jets \cite{ATLAS:2013aph, CMS:2013huw}, and several others.

The parton-level initial state in DPS is described by double parton
distributions (DPDs), which quantify the joint distribution of two partons with
momentum fractions $x_1$ and $x_2$ and a given transverse distance $y$ from each
other.  These are genuinely non-perturbative quantities and remain poorly known
compared with single-parton distributions (PDFs), which have by now entered the
realm of precision physics.  However, at small distance $y$ between the two
partons, the dominant contribution to a DPD comes from graphs in which the two
observed partons originate from the perturbative splitting of a single parton.
For ease of language we refer to this as ``parton splitting in
DPDs'' or ``DPD splitting'' henceforth.\footnote{%
Note that ``parton splitting'' in this sense does \emph{not} include additional
splitting processes that are described by DGLAP evolution.}
This mechanism has been studied long ago in the context of scale evolution
\cite{Kirschner:1979im, Shelest:1982dg, Snigirev:2003cq, Gaunt:2009re,
Ceccopieri:2010kg} and later from more general points of view
\cite{Diehl:2011yj, Gaunt:2011xd, Gaunt:2012dd, Blok:2011bu, Blok:2013bpa,
Ryskin:2011kk, Ryskin:2012qx, Manohar:2012pe, Diehl:2017kgu}.  Depending on the
parton combination and kinematics (parton momentum fractions and hard scales),
the importance of this splitting contribution to the DPS cross section can be
anything between negligible and dominant.  In cases where it is dominant, one
has the remarkable situation that DPS can be largely predicted in terms of
ordinary PDFs and perturbatively calculable quantities.
At leading order (LO) in $\alpha_s$, the kernels for parton splitting in DPDs
have a simple connection to the one-loop splitting functions in the DGLAP
equation.  This connection is lost at higher orders, and the DPD splitting
kernels at next-to-leading order (NLO) have been computed for massless
unpolarised partons in \cite{Diehl:2019rdh, Diehl:2021wpp}.  The impact of heavy
quark masses in the splitting process was investigated in \cite{Diehl:2022dia}.

In physical cross sections, DPDs are evaluated at the characteristic scales
$\mu_1$ and $\mu_2$  of the two hard-scattering processes, whereas the splitting
process just described should be evaluated at its natural scale
$\mu_{\text{init}} \sim 1/y$.  Numerical studies in \cite{Diehl:2022dia} have
shown that the dependence of the cross section on the specific choice of
$\mu_{\text{init}}$ can be substantial if the splitting is evaluated at LO,
reaching an order of magnitude or more in certain kinematics.  In such
situations one barely has a useful theoretical prediction.  The first purpose of
the present study is to see how this changes when the splitting is evaluated at
NLO.

Parton splitting in DPDs is also of primary importance because it gives rise to
double counting between DPS and certain higher-loop graphs for SPS
\cite{Cacciari:2009dp, Diehl:2011yj}.  The scheme developed in
\cite{Diehl:2017kgu} avoids this double counting by imposing a lower cutoff
$\ycut$ on $y$ in DPS and by writing the overall cross section as a sum of
contributions from SPS, DPS, and a subtraction term.  The second purpose of the
present work is to revisit the construction of that subtraction term and to
study the sensitivity of predictions on the choice of $\ycut$ when DPD
splitting is evaluated at LO or NLO.

The paper is organised as follows.  In \sect{\ref{sec:basics}} we recall the
basics of DPS and of DPD splitting in the formalism of \cite{Diehl:2011yj,
Diehl:2017kgu, Buffing:2017mqm} that underlies our study.  In
\sect{\ref{sec:subtraction}} we recall our scheme for avoiding double counting
between DPS and SPS, and we present a new prescription for the subtraction term
in that scheme.  The quantitative impact of NLO corrections to the DPD splitting
kernels is studied in \sect{\ref{sec:massless}} for the case where quark masses
are neglected and in \sect{\ref{sec:massive}} for the case where they are
included in an approximate way.  We summarise our results in
\sect{\ref{sec:summary}}.

%%%%%%%%%%%%%%%%%%%%%%%%%%%%%%%%%%%%%%%%%%%%%%%%%%%%%%%%%%%%%%%%%%%%%%%%%%%%%%%%

\section{Double parton scattering basics}
\label{sec:basics}

In this section we recall the theory description of double parton scattering
developed in \cite{Diehl:2011yj, Buffing:2017mqm, Diehl:2022rxb}, with an
emphasis on the perturbative splitting mechanism for DPDs \cite{Diehl:2017kgu,
Diehl:2019rdh, Diehl:2021wpp}.

Consider the cross section for a process $p + p \to A_1 + A_2 + X$, where $A_i$
denotes the system of observed particles produced in the hard subprocess number
$i$ ($i = 1,2$), and $X$ denotes the unobserved part of the final state.  If the
hard subprocesses are computed at LO, then the DPS cross section is given by
\begin{align}
   \label{dps-Xsect}
   \frac{d \sigma_{\text{DPS}}}{d M_1^2\, d Y_1^{}\, d M_2^2 \, d Y_2^{}}
   &
   =
   \frac{1}{s^2} \;
   \frac{1}{1 + \delta_{A_1 A_2}} \;
   \sum_{R_1 R_2 R_3 R_4} \, \sum_{a_1 a_2\, b_1 b_2}
   \hspace{-0.5em}
   \prn{\Rbar_{1} \Rbar_{3} \ms}{\hat{\sigma}}_{a_1 b_1}(M_1; \mu_1) \;
   \prn{\Rbar_{2} \Rbar_{4} \ms}{\hat{\sigma}}_{a_2 b_2}(M_2; \mu_2)
   \notag \\
   & \quad \times
   \prn{R_1 R_2,\ms R_3 R_4 \ms}{\lum}_{a_1 a_2,\ms b_1 b_2}(
      M_1, Y_1, M_2, Y_2; \mu_1, \mu_2, \nu, s)
   \,.
\end{align}
Kinematic quantities appearing here are the c.m.\ energy $\sqrt{s}$ of the
proton-proton collision, along with the invariant mass $M_i$ and the rapidity
$Y_i$ of the observed system $A_i$.  Depending on the process, the parton-level
cross sections $\hat{\sigma}$ and the physical cross section may depend on
further quantities, such as the transverse momentum in jet production.

\begin{figure}
\begin{center}
\includegraphics[width=0.6\textwidth]{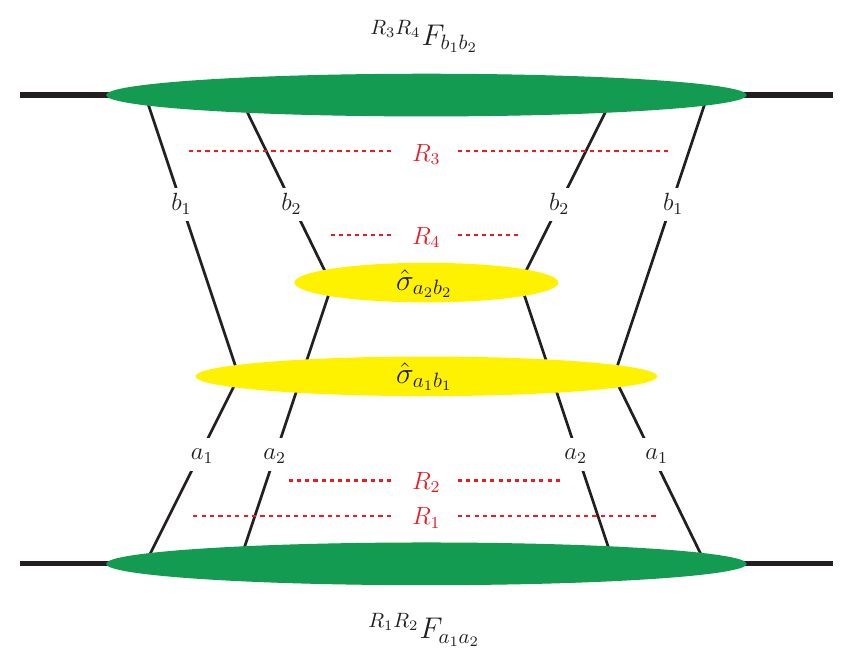}
\caption{\label{fig:double-colour} Graph corresponding to the factorisation
formula for double parton scattering, given by \protect\eqref{dps-Xsect} and
\protect\eqref{dpd-lumi-def}.  Pairs of parton lines have their colour indices
coupled to irreducible representations $R_i$ of SU(3).  The final-state cut
would be a vertical line through the centre of the graph and is omitted for
clarity.  The horizontal lines at the very left and the very right represent the
incoming protons.}
\end{center}
\end{figure}

$R_1$ denotes the colour representation of the pair of quark or gluon lines
associated with the parton label $a_1$, as indicated in
\fig{\ref{fig:double-colour}}, with corresponding associations of $R_2
\leftrightarrow a_2$, $R_3 \leftrightarrow b_1$, and $R_4 \leftrightarrow b_2$.
The relevant representations are $R=1$ and $R=8$ for quarks and antiquarks, and
$R=1, A, S, 10, \overline{10}, 27$ for gluons, where $A$ and $S$ respectively
stand for the antisymmetric and symmetric colour octet.  By $\overline{R}_i$ we
denote the conjugate of the representation $R_i$, which is equal to $R_i$ for
all cases except $10$ and $\overline{10}$.  More details about colour
non-singlet DPDs can be found in \cite{Buffing:2017mqm, Diehl:2021wvd}.

The parton-level cross sections $\hat{\sigma}$ in \eqref{dps-Xsect} are
multiplied by double parton luminosities $\lum$, which are products of two DPDs
integrated over the transverse distance $y$ between the two partons extracted
from each proton:
\begin{align}
   \label{dpd-lumi-def}
   &
   \prn{R_1 R_2,\ms R_3 R_4 \ms}{\lum}_{a_1 a_2,\ms b_1 b_2}(
      M_1, Y_1, M_2, Y_2; \mu_1, \mu_2, \nu, s)
   \notag \\
   & \quad
   =
   2\pi \int_{\ycut}^{\infty} d y\, y\;
   \prn{R_1 R_2}{F}_{a_1 a_2}(x_1, x_2, y; \mu_1, \mu_2, \zeta) \;
   \prn{R_3 R_4}{F}_{b_1 b_2}(\bar{x}_1, \bar{x}_2, y;
      \mu_1, \mu_2, \bar{\zeta})
   \,.
\end{align}
In full generality, the labels $a_1, a_2$ and $b_1, b_2$ also specify the
polarisation of the partons.  For transversely polarised quarks and for linearly
polarised gluons, both DPDs and parton-level cross sections carry transverse
Lorentz indices, and the DPDs depend on the direction of the transverse distance
vector $\mvec{y}$ \cite{Diehl:2011yj}.  Our present study is restricted to
unpolarised partons, since NLO splitting effects have not yet been computed for
the polarised cases.  We therefore have DPDs depending only on $y = |\mvec{y}|$,
and the trivial integration over the azimuth of $\mvec{y}$ has been carried out
in \eqref{dpd-lumi-def}.

The DPDs in \eqref{dpd-lumi-def} are evaluated for longitudinal momentum fractions
\begin{align}
   \label{x-fractions}
   x_i
   &=
   \frac{M_i}{\sqrt{s}}\; e^{Y_i} \,,
   &
   \bar{x}_i
   &=
   \frac{M_i}{\sqrt{s}}\; e^{- Y_i} \,,
   &
   i=1,2,
\end{align}
which are fixed by the process kinematics.  If the parton-level cross sections
are evaluated beyond LO, then the factorisation formula \eqref{dps-Xsect} must
be modified to contain convolution integrals over the parton momentum fractions,
in the same way as is done for single parton scattering.

The DPDs and parton-level cross sections depend on factorisation scales $\mu_1$
and $\mu_2$, and the dependence of the DPDs on these scales is given by DGLAP
equations whose full form can be found in \sect{2.2} of \cite{Diehl:2023jje}.
For colour singlet distributions ($R_1 = R_2 = 1$) these are straightforward
generalisations of the DGLAP equations for PDFs and contain the same DGLAP
splitting functions.  The evolution of colour non-singlet DPDs up to NLO
accuracy has been studied in \cite{Diehl:2022rxb, Diehl:2023jje}.

As will be reviewed in \sect{\ref{sec:subtraction}}, an important property of
factorisation for DPS is that the integral over $y$ has a lower cutoff $\ycut$,
which gives rise to a further momentum scale $\nu$:
\begin{align}
   \label{nu-ycut}
   \ycut
   &=
   b_0 / \nu
   \,,
\end{align}
where
\begin{align}
   \label{b0-def}
   b_0
   &=
   2 e^{-\gamma_E} \approx 1.12
\end{align}
and $\gamma_E$ is the Euler-Mascheroni constant.  As explained in \sect{6} of
\cite{Diehl:2017kgu}, an appropriate choice of $\nu$ is
\begin{align}
   \label{nu-scale}
   \nu
   &
   = \min(\mu_1, \mu_2)
   \,,
\end{align}
which will be adopted as a default throughout this work.
The rapidity parameters $\zeta$ and $\bar{\zeta}$ on the r.h.s.\ of
\eqref{dpd-lumi-def} will be explained in \sect{\ref{sec:rapidity-evo}}.

We write the DPDs in terms of splitting and intrinsic contributions as
\begin{align}
   \label{F-spl-intr}
   F
   &=
   F^{\text{spl}} + F^{\text{intr}}
   ,
\end{align}
where each contribution depends on the same labels and variables as $F$ in
\eqref{dpd-lumi-def}.
This separation is model independent at small $y$, where the two
contributions can be defined via an operator product expansion
\cite{Diehl:2017kgu}.  When evaluated at $\mu_1 \sim \mu_2 \sim \sqrt{\zeta}
\sim 1/y$, they have a power behaviour $F^{\text{spl}} \sim y^{-2}$ and
$F^{\text{intr}} \sim y^{0}$, and they can be evaluated by the convolution of
perturbative kernels with either PDFs (for $F^{\text{spl}}$) or twist-four
distributions (for $F^{\text{intr}}$).  For $y$ in the non-perturbative region,
we retain the separation \eqref{F-spl-intr} in the sense of a model ansatz.

The separation \eqref{F-spl-intr} induces a separation of double parton
luminosities into four parts:
\begin{align}
   \label{partial-lumis}
   \lum_{\text{1v1}}
   & \leftrightarrow
   F^{\text{spl}} \, F^{\text{spl}}
   \,,
   &
   \lum_{\text{1v2}}
   & \leftrightarrow
   F^{\text{spl}} \, F^{\text{intr}}
   \,,
   &
   \lum_{\text{2v1}}
   & \leftrightarrow
   F^{\text{intr}} \, F^{\text{spl}}
   \,,
   &
   \lum_{\text{2v2}}
   & \leftrightarrow
   F^{\text{intr}} \, F^{\text{intr}}
   \,,
\end{align}
where ``1'' indicates that there is a single parton at the origin of the two
partons that enter the DPS process \cite{Gaunt:2012dd}.

The perturbative splitting formula for DPDs has the form
\begin{align}
   \label{split-master}
   \pr{R_1 R_2}{F}^{\text{spl}}_{a_1 a_2}(x_1, x_2, y; \mu, \mu,\zeta)
   &
   =
   \frac{1}{\pi\ms y^2}\, \sum_{a_0}\,
   \int\limits_{x_1+x_2}^1 \!\! \frac{d z}{z^2} \;
   \prn{R_1 R_2}{V}_{a_1 a_2, a_0}
   \biggl( \frac{x_1}{z}, \frac{x_2}{z}, \alpha_s(\mu),
      L_y, L_\zeta \biggr) \,
   f_{a_0}(z; \mu)
   \notag \\
   &
   \overset{\text{def}}{=}
   \frac{1}{\pi\ms y^2}\, \sum_{a_0}\,
   \prn{R_1 R_2}{V}_{a_1 a_2, a_0} \conv{12} f_{a_0}
\end{align}
with logarithms
\begin{align}
   \label{L-defs}
   L_y
   &=
   \ln\frac{y^2 \mu^2}{b_0^2}
   \,,
   &
   L_\zeta
   &=
   \ln \frac{\mu^2}{x_1 x_2\ms \zeta}
   \,.
\end{align}
The splitting kernels $\prn{R_1 R_2}{V}_{a_1 a_2, a_0}$ have a perturbative
expansion starting at order $\alpha_s$.  The masses of heavy quarks ($c$, $b$,
$t$) can be taken into account in the splitting, as we will discuss in
\sect{\ref{sec:massive}}.

%%%%%%%%%%%%%%%%%%%%%%%%%%%%%%%%%%%%%%%%

\subsection{Rapidity evolution}
\label{sec:rapidity-evo}

As explained in
\cite{Buffing:2017mqm}, colour non-singlet DPDs depend not only on the
factorisation scales $\mu_1$ and $\mu_2$ for the two partons, bit also on a
rapidity parameter.   In the double parton luminosity \eqref{dpd-lumi-def}, the
rapidity parameters $\zeta$ and $\bar{\zeta}$ of the two DPDs must satisfy the
kinematic relation
\begin{align}
   \label{zeta-basic-constraint}
   \bigl(\ms \zeta \bar{\zeta} \,\bigr)^{1/2}
   &= s
   \,,
\end{align}
which is realised by the explicit choice
\begin{align}
   \label{zeta-kin-choice}
   \zeta
   &=
   \frac{M_1 M_2}{x_1 x_2}
   \,,
   &
   \bar{\zeta}
   &=
   \frac{M_1 M_2}{\bar{x}_1 \bar{x}_2}
   \,.
\end{align}
The reason for choosing rapidity parameters that depend on the parton momentum
fractions is explained in \sect{3.2} of \cite{Diehl:2023jje}.

The rapidity dependence of DPDs is described by a Collins-Soper equation
\begin{align}
   \label{DPD-CS}
   \frac{d}{d\ln \zeta} \,
   \ln \prn{R_1 R_2}{F}_{a_1 a_2}(x_1, x_2, y; \mu_1, \mu_2, \zeta)
   &=
   \frac{1}{2} \, \prb{R_1}{J}(y; \mu_1, \mu_2)
   \,,
\end{align}
in close analogy to the case of transverse-momentum dependent parton
distributions (TMDs) \cite{Collins:2011zzd}.  The Collins-Soper kernel
$\pr{R}{J}$ satisfies a renormalisation group equation in $\mu_1$
\begin{align}
   \label{RGE-CS}
   \frac{d}{d\ln \mu_1} \pr{R}{J}(y; \mu_1, \mu_2)
   &=
   {}- \prn{R}{\gamma}_J(\mu_1)
\end{align}
and its analogue for the scale $\mu_2$.  The kernel $\pr{R}{J}$ is zero for the
colour singlet ($R=1$) and nonzero for all other colour representations.  For
the colour octet channels, one has the exact relation \cite{Vladimirov:2016qkd,
Buffing:2017mqm}
\begin{align}
   \label{octet-CS-kernel}
   \pr{8}{J}(y; \mu, \mu)
   = \prb{S}{J}(y; \mu, \mu)
   = \pr{A}{J}(y; \mu, \mu)
   &=
   K_g(y, \mu)
   \,,
\end{align}
where $K_g$ is the Collins-Soper kernel for the rapidity evolution of
single-gluon TMDs.

The system \eqref{DPD-CS} and \eqref{RGE-CS} of differential equations is
readily solved analytically.  The DGLAP equations for the $\mu_1$ and $\mu_2$
dependence of DPDs depend on $\zeta$ and $\gamma_J$ as well.  Solving the full
system of evolution equations and using the condition
\eqref{zeta-basic-constraint}, one finds that the explicit dependence of the
DPDs on $\zeta$ or $\bar{\zeta}$ drops out in the double parton
luminosity~\eqref{dpd-lumi-def}.  A detailed discussion is given in \sect{4.2}
of \cite{Diehl:2023jje}.  The particular choice \eqref{zeta-kin-choice} is hence
not relevant for the value of the double parton luminosities, but it simplifies
the discussion of rapidity evolution effects in the DPDs.

In the present work we will only be concerned with colour singlet and colour
octet channels.  We use \eqref{octet-CS-kernel} and assume Casimir scaling
between the Collins-Soper kernels $K_g$ and $K_q$ for gluon and quark TMDs,
which holds at small distances up to order $\alpha_s^3$ \cite{Li:2016ctv,
Echevarria:2016scs}.  This allows us to construct a model for $\pr{8}{J}$ using
input from phenomenological determinations of $K_q$.  We follow the model
construction laid out in \sect{3.1} of \cite{Diehl:2023jje}, which has the form
\begin{align}
   \label{J-gen-form}
   &
   \pr{8}{J}(y; \mu_1, \mu_2)
\notag \\
   &\qquad =
   \pr{8}{J}^{\text{pt}}\bigl( y^*(y); \muy(y), \muy(y) \bigr)
   + \prn{8}{\Delta J}(y)
   - \int_{\muy(y)}^{\mu_1} \frac{d \mu}{\mu}\, \prn{8}{\gamma}_J(\mu)
   - \int_{\muy(y)}^{\mu_2} \frac{d \mu}{\mu}\, \prn{8}{\gamma}_J(\mu)
\end{align}
with functions $\muy(y)$ and $y^*(y)$ specified in \eqref{mu-of-y} and
\eqref{star-choice} below.  The short-distance part $\pr{8}{J}^{\text{pt}}$ is
evaluated in fixed-order perturbation theory, and the long-distance part
$\prn{8}{\Delta J}$ is fixed such that the full kernel $\pr{8}{J}$ is equal to
$C_A / C_F$ times the kernel $K_q$ in the SV19
parametrisation~\cite{Scimemi:2019cmh}.

The combinations of perturbative orders used for $\pr{8}{J}^{\text{pt}}$, for
the anomalous dimension $\prn{8}{\gamma}_J$, and for the QCD $\beta$ function
are discussed in \sect{4.3} of \cite{Diehl:2023jje}.  In the present work we use
the two combinations given in \tab{\ref{tab:orders}}.  We note that the $\beta$
function is taken at one order less than what is required for NLL or NNLL
accuracy in the nomenclature of resummation.  The reason for this is that we
take the PDFs in the DPD splitting formula from global fits, which use the same
perturbative order for the DGLAP splitting functions and for $\beta / \alpha_s$.
For further discussion, we refer to \sect{2.4} in \cite{Bacchetta:2019sam} and
to \sect{4} in \cite{Stewart:2010qs}.

\begin{table}
\begin{center}
\begin{tabular}{c c c c}
   \hline
   $P$ & $\beta / \alpha_s$ & $\gamma_J$ & $J$ \\
   \hline
   $\alpha_s$ & $\alpha_s$ & $\alpha_s^2$  & $\alpha_s$\\
   $\alpha_s^2$ & $\alpha_s^2$  & $\alpha_s^3$ & $\alpha_s^2$ \\
   \hline
\end{tabular}
\caption{\label{tab:orders} Combination of perturbative orders used in the
present work for scale and rapidity evolution.  The entries for $P$ refer to the
DGLAP kernels for the evolution of the DPDs and of the PDFs in the DPD splitting
formula, those for $\beta / \alpha_s$ to the running of $\alpha_s$, and those
for $J$ to the fixed-order perturbative input $\pr{8}{J}^{\text{pt}}$ of the
Collins-Soper kernel in \protect\eqref{J-gen-form}.}
\end{center}
\end{table}

%%%%%%%%%%%%%%%%%%%%%%%%%%%%%%%%%%%%%%%%

\subsection{Initial scales for DPD evolution}
\label{sec:init-scale}

In DPS cross sections, DPDs are needed at the scales $\mu_1$ and $\mu_2$ of the
two parton-level subprocesses.  We obtain them by DGLAP evolution from a
suitable starting scale $\mu_{\text{init}}$, which is always taken equal for the
two partons.

To evaluate the initial condition for the splitting part $F^{\text{spl}}$ using
a fixed-order formula, we take a $y$ dependent initial scale
$\mu_{\text{init}}(y)$ so as to avoid large logarithms of $y \mu_{\text{init}}$
that would spoil the convergence of the perturbative series. Our default choice
is $\mu_{\text{init}}(y) = \muy(y)$, where
\begin{align}
   \label{mu-of-y}
   \muy(y)
   &=
   b_0 / y^*(y)
\end{align}
with a function $y^*(y)$ that satisfies
\begin{align}
   \label{y-star-cond}
   y^*(y)
   \to
   \begin{cases}
      y & \text{ for } y \ll 1/\Lambda
      \,,
      \\
      y_{\text{max}}
      &
      \text{ for } y \gg 1/\Lambda
      \,,
   \end{cases}
\end{align}
where $\Lambda$ represents a typical non-perturbative scale.  As discussed in
\sect{3.2} of \cite{Diehl:2023jje}, it is natural to use $\mu_{\text{init}}(y)$
as an initial scale also for the intrinsic part $F^{\text{intr}}$.
In our numerical studies we take
\begin{align}
   \label{star-choice}
   y^*(y) &= \frac{y}{\sqrt[4]{1 + y^4 / y_{\text{max}}^4}}
   \,,
\end{align}
which ensures that $\muy$ is always greater than $\mu_{\text{min}} = b_0 /
y_{\text{max}}$.  We set
\begin{align}
   \label{mu-min}
   \mu_{\text{min}}
   &=
   2 \gev
\end{align}
as we did in our evolution study \cite{Diehl:2023jje}, mainly in order to avoid
the gluon PDF in the DPD splitting formula becoming negative at low $x$ and low
scales.

For later use, we note that the choice \eqref{star-choice} implies
\begin{align}
   \label{ycut-vs-nu}
   \muy(\ycut)
   &=
   \nu \; \sqrt[4]{1 + \mu_{\smash{\text{min}}}^4 / \nu^4}
   \notag \\[0.2em]
   &\approx
   \nu
   &
   \text{for } \nu \gg \mu_{\text{min}}
   \,.
   \end{align}
The corrections to the approximation in the second line are of order
$(\mu_{\text{min}} / \nu)^4$ and thus tiny in the range $\nu \ge 10 \gev$ we
will consider later in this work.

Again in the spirit of avoiding large logarithms in the fixed-order splitting
formula, we take the initial scale of rapidity evolution as
\begin{align}
   \label{zeta-init}
   \zeta_{\text{init}}(y)
   &=
   \frac{\mu_{\text{init}}^2(y)}{x_1 \ms x_2}
   \,,
   &
   \bar{\zeta}_{\text{init}}(y)
   &=
   \frac{\mu_{\text{init}}^2(y)}{\bar{x}_1 \ms \bar{x}_2}
\end{align}
in the relevant DPD.
To assess the dependence of our results on the choice of $\mu_{\text{init}}(y)$,
we will vary that scale within the limits
\begin{align}
   \label{mu-init-variation}
   \min\bigl\{ \mu_{\text{min}} ,\, \muy(y) \big/ 2 \bigr\}
   \le \mu_{\text{init}}(y)
   \le 2 \muy(y)
\end{align}
around our default choice $\mu_{\text{init}}(y) = \muy(y)$.  This corresponds to
the conventional variation by a factor of 2 up and down when $\muy(y) > 2
\mu_{\text{min}}$, while avoiding scales below $\mu_{\text{min}}$ at large $y$.
We regard any initial scale within these limits as a reasonable theoretical
choice.

We will shortly show that splitting DPDs obtained with initial scales in the
region \eqref{mu-init-variation} differ by an amount that is of higher
perturbative order than the order used in the splitting formula.  Variation of
scales that are internal to perturbative calculations is often used to estimate
the uncertainty due to missing higher orders, but the limitations of such
an approach have long been known, and a detailed recent discussion may be found
in \sect{2.3} of \cite{Tackmann:2024kci}.  Nevertheless, the impact of varying
$\mu_{\text{init}}$ within the limits \eqref{mu-init-variation} carries
important information.  If this results in a large variation of double parton
luminosities, higher-order corrections are manifestly large.  Conversely, a
moderate or small impact of the scale variation on double parton luminosities is
a necessary although not a sufficient condition for having a reliable theory
result.

Let us assume that the DPD splitting kernels are evaluated up to order
$\alpha_s^k$, with $k=1$ at LO, $k=2$ at NLO, etc.  We introduce the notation
\begin{align}
   \label{DPD-with-scales}
   F^{\text{spl}\ms (k)}(x_1, x_2, y; \mu_1, \mu_2, \zeta |
      \mu_0, \zeta_0)
\end{align}
for a DPD that is initialised from the $k$-th order splitting formula at scale
$\mu_0$ and rapidity parameter $\zeta_0$ and then evolved to the final scales
$\mu_1, \mu_2$ and the final rapidity parameter $\zeta$.  For brevity, we omit
parton and colour labels in \eqref{DPD-with-scales} and the following equations.

We must of course require that $y$ is in the perturbative regime, where the
splitting formula for DPDs is applicable and where the Collins-Soper kernel
$\pr{R}{J}$ can be expanded in $\alpha_s$. We also require that $\mu_0 \sim
\sqrt{\zeta_0} \sim 1/y$, so that no large logarithms in the ratio of these
scales can spoil counting powers of $\alpha_s$ in the following perturbative
estimates.\footnote{%
This statement holds up to possible large logarithms $\ln x_1$ and $\ln x_2$,
which are of dynamical origin and beyond the scope of our analysis.}

In terms of an evolution operator $U$, which is the Green's function of the
combined system of DGLAP and Collins-Soper equations, we can write
\begin{align}
   \label{DPD-evol-op}
   F^{\text{spl}\ms (k)}(x_1, x_2, y; \mu_1, \mu_2, \zeta |
      \mu_0, \zeta_0)
   &=
   \int_{x_1}^1 \frac{d z_1}{z_1} \, \int_{x_2}^1 \frac{d z_2}{z_2} \,
   U(x_1, x_2, \mu_1, \mu_2, \zeta
      | z_1, z_2, \mu_0, \mu_0, \zeta_0)
   \notag \\[0.3em]
   &\quad \times
      F^{\text{spl}\ms (k)}(z_1, z_2, y;
      \mu_0, \mu_0, \zeta_0 |
      \mu_0, \zeta_0)
   \,,
\end{align}
where the DPD on the r.h.s.\ is given by the DPD splitting formula
\eqref{split-master} at order $\alpha_s^{k}$.

It is sufficient for the following results that the evolution of DPDs and of the
PDFs in the splitting formula is evaluated with DGLAP and Collins-Soper kernels
of order $\alpha_s^{k-1}$.  This is consistent because the explicit logarithms
of $\mu$ or $\zeta$ in the $k$-th order splitting formula are accompanied by
evolution kernels of at least one order less, given that the splitting DPD
itself starts at order $\alpha_s$ \cite{Diehl:2019rdh,Diehl:2021wpp}.  In
practice one will typically take the evolution kernels at order $\alpha_s^{k}$,
and $\gamma_J$ even at order $\alpha_s^{k+1}$ (see also \tab{\ref{tab:orders}}).

The splitting DPD evaluated up to order $k$ satisfies the Collins-Soper equation
up to terms of order $\alpha_s^{k+1}$ or higher, i.e.
\begin{align}
   &
   \frac{d}{d \ln \zeta_0} \,
   F^{\text{spl}\ms (k)}(x_1, x_2, y;
                         \mu_0, \mu_0, \zeta_0 \ms|\ms \mu_0 \zeta_0)
   \notag \\
   &\quad =
   \frac{1}{2} \, J^{(k-1)}(y; \mu_0, \mu_0) \,
      F^{\text{spl}\ms (k)}(x_1, x_2, y;
                           \mu_0, \mu_0, \zeta_0 \ms|\ms \mu_0 \zeta_0)
   + \mathcal{O}\bigl( \alpha_s^{k+1}(\mu_0) \bigr)
\end{align}
where $J^{(i)}$ is the Collins-Soper kernel evaluated up to order
$\alpha_s^{i}$. A corresponding statement holds for the DGLAP equation.
As a consequence, the dependence on $\mu_0$ and $\zeta_0$ does not exactly
cancel between $U$ and $F$ on the r.h.s.\ of \eqref{DPD-evol-op}, and one has
\begin{align}
   \label{DPD-scale-ODE}
   \frac{d}{d \ln \mu_0} \,
      F^{\text{spl}\ms (k)}(x_1, x_2, y; \mu_1, \mu_2, \zeta |
      \mu_0, \zeta_0)
   =
   \mathcal{O}\bigl( \alpha_s^{k+1}(\mu_0) \bigr)
   \notag \\
   \frac{d}{d \ln \zeta_0} \,
      F^{\text{spl}\ms (k)}(x_1, x_2, y; \mu_1, \mu_2, \zeta |
      \mu_0, \zeta_0)
   =
   \mathcal{O}\bigl( \alpha_s^{k+1}(\mu_0) \bigr)
\end{align}
We emphasise that the orders on the r.h.s.\ are dictated by the truncation of
the DPD splitting kernel and not by the orders at which evolution is performed,
as long as the latter includes the order $\alpha_s^{k-1}$.

Integrating the differential equations \eqref{DPD-scale-ODE} between two scales
$\mu_a$ and $\mu_b$, we obtain
\begin{align}
   \label{DPD-resid-scale-dep}
   &
   F^{\text{spl}\ms (k)}\biggl( x_1, x_2, y; \mu_1, \mu_2, \zeta \,\bigg|\,
      \mu_{a}, \frac{\mu_{a}^2}{x_1 x_2} \biggr)
   - F^{\text{spl}\ms (k)}\biggl( x_1, x_2, y; \mu_1, \mu_2, \zeta \,\bigg|\,
      \mu_{b}, \frac{\mu_{b}^2}{x_1 x_2} \biggr)
   \notag \\[0.2em]
   &\quad
   =
   \mathcal{O}\bigl( \alpha_s^{k+1}(\mu_a) \bigr)
   \hspace{20em}
   \text{for } \mu_a \sim \mu_b \sim 1/y
   \,.
\end{align}
The dependence of an evolved splitting DPD on the initial scale is thus one
order higher than the perturbative accuracy at which the DPD is evaluated, as we
claimed above.

\section{Double counting subtraction}
\label{sec:subtraction}

\begin{figure}
\begin{center}
\subfloat[\label{fig:dps-1v1-sps} SPS]{\includegraphics[height=0.3\textwidth]{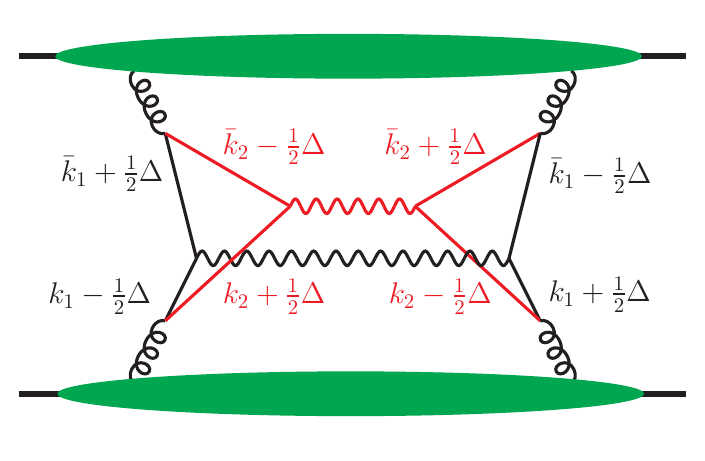}
}
\hspace{1em}
\subfloat[DPS]{\includegraphics[height=0.3\textwidth]{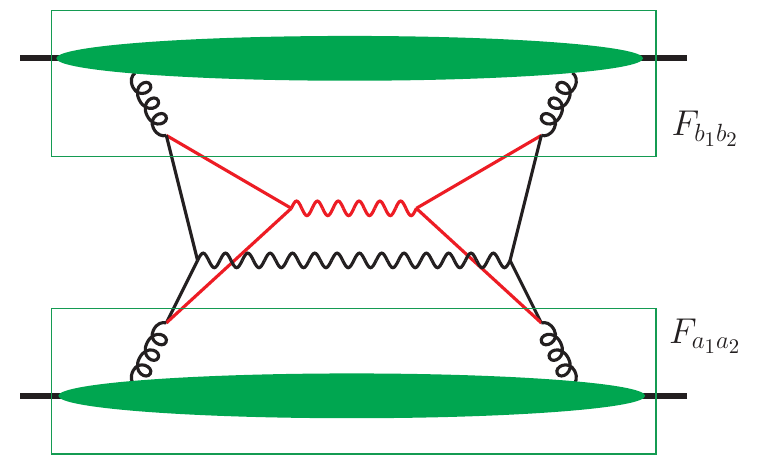}
}
\caption{\label{fig:dps-1v1} (a): Box graph for the production of two
electroweak gauge bosons via SPS.  All loop momenta are routed such that they
flow towards the gauge boson vertex.  (b): The same graph, interpreted as a
contribution to DPS with two splitting DPDs.  The final state cut would be a
vertical line through the centre of the graph and is omitted for clarity.}
\end{center}
\end{figure}

Since the two partons in a DPD can originate from the splitting of a single
parton, there is an overlap between DPS with two splitting DPDs and higher-loop
corrections to single parton scattering \cite{Diehl:2011yj}.  In different
regions of the loop momenta, the same graph can be regarded as either DPS or
SPS, as shown in \fig{\ref{fig:dps-1v1}}.  A scheme to delineate the two
mechanisms was developed in \cite{Diehl:2017kgu}.  In this scheme, double
counting between SPS and DPS is avoided by a subtraction term.  It is the
purpose of this section to revisit the construction of this term.

In the scheme of \cite{Diehl:2017kgu}, the full cross section for a process to
which DPS can contribute reads
\begin{align}
   \label{full-Xsect}
   \sigma
   &=
   \sigma_{\text{DPS}}(\nu) - \sigma_{\text{sub}}(\nu)
   + \sigma_{\text{SPS}}
   \,.
\end{align}
To understand this construction, we write all three terms in \eqref{full-Xsect}
as integrals over the transverse distance between the two partons:
\begin{align}
   \label{y-integrals}
   \sigma_{\text{DPS}}(\nu)
   &=
   \int d^2 y\; \theta(y - \ycut) \, I_{\text{DPS}}(y)
   \,,
   \notag \\
   \sigma_{\text{sub}}(\nu)
   &=
   \int d^2 y\; \theta(y - \ycut) \, I_{\text{sub}}(y)
   \,,
   \notag \\
   \sigma_{\text{SPS}}
   &=
   \sigma_{\text{SPS,0}}
   + \int d^2 y\; I_{\text{SPS}}(y)
   \,,
\end{align}
where we use $\ycut$ and $\nu = b_0 / \ycut$ interchangeably.  Throughout this
section, it is understood that $\nu \sim \min(\mu_1, \mu_2)$.

The expression of $I_{\text{DPS}}$ in terms of DPDs and parton-level cross
sections is readily obtained from \eqref{dps-Xsect} and \eqref{dpd-lumi-def}. In
analogy to \eqref{partial-lumis} it can be split into four contributions that
involve the splitting and intrinsic parts of the two DPDs in all combinations:
$I_{\text{DPS}} = I_{\text{1v1}} + I_{\text{1v2}} + I_{\text{2v1}} +
I_{\text{2v2}}$.  At small $y$ this decomposition is model independent.

The full SPS cross section $\sigma_{\text{SPS}}$ can be expressed by a standard
factorisation formula involving PDFs and parton-level cross sections. The part
$\sigma_{\text{SPS,0}}$ is due to graphs that do not have any overlap with DPS,
whereas $I_{\text{SPS}}$ is obtained from the loop graphs that do have a DPS
region.  In these graphs, a particular loop momentum $\Delta$ can be identified
such that its transverse component is Fourier conjugate to the transverse
distance $\mvec{y}$.  An example is given in \fig{\ref{fig:dps-1v1-sps}}.

The integrand of the double counting subtraction term $\sigma_{\text{sub}}$ must
satisfy the conditions
\begin{align}
   \label{sub-conditions}
   I_{\text{sub}}(y)
   &\to
   \begin{cases}
      I_{\text{1v1}}(y) & \text{ for } y \sim \ycut \,,
      \\
      I_{\text{SPS}}(y) & \text{ for } y \gg \ycut \,,
   \end{cases}
\end{align}
in order to ensure that
\begin{enumerate}
\item for $y \sim \ycut$ the subtraction term cancels the 1v1 contribution to
DPS in the sum \eqref{full-Xsect}, such that one is left with the SPS term,
which correctly describes the process in that region of $y$,
\item the dependence on $\nu$ cancels between the DPS and subtraction terms,
\item for $y \gg \ycut$ the subtraction term cancels the part of SPS that has
overlap with DPS.  One is then left with the DPS term and with the part of SPS
that has no overlap with DPS.
\end{enumerate}
In all three points, the cancellation is understood to be accurate up to
remainder terms that are beyond the accuracy of the calculation.

Our discussion so far is oversimplified in the sense that the double counting
subtraction in \eqref{full-Xsect} only addresses the 1v1 part of the DPS cross
section.  The 1v2 and 2v1 parts of $\sigma_{\text{DPS}}$ also have an overlap,
namely with contributions that involve a twist-four distribution in one of the
colliding protons and a PDF in the other proton.  In principle one should
include these contributions, along with a further double counting subtraction
term, but this would be quite impractical, given the paucity of knowledge about
twist-four distributions and the fact that the corresponding hard-scattering
cross sections have not been computed.  As explained in \sect{4.1} of
\cite{Diehl:2017kgu}, omitting these twist-four contributions while keeping the
1v2 and 2v1 parts of $\sigma_{\text{DPS}}$, one retains the leading logarithmic
part of graphs in which the hard scattering is initiated by two partons in one
proton and one parton in the other proton.  The $\nu$ dependence of the 1v2 and
2v1 contributions to $\sigma_{\text{DPS}}$ goes like $\ln \nu$ and is much
weaker than the quadratic $\nu$ dependence of the 1v1 contribution.

Returning to the construction of $I_{\text{sub}}$, we recall that the integrand
$I_{\text{DPS}}$ is constructed from DPDs
\begin{align}
   \label{DPD-in-sigma-DPS}
   F\biggl(x_1, x_2, y; \mu_1, \mu_2, \frac{M_1 M_2}{x_1 x_2}
      \,\bigg|\, \mu_{\text{init}}(y), \frac{\mu_{\text{init}}^2(y)}{x_1 x_2}
   \biggr)
\end{align}
with $F = F^{\text{spl}} + F^{\text{intr}}$.  Here we use again the notation
introduced in \eqref{DPD-with-scales} that specifies both the initial and final
scales of evolution.
In the colour singlet sector, the construction of $I_{\text{sub}}$ is very
simple if the two hard scales are equal, $\mu_1 = \mu_2 = \mu$ and if one takes
$\mu_h = \mu$ as factorisation scale in the SPS cross section.  In that case,
$I_{\text{sub}}$ is obtained from $I_{\text{DPS}}$ by replacing the DPDs in
\eqref{DPD-in-sigma-DPS} by $F^{\text{spl}}$ evaluated with the splitting
formula at the scale~$\mu$.

For unequal scales $\mu_1 \neq \mu_2$, \sect{6.3} of the original work
\cite{Diehl:2017kgu} outlined a construction in which the splitting DPDs in the
subtraction term have $y$ dependent initial \emph{and} final scales $\mu_1$,
$\mu_2$, and $\mu_0$.  In the present work we present an alternative that we
find both simpler and more flexible.  We also include the case of colour
non-singlet DPDs, which was not considered in \cite{Diehl:2017kgu}.  In our new
scheme, $I_{\text{sub}}$ is constructed from two components $I_{\text{sub,DPS}}$
and $I_{\text{sub,SPS}}$ in the following manner.
\begin{itemize}
\item $I_{\text{sub,DPS}}(y)$ is constructed with DPDs
\begin{align}
   F^{\text{spl}}\biggl(x_1, x_2, y; \mu_1, \mu_2, \frac{M_1 M_2}{x_1 x_2}
      \,\bigg|\, \mu_{\text{sub,init}}, \frac{\mu_{\text{sub,init}}^2}{x_1 x_2}
   \biggr)
\end{align}
instead of those in \eqref{DPD-in-sigma-DPS}, where
\begin{align}
   \label{DPS-sub-init-scale}
   \mu_{\text{sub,init}}
   &=
   \mu_{\text{init}}(\ycut)
   \,.
\end{align}

By construction, $I_{\text{sub,DPS}}(y)$ and $I_{\text{1v1}}(y)$ are equal at $y
= \ycut$ and gradually deviate from each other as $y$ increases.  This holds for
the default choices of $\mu_{\text{init}}$ and $\nu = b_0 / \ycut$ as well as
for the cases where these scales are varied.

Note that for our central choice of initial scale one has
\begin{align}
   \label{DPS-sub-init-approx}
   \mu_{\text{sub,init}}
   &=
   \muy(\ycut)
   \approx
   \nu
   \,,
\end{align}
where the approximation is very accurate for the range of $\nu$ in our numerical
studies, as mentioned after \eqn{\eqref{ycut-vs-nu}}.

If we also take our default choice \eqref{nu-scale} for $\nu$, we get
\begin{align}
   \label{DPS-sub-init-default}
   \mu_{\text{sub,init}}
   & \approx
   \min(\mu_1, \mu_2)
   \,.
\end{align}
For $\mu_1 < \mu_2$ the DPDs in $I_{\text{sub,DPS}}(y)$ are then initialised
with both scales equal to $\mu_1$, after which the second scale is evolved to
$\mu_2$.  There is evolution in $\zeta$ as well.  An analogous statement holds
for $\mu_1 > \mu_2$.  If $\mu_1 = \mu_2$, then the DPDs are essentially not
evolved at all.
\item $I_{\text{sub,SPS}}(y)$ is constructed from DPDs
\begin{align}
   F^{\text{spl}}\biggl(x_1, x_2, y; \mu_h, \mu_h, \frac{M_1 M_2}{x_1 x_2}
      \,\bigg|\, \mu_h, \frac{M_1 M_2}{x_1 x_2} \biggr)
\end{align}
that are not evolved in either the scale or the rapidity parameter.  Here the
factorisation scale $\mu_h$ is the same as the one taken in the SPS
factorisation formula.  We take
\begin{align}
   \label{SPS-scale}
   \mu_h
   &=
   \frac{1}{2} \ms (\mu_1 + \mu_2)
\end{align}
in our numerical studies; alternative choices such as $\mu_h = \sqrt{\mu_1 \,
\mu_2\rule{0pt}{1.5ex}}$ would work as well.
\item The full subtraction term interpolates between these two functions as
\begin{align}
   \label{interpol-I}
   I_{\text{sub}}(y)
   &=
   \bigl[ 1 - \rho(y) \bigr] \, I_{\text{sub,DPS}}(y)
   + \rho(y) \, I_{\text{sub,SPS}}(y)
\end{align}
where $\rho$ is a function that smoothly interpolates between 0 and 1 in the
interval $y \in [r_1 \ms y_\rho, r_2 \ms y_\rho]$ with $r_1 < r_2$ and $y_\rho
\sim \ycut$.  We chose
\begin{align}
   \label{rho-scaled}
   \rho(y)
   &= \tilde{\rho}\biggl(
      \frac{2 y - (r_2 + r_1) \ms y_\rho}{(r_2 - r_1) \ms y_\rho}
      \biggr)
\end{align}
with
\begin{align}
   \label{rho-zero}
   \tilde{\rho}(x)
   &=
   \begin{cases}
      0
      & \text{ for } x \le -1 ,
      \\
      \frac{1}{2} + \frac{1}{2}
         \sin \Bigl( \frac{\pi}{2} \sin \bigl( \frac{\pi}{2} \ms x \bigr) \Bigr)
      & \text{ for } -1 < x < 1 ,
      \\
      1
      & \text{ for } x \ge 1 .
   \end{cases}
\end{align}
$\tilde{\rho}(x)$ has the feature that its first and second derivatives are
continuous and zero at $|x| = 1$.  Alternatively, one may replace $\sin
\frac{\pi}{2} z$ with $\frac{3}{2} z \ms \bigl( 1 - \frac{1}{3} z^2 \bigr)$ for
one or both sine functions in \eqref{rho-zero}.

For the results derived in \sect{\ref{sec:cutoff-dep}} it is required $(i)$ that
$r_2 \ms y_\rho$ is in the perturbative region and $(ii)$ that the size of $r_2$
is limited by the condition $\alpha_s(\nu) \, \ln r_2 \ll 1$.
\end{itemize}

With this new scheme it is rather easy to assess the dependence of results on
the specific form of the interpolation function \eqref{rho-zero}, because a
change of that function does not require re-computation of DPDs.  This is not
the case for the construction in \cite{Diehl:2017kgu}, where the interpolation
between the SPS and DPS regimes is achieved at the level of the initial and
final scales of the splitting DPDs.

In the case $\mu_1 = \mu_2$ our default choice of $\nu$ gives
\mbox{$\mu_{\text{sub,init}} \approx \mu_1 = \mu_h$} according to
\eqref{DPS-sub-init-default} and \eqref{SPS-scale}.  In the colour singlet
sector, one then has $I_{\text{sub}} \approx I_{\text{sub,DPS}} \approx
I_{\text{sub,SPS}}$ to high accuracy and recovers our simple original
construction for equal scales.  The same holds for colour non-singlet channels
if one additionally takes $\mu_i = M_i$.  In these cases, the choice of the
interpolation function $\rho(y)$ has no impact on the overall double counting
subtraction.  It does, however, determine the individual terms in the sum
\eqref{sub-lumis} below.

As will be discussed in the following sections, the DPD splitting formula should
be evaluated for a number $n_f$ of active quark flavours that depends on $y$,
possibly with the inclusion of heavy quark masses.  The logic of the subtraction
mechanism then requires that
\begin{itemize}
\item $I_{\text{sub,DPS}}$ should be evaluated with the flavour settings used
for DPDs at $y = \ycut$,
\item the value of $n_f$ and the treatment of heavy quark masses in
$I_{\text{sub,SPS}}$ should be the same as in the computation of the SPS cross
section.
\end{itemize}

Since both the DPS cross section and the double counting subtraction involve
products of DPDs, their relative size and their combination can be discussed at
the level of double parton luminosities, which are independent of specific
hard-scattering cross sections.  The counterpart of the double parton luminosity
$\lum$ that goes into $\sigma_{\text{DPS}}$ is then the double parton luminosity
$\lum_{\text{sub}}$ from which $\sigma_{\text{sub}}$ is constructed. Separating
the two terms in the  interpolation prescription \eqref{interpol-I}, we can
further write
\begin{align}
   \label{sub-lumis}
   \lum_{\text{sub}}(\nu)
   &=
   \lum_{\text{sub,DPS}}(\nu) + \lum_{\text{sub,SPS}}(\nu)
   \,,
\end{align}
where the first term corresponds to the $y$-integral of $(1 - \rho)\,
I_{\text{sub,DPS}}$ and the second one to the $y$-integral of $\rho\,
I_{\text{sub,SPS}}$.  Both terms involve the product of two DPDs without
hard-scattering cross sections.

%%%%%%%%%%%%%%%%%%%%%%%%%%%%%%%%%%%%%%%%

\subsection{Dependence on the cutoff scale}
\label{sec:cutoff-dep}

Let us now show to which formal accuracy the $\nu$ dependence cancels in the
overall cross section \eqref{full-Xsect} with our new construction.
As in \sect{\ref{sec:init-scale}}, we assume that splitting DPDs are computed up
to order $\alpha_s^k$ and that evolution is performed at order $\alpha_s^{k-1}$
or higher.
As a compact notation let us write
\begin{align}
   P(y; \mu)
   &=
   2\pi y\,
   F^{\text{spl}\ms (k)}\biggl(x_1, x_2, y;
      \mu_1, \mu_2, \frac{M_1 M_2}{x_1 x_2}
      \,\bigg|\, \mu, \frac{\mu^2}{x_1 x_2} \biggr)
   \notag \\
   &\quad \times
   F^{\text{spl}\ms (k)}\biggl(\bar{x}_1, \bar{x}_2, y;
      \mu_1, \mu_2, \frac{M_1 M_2}{\bar{x}_1 \bar{x}_2}
      \,\bigg|\, \mu, \frac{\mu^2}{\bar{x}_1 \bar{x}_2} \biggr)
   \,,
\end{align}
where for brevity we suppress parton and colour labels, as well as the
dependence of $P$ on all arguments that are left constant in the following
derivation.  We then have
\begin{align}
   \lum_{\text{1v1}}(\nu)
   &=
   \int_{b_0 / \nu}^{\infty} d y\,
      P\bigl( y; \mu_{\text{init}}(y) \bigr)
   \,,
\notag \\
   \lum_{\text{sub,DPS}}(\nu)
   &=
   \int_{b_0 / \nu}^{\infty} d y\, \bigl[ 1 - \rho(y) \bigr] \,
      P\bigl( y; \mu_{\text{init}}(b_0 / \nu) \bigr)
   \,.
\end{align}

We now study the difference of subtracted 1v1 luminosities at two cutoff scales
\mbox{$\nu_1 \sim \nu_2$}.  We take the \emph{same} interpolating function
$\rho(y)$ and require that the parameter $y_{\rho}$ in \eqref{rho-scaled}
satisfies $y_{\rho} \sim 1 / \nu_1 \sim 1 / \nu_2$ and $r_1\ms y_{\rho} \ge
\max(b_0 / \nu_1, b_0 / \nu_2)$.  The second condition ensures that $\rho(y) =
0$ for $y$ between $b_0 / \nu_1$ and $b_0 / \nu_2$.  We then have
\begin{align}
   \label{lumi-nu-dep}
   &
   \bigl[ \lum_{\text{1v1}}(\nu_1) - \lum_{\text{sub}}(\nu_1) \bigr]
   - \bigl[ \lum_{\text{1v1}}(\nu_2) - \lum_{\text{sub}}(\nu_2) \bigr]
   \notag \\[0.2em]
   &\quad =
   \int_{b_0/ \nu_1}^{\infty} d y\;
      \Bigl\{ P(y, \mu_{\text{init}}(y))
         - \bigl[ 1 - \rho(y) \bigr] \,
            P( y; \mu_{\text{init}}(b_0 / \nu_1) \bigr) \Bigr\}
   \notag \\
   &\qquad
   - \int_{b_0/ \nu_2}^{\infty} d y\;
      \Bigl\{ P(y, \mu_{\text{init}}(y))
         - \bigl[ 1 - \rho(y) \bigr] \,
            P( y; \mu_{\text{init}}(b_0 / \nu_2) \bigr) \Bigr\}
   \notag \\
   &\quad =
   \int_{b_0/ \nu_1}^{b_0/ \nu_2} d y\;
      \Bigl\{ P(y, \mu_{\text{init}}(y))
            - P( y; \mu_{\text{init}}(b_0 / \nu_1) \bigr) \Bigr\}
   \notag \\
   &\qquad
   + \int_{b_0/ \nu_2}^{\infty} d y\, \bigl[ 1 - \rho(y) \bigr] \;
      \Bigl\{ P( y; \mu_{\text{init}}(b_0 / \nu_2) \bigr)
            - P( y; \mu_{\text{init}}(b_0 / \nu_1) \bigr) \Bigr\}
   \,.
\end{align}
Note that the contribution from $\lum_{\text{sub,SPS}}$ has dropped out because
it is independent of the cutoff parameter $\nu$.  Because of the factor
$1 - \rho(y)$, the last integral in \eqref{lumi-nu-dep} is limited to $y \le
r_2\ms y_{\rho}$, and the integrand is strongly damped in the vicinity of its
upper limit. Under the conditions stated after \eqn{\eqref{rho-zero}}, we can
apply the result \eqref{DPD-resid-scale-dep} to both integrals on the r.h.s.\ of
\eqref{lumi-nu-dep}.  This gives
\begin{align}
  \label{lumi-nu-dep-final}
   \bigl[ \lum_{\text{1v1}}(\nu_1) - \lum_{\text{sub}}(\nu_1) \bigr]
   - \bigl[ \lum_{\text{1v1}}(\nu_2) - \lum_{\text{sub}}(\nu_2) \bigr]
   &=
   \mathcal{O}\bigl( \alpha_s^{k+2}(\nu_1) \bigr)
   \,,
\end{align}
where the power of $\alpha_s$ results from the fact that a splitting DPD starts
at order $\alpha_s$.  The product of two splitting DPDs computed at order $k$
contains significant terms from order $\alpha_s^2$ up to order $\alpha_s^{k+1}$,
so that the difference in \eqref{lumi-nu-dep-final} is beyond the accuracy of
the computation, as it should be.

%%%%%%%%%%%%%%%%%%%%%%%%%%%%%%%%%%%%%%%%

\subsection{Summary of the subtraction formalism}
\label{sec:sub-summary}

At this point we can review the overall logic of our subtraction formalism.
\begin{itemize}
\item For $y \sim \ycut \sim 1 / \min(\mu_1, \mu_2)$, the DPS formalism is not
valid, and the correct description of the process is given by SPS alone.  In the
region $y \le r_1 y_\rho$, the subtraction term we constructed removes the 1v1
part of DPS up to contributions that are at least one order of $\alpha_s$ higher
than the accuracy of the double parton luminosities.  These contributions may be
regarded as an artefact of our procedure, and their size can be estimated by
varying $\ycut$ in $\sigma_{\text{1v1}} - \sigma_{\text{sub}}$ around a suitably
chosen central value.
\item For $y \gg \ycut$, the DPS formalism is valid.  In the region where $y$ is
still perturbative, the 1v1 contribution is fully calculable in terms of PDFs
and DPD splitting kernels, and it correctly resums large logarithms of $y \mu_1$
and $y \mu_2$ (as well as large Sudakov double logarithms in colour non-singlet
channels) that are missed when SPS is evaluated at fixed order.  In the region
$y \ge r_2 \ms y_\rho$, our subtraction term removes the contribution of SPS
graphs that have DPS topology, up to power corrections in $\ycut / y$ that are
due to applying the DPS approximation to the graphs.
\item In the region $r_1 \ms y_\rho \le y \le r_2 \ms y_\rho$, our subtraction
term smoothly interpolates between the two cases just described.  As is typical
of the interpolation between two regimes with different approximations, there is
no simple parametric estimate of the error incurred by this procedure.  This
error can be estimated by varying the parameters of the interpolating function
(i.e.\ $r_1$ and $r_2$ in our case) within appropriate limits.
\end{itemize}

We recall from the discussion above \eqn{\eqref{DPD-in-sigma-DPS}} that our
master formula \eqref{full-Xsect} misses explicit contributions with a
twist-four distribution in one proton and a PDF in the other, but that the 1v2
and 2v1 parts of DPS include the part of these contributions that is enhanced by
$\ln(\nu / \Lambda)$.  The variation of the 1v2 and 2v1 parts of DPS with
$\ycut$ around an appropriate central value may hence be regarded as an estimate
of the uncertainty of this specific leading logarithmic approximation.  The
variation of the 2v2 part of DPS with $\ycut$ is negligible, because this
contribution is dominated by a broad range of $y$ values in the non-perturbative
region.

Overall, the variation of $\sigma_{\text{DPS}} - \sigma_{\text{sub}}$ or
$\lum_{\text{DPS}} - \lum_{\text{sub}}$ with $\ycut$ may thus be regarded as a
measure for the theoretical uncertainty of our scheme for handling the overlap
between DPS and other contributions to a given physical process.

%%%%%%%%%%%%%%%%%%%%%%%%%%%%%%%%%%%%%%%%%%%%%%%%%%%%%%%%%%%%%%%%%%%%%%%%%%%%%%%%

\section{Splitting with massless quarks}
\label{sec:massless}

In this section, we study DPD splitting without heavy quark masses.  We use the
general perturbative expansion \eqref{split-master} with splitting kernels up to
NLO:
\begin{align}
   \label{split-expand}
   \prn{R_1 R_2}{V}_{a_1 a_2, a_0}(z_1, z_2, \alpha_s(\mu), L_y, L_\zeta)
   &
   =
   \frac{\alpha_s(\mu)}{2\pi} \;
   \prn{R_1 R_2}{V}^{(1)}_{a_1 a_2, a_0}(z_1, z_2)
   \notag \\
   & \quad
   +
   \biggl( \frac{\alpha_s(\mu)}{2\pi} \biggr)^{2} \;
   \prn{R_1 R_2}{V}^{(2)}_{a_1 a_2, a_0}(z_1, z_2, L_y, L_\zeta)
   + \mathcal{O}(\alpha_s^3)
   \,.
\end{align}
The lowest-order splitting kernels are
\begin{align}
   \label{split-LO}
   \prn{R_1 R_2}{V}^{(1)}_{a_1 a_2, a_0}(z_1, z_2)
   &=
   c_{a_1 a_2, a_0}(R_1 R_2) \;
   \delta(1 - z_1 - z_2) \,
   P^{(0)}_{a_1 a_0}(z_1)
   \,,
\end{align}
where $P^{(0)}_{a_1 a_0}(z)$ is the leading-order DGLAP splitting function for
PDFs without its distributional part at $z=1$.  The colour factors $c_{a_1 a_2,
a_0}(R_1 R_2)$ can for instance be found in \sect{4} of \cite{Diehl:2021wpp}.
The NLO splitting kernels have the form
\begin{align}
   \label{split-NLO-logs}
   \prn{R_1 R_2}{V}^{(2)}_{a_1 a_2, a_0}(z_1, z_2, L_y, L_\zeta)
   &
   =
   \prn{R_1 R_2}{V}^{[2, 0]}_{a_1 a_2, a_0}(z_1, z_2)
   + L_y \, \prn{R_1 R_2}{V}^{[2, 1]}_{a_1 a_2, a_0}(z_1, z_2)
   \notag \\
   & \quad
   + \frac{\prn{R_1}{\gamma}_J^{(0)}}{2} \,
     \biggl( L_y L_\zeta - \frac{L_y^2}{2} - \frac{\pi^2}{12} \biggr) \,
     \prn{R_1 R_2}{V}^{(1)}_{a_1 a_2, a_0}(z_1, z_2)
   \,,
\end{align}
where
\begin{align}
   \prn{1}{\gamma}_J^{(0)}
   &= 0 \,,
   &
   \prn{8}{\gamma}_J^{(0)} = \prn{S}{\gamma}_J^{(0)} = \prn{A}{\gamma}_J^{(0)}
   &= 2 C_A
\end{align}
are the LO coefficients of the anomalous dimension for rapidity evolution.

The number $n_f$ of active quark flavours for both the DPDs and the PDFs in the
splitting formula \eqref{split-master} is taken according to the characteristic
scale
\begin{align}
   \label{mu-y-def}
   \mu_y
   &=
   b_0 / y
\end{align}
of the splitting process.  We include a heavy quark $Q$ in the active flavours
if $\mu_y > \gamma m_Q$ with a parameter $\gamma \sim 1$.  Specifically, we take
\begin{align}
   \label{nf-choice}
   n_f
   &=
   \begin{cases}
      3 & \text{ for } \mu_y < \gamma \ms m_c \,,
      \\
      4 & \text{ for } \gamma m_c < \mu_y < \gamma \ms m_b \,,
      \\
      5 & \text{ for } \gamma \ms m_b < \mu_y < \gamma \ms m_t \,,
      \\
      6 & \text{ for } \gamma \ms m_t < \mu_y \,.
   \end{cases}
\end{align}
In the present section we set $\gamma=1$; different values will be explored in
\sect{\ref{sec:massive-lumis}}.  We follow here the procedure of our earlier
work \cite{Diehl:2022dia} and take the scale $\mu_y$ as criterion for
determining $n_f$.  One could alternatively take the scale $\muy(y)$ from
\eqref{mu-of-y}, in which case the lowest number of active flavours for
DPDs would be $n_f = 4$ because $\muy(y) > \mu_{\text{min}} = 2 \gev$.

Starting from initial conditions with $n_f$ from \eqref{nf-choice}, DPDs with
additional active flavours are obtained by matching, adding the heavy quark $Q$
at $\mu_1 = m_Q$ for the first parton and at $\mu_2 = m_Q$ for the second one.
The matching proceeds in the same way as for PDFs and is described in \sect{2.2}
of \cite{Diehl:2022dia} for DPDs in the colour singlet channel.  The one-loop
flavour matching kernels for colour octet DPDs are related to their colour
singlet analogues as
\begin{align}
   \label{gg-matching}
   \pr{A A}{A}^{Q(1)}_{g g} = \pr{S S}{A}^{Q(1)}_{g g}
   &=
   \prb{1 1}{A}^{Q(1)}_{g g}
   \,,
   &
   \pr{A S}{A}^{Q(1)}_{g g} = \pr{S A}{A}^{Q(1)}_{g g}
   &=
   0
\end{align}
for gluon-gluon transitions, where the relevant graphs are purely virtual, and
as
\begin{align}
   \pr{8 R}{A}^{Q(1)}_{Q g}
   &=
   c_{q g}(8 R) \; \prb{1 1}{A}^{Q(1)}_{Q g}
   &&
   \text{ with } R = A, S
\end{align}
for the transition from a gluon to a heavy quark, where the relevant graphs are
purely real and involve the same colour factors
\begin{align}
   c_{q g}(8 A)
   &=
   \sqrt{\frac{N_c^2}{2 (N_c^2 - 1)}}
   \,,
   &
   c_{q g}(8 S)
   &=
   \sqrt{\frac{N_c^2 - 4}{2 (N_c^2 - 1)}}
\end{align}
as the leading-order DGLAP kernels $\pr{8 R}{P}^{(0)}_{q g}$ \cite{Diehl:2011yj,
Diehl:2022rxb}.

%%%%%%%%%%%%%%%%%%%%%%%%%%%%%%%%%%%%%%%%

\subsection{Parton kinematics and scales}
\label{sec:parton-kin}

In the following numerical studies we consider DPDs and DPD luminosities for
$x_i$ and $\bar{x_i}$ evaluated from \eqref{x-fractions} with
\begin{align}
   \sqrt{s} &= 14 \tev
   \,,
   &
   Y_1 &= - Y_2 = Y
   \,,
\end{align}
where $Y \in [-4, 4]$.  For the invariant masses of the produced systems, we
take one of the two combinations
\begin{align}
   \label{equal-scales}
   M_1 &= M_2 = 80 \gev
\end{align}
or
\begin{align}
   \label{diff-scales}
   M_1 &= 80 \gev
   \,,
   \qquad
   M_2 = 10 \gev
   \,.
\end{align}
An invariant mass of $80 \gev$ is for instance relevant for the production of a
$W$ or a dijet.  An invariant mass of $10 \gev$ could be a typical scale for
producing a pair of mini-jets, or for a $J/\Psi$ with some transverse momentum.
The corresponding parton momentum fractions are
\begin{align}
   \label{x-central}
   x_1
   =
   \bar{x}_1
   &\approx
   \begin{cases}
      7.1 \times 10^{-4}  & \text{ for $M_1 = 10 \gev$} \\
      5.7 \times 10^{-3}  & \text{ for $M_1 = 80 \gev$}
   \end{cases}
   &
   \text{ if } Y_1 = 0
\end{align}
and
\begin{align}
   \label{x-forward}
   x_1
   &\approx
   \begin{cases}
      3.9 \times 10^{-2} \\
      3.1 \times 10^{-1}
   \end{cases}
   \quad
   \bar{x}_1
   \approx
   \begin{cases}
      1.3 \times 10^{-5}     & \text{ for $M_1 = 10 \gev$} \\
      1.0 \times 10^{-4}     & \text{ for $M_1 = 80 \gev$}
   \end{cases}
   &
   \text{ if } Y_1 = 4
   \,,
   \end{align}
with corresponding values for $x_2$ and $\bar{x}_2$ obtained by symmetry
considerations.

The factorisation scales in the double parton luminosities are always taken as
\begin{align}
   \mu_i &= M_i
   \,,
   &&
   i = 1,2
\end{align}
and the rapidity parameters as specified in \eqref{zeta-kin-choice}.  We will
not vary the scales $\mu_i$, which would be more instructive when DPDs are
multiplied by parton-level cross sections evaluated at the same scales.

%%%%%%%%%%%%%%%%%%%%%%%%%%%%%%%%%%%%%%%%

\subsection{Splitting DPDs}
\label{sec:massless-dpds}

The DPD splitting formula \eqref{split-master} is valid at small $y$, whereas
DPD luminosities involve an integral over $y$ up to infinity. Clearly, DPDs at
large $y$ are outside the reach of perturbative computations, and we need to
model them.  Following our strategy in earlier work \cite{Diehl:2017kgu,
Diehl:2020xyg} we use the decomposition $F = F^{\text{spl}} + F^{\text{intr}}$
at all values of $y$, modelling both terms when $y$ is large.  For the splitting
part, we do so by multiplying the perturbative form $F^{\text{spl,\,pt}}$ in
\eqref{split-master} with a Gaussian factor:
\begin{align}
   \label{splitting-DPD}
   &
   \prn{R_1 R_2}{F}_{a_1 a_2}^{\text{spl}}(x_1, x_2, {y};
      \mu_{\text{init}}, \mu_{\text{init}}, \zeta_{\text{init}})
\notag \\
   & \qquad =
   \exp \biggl[ \frac{-y^2}{4 h_{a_1 a_2}} \biggr] \;
   \prn{R_1 R_2}{F}_{a_1 a_2}^{\text{spl,\,pt}}(x_1, x_2, {y};
      \mu_{\text{init}}, \mu_{\text{init}}, \zeta_{\text{init}})
   \,,
\end{align}
where $\zeta_{\text{init}}$ is given in \eqref{zeta-init} and the parameters are
taken from \sect{9.2.1} of \cite{Diehl:2017kgu}:
\begin{align}
   \label{damping-parameters}
   h_{g g} = 4.66 \gev^{-2}
   \,,
   && h_{g q} = h_{q g} = 5.86 \gev^{-2}
   \,,
   && h_{q q} = 7.06 \gev^{-2}
   \,,
\end{align}
with equal values for quarks and antiquarks.

For the PDFs in the splitting formula we use the default LO or NLO sets of the
MSHT20 fit \cite{Bailey:2020ooq}, depending on the order at which we evolve the
DPDs.  The strong coupling used in these sets is
\begin{align}
   \label{as-lo}
   \smash{\alpha_s^{(n_f=5)}(m_Z)} &= 0.13
   &&
   \text { at LO}
   \intertext{and}
   \label{as-nlo}
   \smash{\alpha_s^{(n_f=5)}(m_Z)} &= 0.118
   &&
   \text{ at NLO.}
\end{align}
The PDFs are initialised at $\mu_0 = 1 \gev$ from the LHAPDF interface
\cite{Buckley:2014ana}, where the respective sets have the keys
\texttt{MSHT20lo\_as130} and \texttt{MSHT20nlo\_as118}.  The quark masses are
\begin{align}
   \label{quark-masses}
   m_c &= 1.4 \gev
   \,,
   &
   m_b &= 4.75 \gev
\end{align}
for both sets.

%%%%%%%%%%%%%%%%%%%%%%%%%%%%%%%%%%%%%%%%

\begin{figure}
\centering
\subfloat[$g g$, colour singlet]{
   \includegraphics[height=0.35\textwidth]{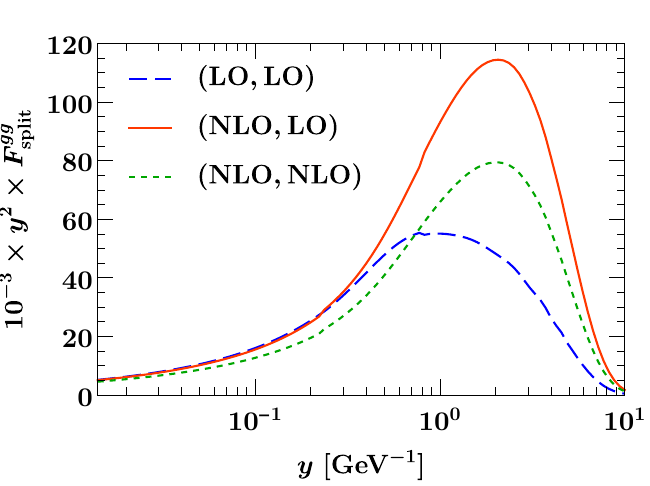}
}
\hfill
\subfloat[$g g$, $A A$]{
   \includegraphics[height=0.35\textwidth]{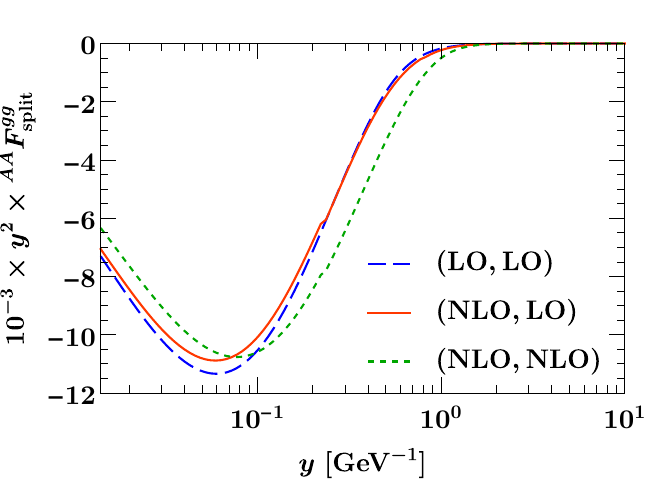}
}
\\[1.5em]
\subfloat[$u g$, colour singlet]{
   \includegraphics[height=0.35\textwidth]{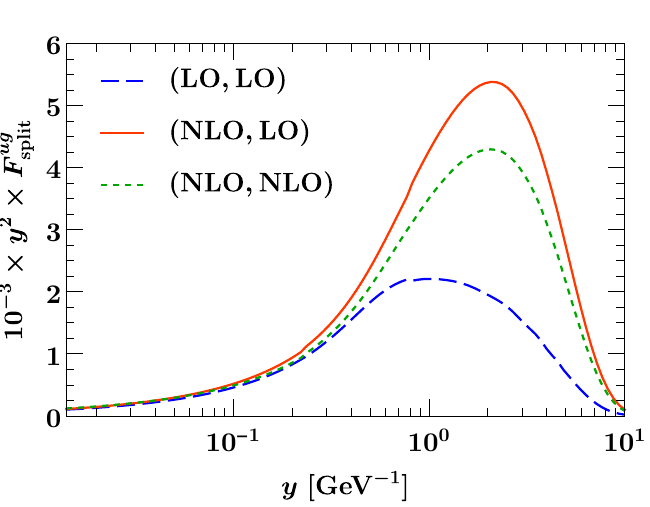}
}
\hfill
\subfloat[$u g$, $8 A$]{
   \includegraphics[height=0.35\textwidth]{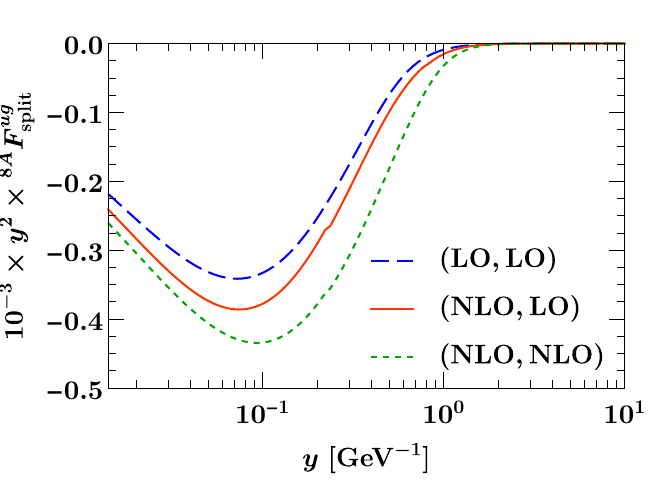}
}
\caption{\label{fig:dpds-g} Splitting DPDs for $g g$ and $u g$ at $x_1 = x_2
\approx 5.7 \times 10^{-3}$, evolved to scales $\mu_1 = \mu_2 = 80 \gev$.  The
first specification of ``LO'' or ``NLO'' in the plots refers to the DPD
splitting kernels, and the second one refers to the order of DGLAP evolution of
the DPDs and of the  PDFs in the splitting formula.}
\end{figure}

\begin{figure}
\centering
\subfloat[$u \bar{d}$, colour singlet]{
   \includegraphics[height=0.35\textwidth]{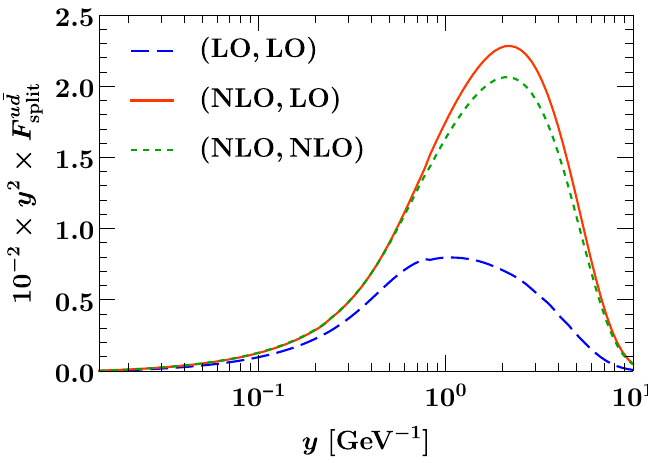}
}
\subfloat[$u \bar{d}$, colour octet]{
   \includegraphics[height=0.35\textwidth]{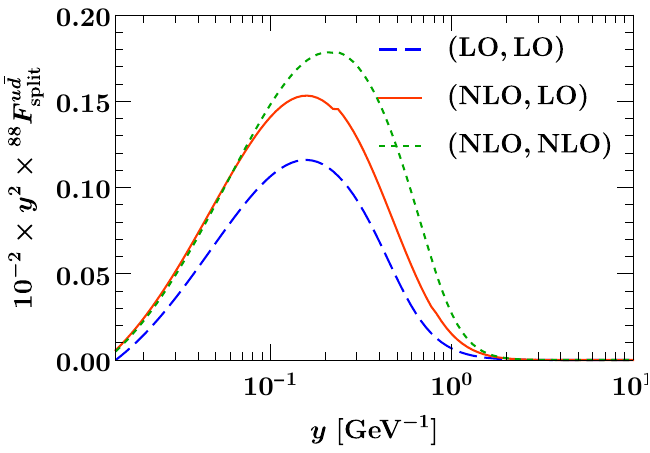}
}
\\[1.5em]
\subfloat[\label{fig:dpds-uubar} $u \bar{u}$, colour singlet, $x_1 = x_2$]{
   \includegraphics[height=0.35\textwidth]{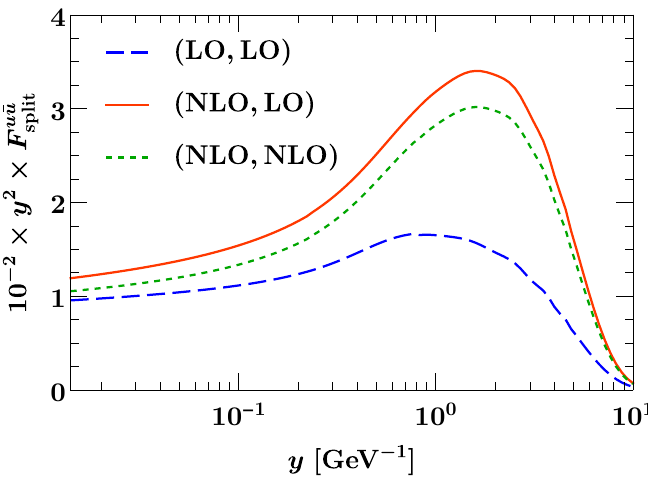}
}
\subfloat[$u \bar{u}$, colour octet, $x_1 = x_2$]{
   \includegraphics[height=0.35\textwidth]{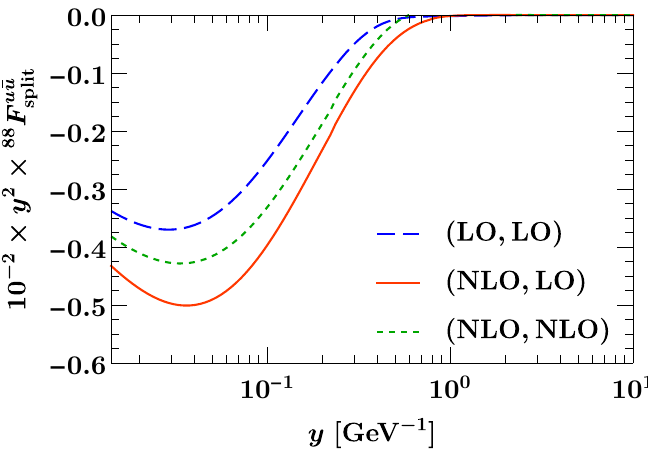}
}
\caption{\label{fig:dpds-q} As \fig{\protect\ref{fig:dpds-g}}, but for $u
\bar{d}$ and $u \bar{u}$.}
\end{figure}

\begin{figure}
\centering
\subfloat[\label{fig:dps-uubar-asy} $u \bar{u}$, colour singlet, $x_1 \gg x_2$]{
   \includegraphics[height=0.35\textwidth]
   {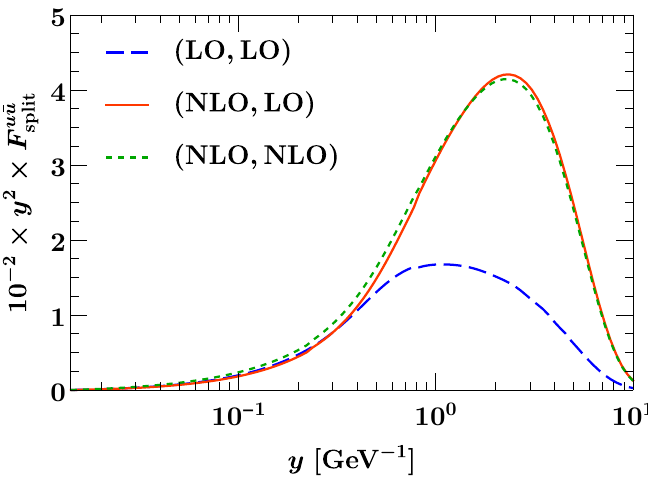}
}
\subfloat[$u \bar{u}$, colour octet, $x_1 \gg x_2$]{
   \includegraphics[height=0.35\textwidth]{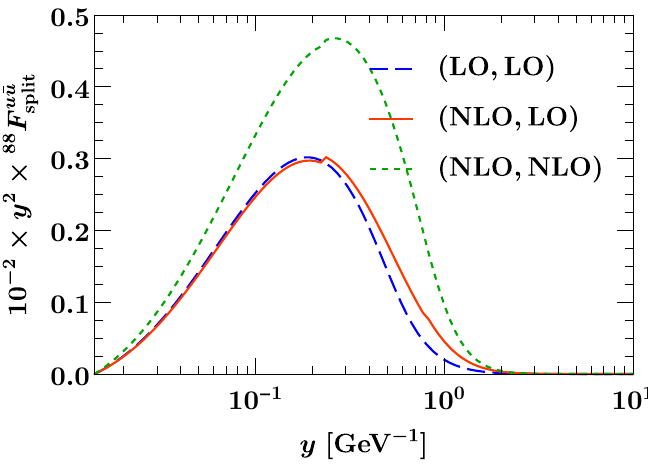}
}
\caption{\label{fig:dpds-q-asy} As the bottom row of
\fig{\protect\ref{fig:dpds-q}}, but for asymmetric momentum fractions $x_1
\approx 0.31$ and $x_2 \approx 10^{-4}$.}
\end{figure}

\paragraph{Comparison of different perturbative orders.}
In \figs{\ref{fig:dpds-g}} and \ref{fig:dpds-q} we show DPDs for the kinematic
setting with $M_1 = M_2 = 80 \gev$ and $Y=0$, which gives $x_1 = x_2 \approx 5.7
\times 10^{-3}$ according to \eqref{x-central}.  Figure~\ref{fig:dpds-q-asy} is
for DPDs at $M_1 = M_2 = 80 \gev$ and $Y = 4$, which gives $x_1 \approx 0.31$
and $x_2 \approx 10^{-4}$ according to \eqref{x-forward}.  All DPDs are evolved
from $\mu_{\text{init}} = \muy(y)$ to $\mu_i = M_i$ and from
$\zeta_{\text{init}}$ in \eqref{zeta-init} to $\zeta$ in
\eqref{zeta-kin-choice}.

The lowest $y$ value in the plots is $y_{\text{min}} = b_0 / \mu_1$.  At this
point one has $\muy(y) \approx \mu_1$, so that there is essentially no evolution
of the DPD after its initialisation.  As $y$ becomes bigger than
$y_{\text{min}}$, the amount of evolution to higher scales increases.
The DPDs in the figures are multiplied with $y^2$, so as to compensate the
$1/y^2$ behaviour of splitting DPDs at the scale where they are initialised.

We observe that in the colour singlet channel, the scaled DPDs grow
significantly with $y$ as a consequence of evolution, which was already observed
in \cite{Diehl:2017kgu}.  A notable exception are $q \bar{q}$ distributions: we
find strong evolution effects in the $y$ dependence for asymmetric momentum
fractions (\fig{\ref{fig:dps-uubar-asy}}) but only mild ones in our setting with
$x_1 = x_2$ (\fig{\ref{fig:dpds-uubar}}).
With the global factor $1/y^2$ included, $q\bar{q}$ colour singlet DPDs in this
setting are hence more strongly concentrated at $y \sim \ycut$ than all other
parton combinations.  This has important consequences for double parton
luminosities at central rapidities, as we shall see.

Scaled colour octet DPDs behave very differently: at small $y$ they are
similar in magnitude to their colour singlet counterparts, whilst at large $y$ they
are strongly suppressed due to Collins-Soper evolution.  Our study in
\cite{Diehl:2023jje} showed that this conclusion does not depend on the specific
model for the non-perturbative part of the Collins-Soper kernel.  For gluons in
the decuplet or the $27$ representation (not shown here), the suppression at
large $y$ is even stronger.

We now turn to the role of higher perturbative orders.  The difference between
the solid red and dashed blue curves in the figures shows the impact of using
DPD splitting kernels at NLO instead of LO, with LO PDFs in the splitting
formula in both cases:
\begin{itemize}
\item We find large differences at intermediate or large $y$ for colour singlet
DPDs, with the NLO result being bigger than the one at LO for all parton
combinations considered.
\item For colour octet distributions at $y$ below $1 \gev^{-1}$, the impact of
NLO corrections to the DPD splitting is clearly visible.  At larger $y$ these
distributions are negligibly small.
\end{itemize}

The difference between the solid red and dotted green lines in the figures shows
the impact of switching from LO to NLO DGLAP evolution for DPDs and PDFs (along
with changing the PDF set).  In most cases, this impact is significant but more
moderate than the change from LO to NLO in DPD splitting.

In the remainder of this paper we will focus on the impact of the NLO kernels
for DPD splitting.  We will therefore always evolve DPDs and PDFs at LO, even
when the DPD splitting is computed at NLO.  This may appear unusual, but we
recall from \sect{\ref{sec:init-scale}} that the combination of NLO splitting
and LO evolution kernels is theoretically consistent.
We also note that evolving DPDs at LO is adequate if one uses the DPS formula
\eqref{dps-Xsect} with LO hard-scattering cross sections $\hat{\sigma}$.  If the
latter are evaluated at NLO, one gets convolution integrals rather than a simple
multiplication of $\hat{\sigma}$ with double parton luminosities.

%%%%%%%%%%%%%%%%%%%%%%%%%%%%%%%%%%%%%%%%

\subsection{Double parton luminosities and subtraction terms}
\label{sec:massless-lumis}

To investigate full double parton luminosities, we need not only the 1v1 but
also the 1v2, 2v1, and 2v2 contributions and therefore require a model for the
intrinsic part of the DPDs.  In the colour singlet case we adopt our previous
choice from \cite{Diehl:2023jje}, which is
\begin{align}
   \label{int-DPD-singlet}
   &
   \prn{11}{F}_{a_1 a_2}^{\text{intr}}(x_1, x_2, {y};
      \mu_{\text{init}}, \mu_{\text{init}})
   \notag \\
   & \qquad =
   n_{a_1 a_2} \;
   \frac{1}{4 \pi h_{a_1 a_2}}
   \exp \biggl[ \frac{-y^2}{4 h_{a_1 a_2}} \biggr] \;
   \frac{(1 - x_1 - x_2)^2}{(1 - x_1)^2 \ms (1 - x_2)^2} \;
   f_{a_1}(x_1; \mu_{\text{init}}) \, f_{a_2}(x_2; \mu_{\text{init}})
\end{align}
with
\begin{align}
   n_{a_1 a_2}
   &=
   \begin{cases}
      0   & \text{ if } a_1 = a_2 = d - \bar{d}, \\
      1/2 & \text{ if } a_1 = a_2 = u - \bar{u}, \\
      1   & \text{ otherwise. }
   \end{cases}
\end{align}
Here the parton labels $a_1$ and $a_2$ have been transformed to the valence-sea
basis, i.e.\ to the linear combinations $q - \bar{q}$ and $\bar{q}$ for each
quark flavour.  The parameters $h_{a_1 a_2}$ are the same as in
\eqref{damping-parameters}.

In \cite{Diehl:2023jje} we also explored models for $F_{\text{intr}}$ in colour
non-singlet channels.  We found that the resulting double parton luminosities
for the 1v2, 2v1, and 2v2 combinations are always considerably smaller than
those for the pure splitting combination 1v1.  This is easily understood:
rapidity evolution strongly suppresses large $y$ in the luminosity integral,
and in the remaining region of small $y$ the splitting part of DPDs is much
larger than their intrinsic part.  We will hence not include the latter in the
present study, also because modelling intrinsic colour non-singlet DPDs is
poorly constrained and thus affected by huge uncertainties.

In the colour singlet channel there is no Collins-Soper suppression at large
$y$, and depending on parton combinations and kinematic settings, both the
intrinsic and the splitting parts of DPDs can be important in the overall double
parton luminosities.  We will see examples for this in the following.

%%%%%%%%%%%%%%%%%%%%%%%%%%%%%%%%%%%%%%%%

\paragraph{Impact of the initial scale.}

Let us take a first look at double parton luminosities and their dependence on
the initial scale $\mu_{\text{init}}$ at which DPDs are computed from the
splitting formula \eqref{splitting-DPD} or from the model
\eqref{int-DPD-singlet} for the intrinsic part of DPDs.

In the remainder of this paper we use the following shorthand notation for
colour representations in double parton luminosities:
\begin{align}
   \label{colour-short}
   \pr{11, 11}{\lum}
   & \to
   \lum
   \,,
   &
   \pr{88, 88}{\lum}
   & \to
   \pr{88}{\lum}
   \,,
   &
   \pr{8A, 8A}{\lum}
   & \to
   \pr{8A}{\lum}
   \,,
   &
   \pr{AA, AA}{\lum}
   & \to
   \pr{AA}{\lum}
   \,.
\end{align}
Note in particular that we simply omit colour labels to designate the colour
singlet.

\begin{figure}
\centering
\subfloat[$u g, \bar{d} g$, 1v1]{
   \includegraphics[width=0.48\textwidth]{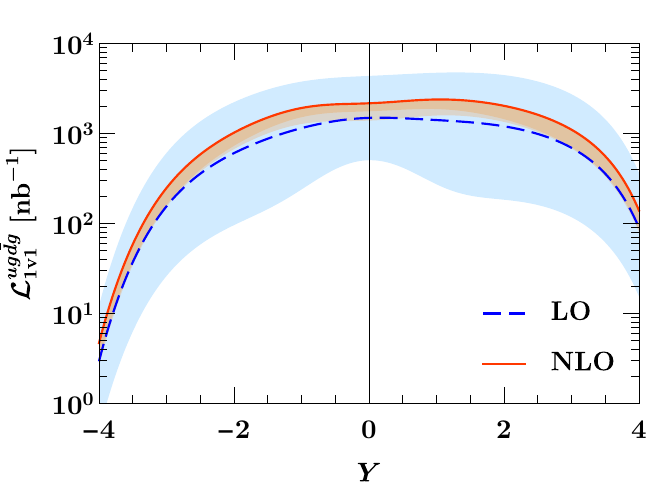}
}
\subfloat[$u g, \bar{d} g$, $8 A, 8 A$, 1v1]{
   \includegraphics[width=0.48\textwidth]{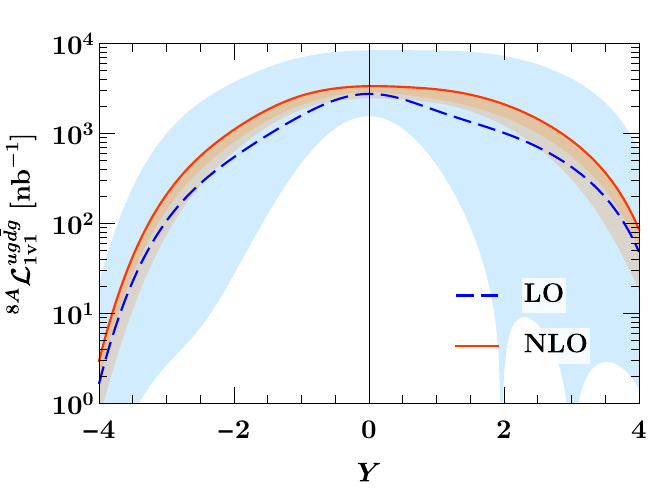}
}
\\[1.5em]
\subfloat[$u g, \bar{d} g$, 1v2+2v1]{
   \includegraphics[width=0.48\textwidth]{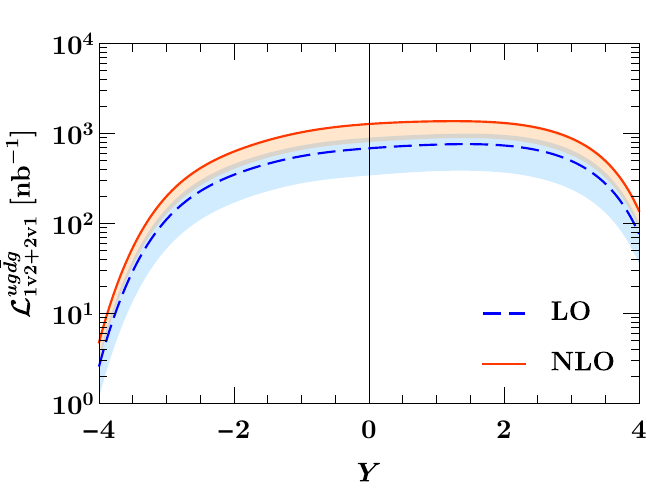}
}
\subfloat[$u g, \bar{d} g$, 2v2]{
   \includegraphics[width=0.48\textwidth]{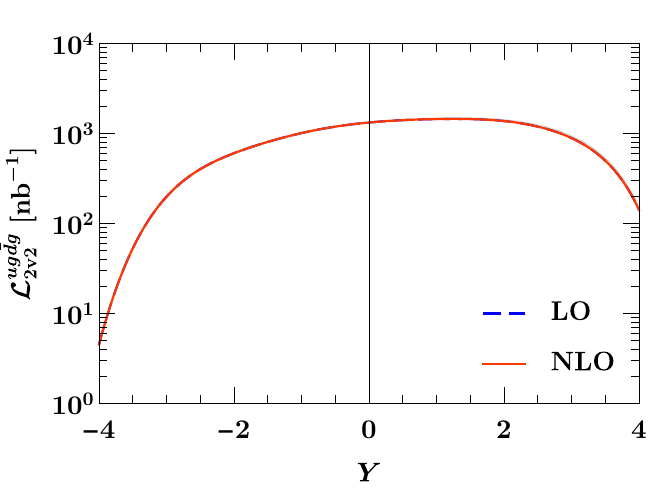}
}
\caption{\label{fig:lumis-g} Double parton luminosities for the product of DPDs
$F_{u g} \, F_{\smash{\bar{d}} g}$ with parton kinematics and scales set for
$M_1 = M_2 = 80 \gev$ as specified in \sect{\protect\ref{sec:parton-kin}}.  The
simplified notation \protect\eqref{colour-short} is used for colour
representations.  Bands correspond to a variation of the initial scale
$\mu_{\text{init}}$ of the DPDs by a factor 2 as specified in
\protect\eqref{mu-init-variation}.  Here and in all following plots, the labels
``LO'' and ``NLO'' indicate the highest order of the DPD splitting kernels
included in the calculation; with evolution being carried out at LO.}
\end{figure}

\begin{figure}[t]
\centering
\subfloat[$u \bar{d}, \bar{d} u$, 1v1]{
   \includegraphics[width=0.48\textwidth]{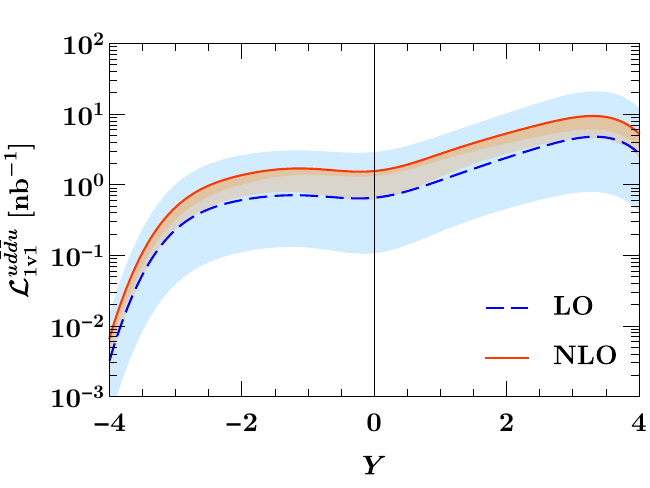}
}
\subfloat[$u \bar{d}, \bar{d} u$, colour octet, 1v1]{
   \includegraphics[width=0.48\textwidth]{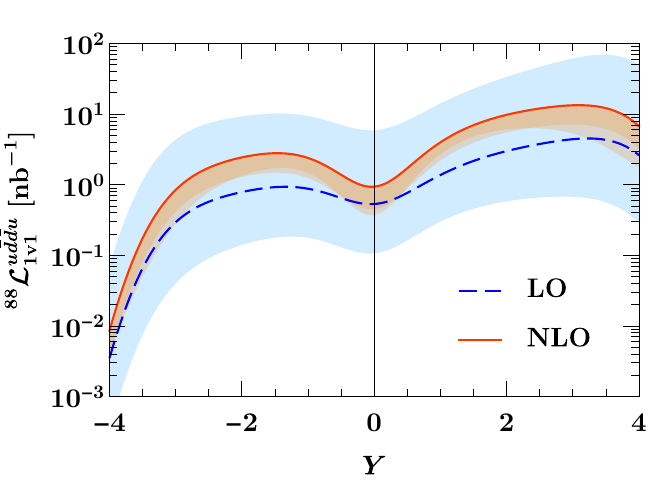}
}
\\[1.5em]
\subfloat[$u \bar{d}, \bar{d} u$, 1v2+2v1]{
   \includegraphics[width=0.48\textwidth]{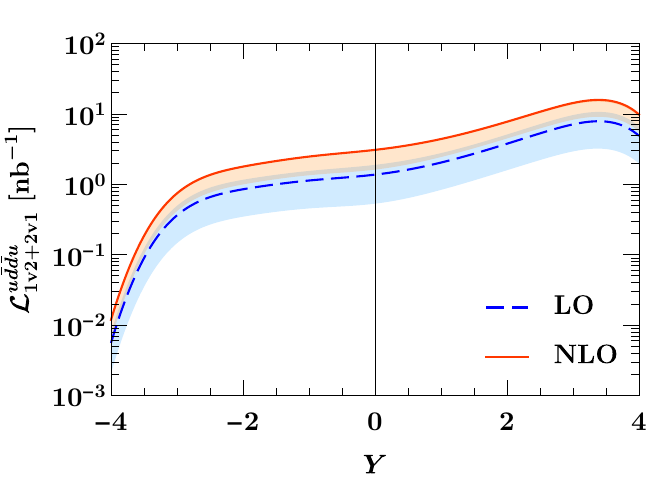}
}
\subfloat[$u \bar{d}, \bar{d} u$, 2v2]{
   \includegraphics[width=0.48\textwidth]{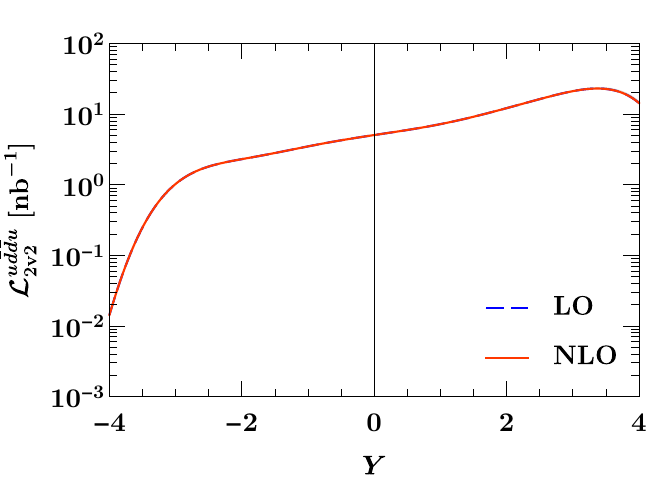}
}
\caption{\label{fig:lumis-q} The same as \fig{\protect\ref{fig:lumis-g}}, but
for the product of DPDs $F_{u \smash{\bar{d}}} \, F_{\smash{\bar{d}} u}$.}
\end{figure}

In \figs{\ref{fig:lumis-g}} and \ref{fig:lumis-q} we show luminosities for the
parton combinations $u g, \bar{d} g$ and $u \bar{d}, \bar{d} u$ at $M_1 = M_2 =
80 \gev$, which are respectively relevant for $W +$ dijet and for $W^+ W^+$
production.  Note that splitting DPDs for the $u \bar{d}, \bar{d} u$ channel are
generated only by NLO corrections to the splitting formula and by evolution from
the initial to the final scale.

Comparing the 1v1, 1v2 + 2v1, and 2v2 contributions in the colour singlet
(panels a, c, and d in both figures) we see that splitting and intrinsic
contributions are comparable in size for both parton combinations.  The change
from LO to NLO splitting kernels leads to a noticeable upward shift of the
central curves (except of course for the 2v2 contribution, where splitting DPDs
do not contribute).

Comparing panels a and b in both \figs{\ref{fig:lumis-g}} and \ref{fig:lumis-q}
we see that the 1v1 contribution in the octet channels is of similar magnitude
as for the singlet, and that the impact of including NLO splitting is similar as
well.

The bands in the two figures show the result of varying the initial scale
$\mu_{\text{init}}$ by a factor 2 up and down (with the restriction specified in
\eqref{mu-init-variation}).  For the intrinsic DPDs and hence for the 2v2
contribution this has almost no effect.  This is expected for our ansatz
\eqref{int-DPD-singlet}: it is known that for small momentum fractions $x_i$
there is little difference between the scale evolution of a single DPD and of
the product of two PDFs \cite{Diehl:2014vaa}.  For the splitting DPDs the
situation is very different.  With LO splitting kernels, the dependence of
double parton luminosities on $\mu_{\text{init}}$ is huge, covering an order of
magnitude or more in several cases.  With NLO kernels this dependence is
significantly reduced, although still visible in our logarithmic plots.  We also
note that at NLO the maximum of the luminosities is typically acquired near our
central scale choice $\mu_{\text{init}} = \muy(y)$.  We do not have an
explanation for this quite remarkable finding.

In order to see details on a linear rather than logarithmic scale, we switch to
plotting normalised double parton luminosities
\begin{align}
   \label{lumi-ratio-def}
   &
   \prn{R_1 R_2,\ms R_3 R_4 \ms}{\mathcal{R}}_{a_1 a_2,\ms b_1 b_2}(
      M_1, Y_1, M_2, Y_2; \mu_1, \mu_2, \nu, s)
   \notag \\[0.2em]
   & \quad
   =
   \frac{\prn{R_1 R_2,\ms R_3 R_4 \ms}{\lum}_{a_1 a_2,\ms b_1 b_2}(
      M_1, Y_1, M_2, Y_2; \mu_1, \mu_2, \nu, s)}{
      f_{a_1}(x_1; \mu_1)\, f_{a_2}(x_2; \mu_2) \,
      f_{b_1}(\bar{x}_1; \mu_1)\, f_{b_2}(\bar{x}_2; \mu_2)}
   \notag \\[0.2em]
   & \quad
   =
   2\pi \int_{\ycut}^{\infty} d y\, y\;
   \frac{\prn{R_1 R_2}{F}_{a_1 a_2}(x_1, x_2, y; \mu_1, \mu_2, \zeta)}{
      f_{a_1}(x_1; \mu_1)\, f_{a_2}(x_2; \mu_2)} \;
   \frac{\prn{R_3 R_4}{F}_{b_1 b_2}(\bar{x}_1, \bar{x}_2, y;
      \mu_1, \mu_2, \bar{\zeta})}{
      f_{b_1}(\bar{x}_1; \mu_1)\, f_{b_2}(\bar{x}_2; \mu_2)}
   \,,
\end{align}
where the momentum fractions of the PDFs and DPDs are given in
\eqref{x-fractions}.
This normalisation actually carries interesting physics information.  If one
makes the simplistic ansatz
\begin{align}
   \label{pocket-ansatz}
   \prn{1 1}{F}_{a_1 a_2}(x_1, x_2, y; \mu_1, \mu_2)
   &\to
   f_{a_1}(x_1; \mu_1) f_{a_2}(x_2; \mu_2) \, G(y)
\end{align}
with a function $G(y)$ that varies over a typical hadronic scale, and if one
neglects all colour non-singlet and all polarised contributions, then one gets
the so-called pocket formula for the DPS cross section
\begin{align}
   \frac{d \sigma_{\text{DPS}}}{d M_1^2\, d Y_1^{}\, d M_2^2 \, d Y_2^{}}
   &\to
   \frac{1}{1 + \delta_{A_1 A_2}}
   \frac{1}{\sigma_{\text{eff}}} \,
   \frac{d \sigma_{A_1}}{d M_1^2\, d Y_1^{}} \,
   \frac{d \sigma_{A_2}}{d M_2^2\, d Y_2^{}}
\end{align}
with
\begin{align}
   \label{sigma-eff}
   \sigma_{\text{eff}}^{-1}
   =
   2\pi \int_0^{\infty} dy\, y\; G^2(y)
   \,,
\end{align}
where $\sigma_{A_i}$ is the cross section for producing the system $A_i$ by
single parton scattering.  In~\eqref{sigma-eff} we have taken $0$ instead of
$\ycut$ as lower integration boundary, which is a good approximation for small
$\ycut$.  Under the same assumptions, the ratio in \eqref{lumi-ratio-def}
becomes
\begin{align}
   \prn{1 1, 1 1}{\mathcal{R}}_{a_1 a_2,\ms b_1 b_2}
   &\to
   \sigma_{\text{eff}}^{-1}
   \,.
\end{align}
The amount of variation of $\mathcal{R}$ with kinematic variables or parton
channels hence indicates the extent to which the simple ansatz
\eqref{pocket-ansatz} works or fails.  In turn, the size of $\mathcal{R}$ may be
compared with the range $\sigma_{\text{eff}}^{-1} \sim 50 \ldots 200
\operatorname{b}^{-1}$ of values that have been extracted from measurements
using the above pocket formula (see e.g.~\cite{Adam:2019krg, CMS:2021qsn} for
compilations of $\sigma_{\text{eff}}$).

%%%%%%%%%%%%%%%%%%%%%%%%%%%%%%%%%%%%%%%%

\paragraph{Impact of the subtraction term.}
In the following plots we compare the 1v1 part of the double parton luminosity
with the full double counting subtraction term $\lum_{\text{sub}}$ and with the
partial contribution $\lum_{\text{sub,SPS}}$ in \eqref{sub-lumis}.  We recall
that the latter is concentrated in the region $y \gg \ycut$, where it is
designed to compensate contributions from SPS graphs that have an overlap with
DPS.

\begin{figure}
\centering
\subfloat[$g g, g g$, LO]{
   \includegraphics[width=0.48\textwidth]{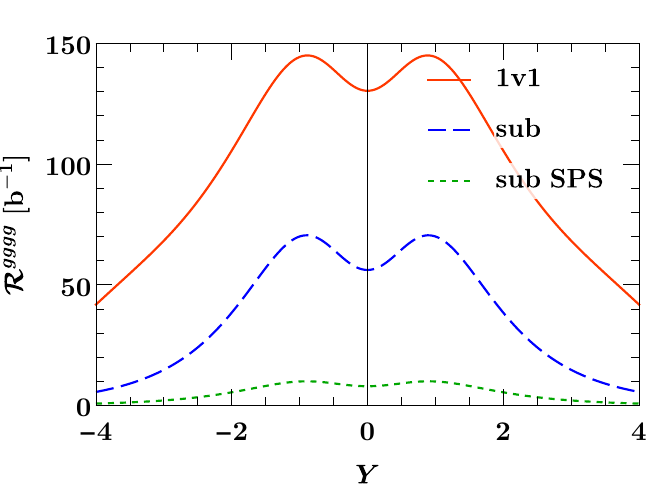}
}
\subfloat[$g g, g g$, NLO]{
   \includegraphics[width=0.48\textwidth]{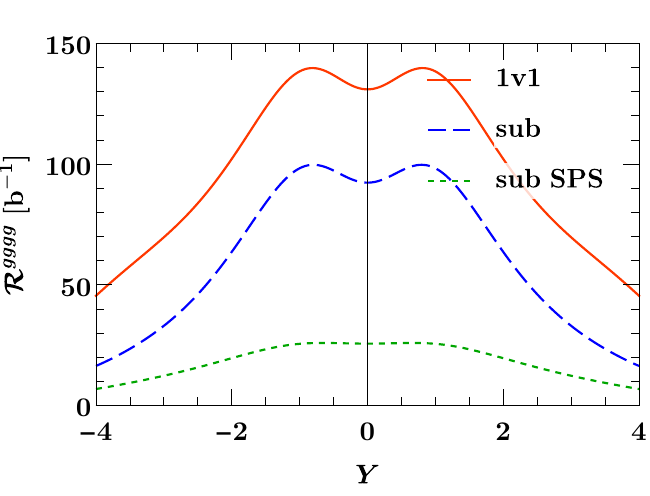}
}
\\[1.5em]
\subfloat[$u g, \bar{d} g$, LO]{
   \includegraphics[width=0.48\textwidth]{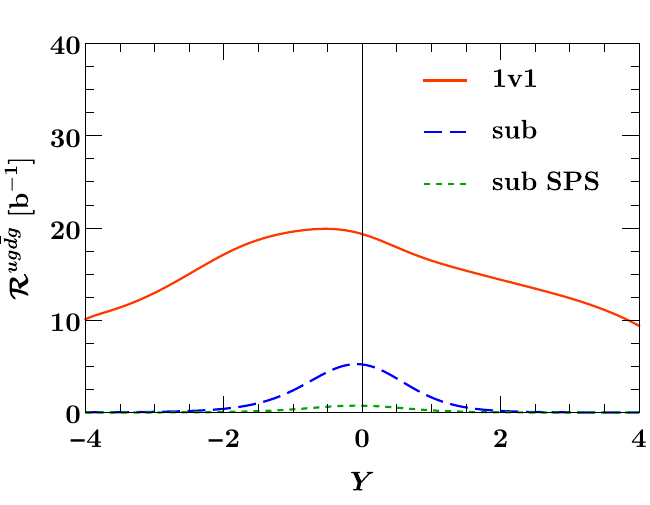}
}
\subfloat[$u g, \bar{d} g$, NLO]{
   \includegraphics[width=0.48\textwidth]{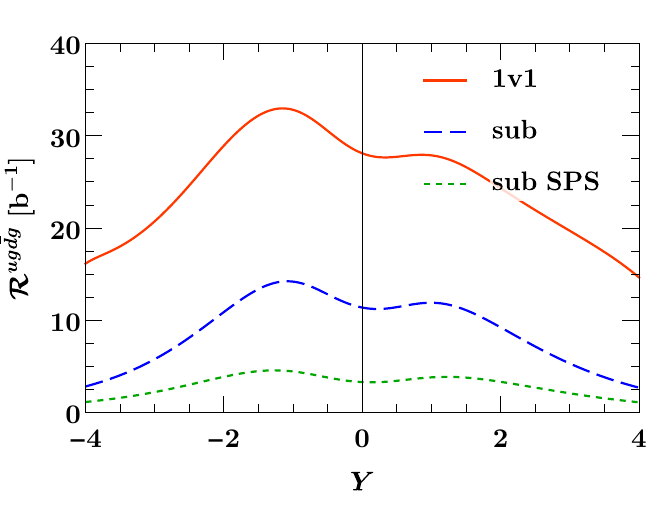}
}
\\[1.5em]
\subfloat[$u \bar{u}, \bar{d} d$, LO]{
   \includegraphics[width=0.48\textwidth]{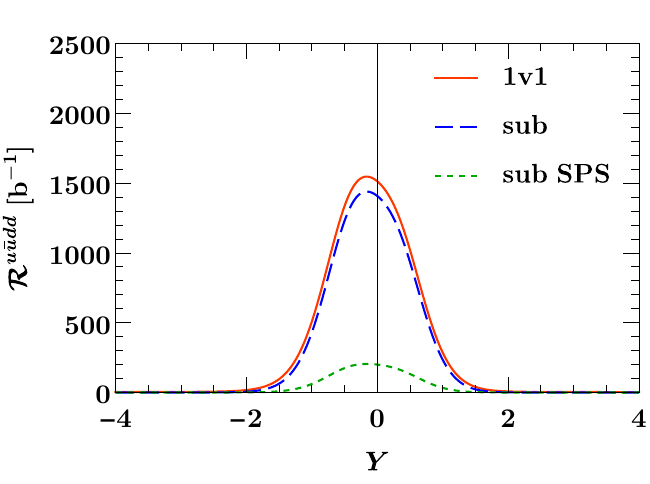}
}
\subfloat[$u \bar{u}, \bar{d} d$, NLO]{
   \includegraphics[width=0.48\textwidth]{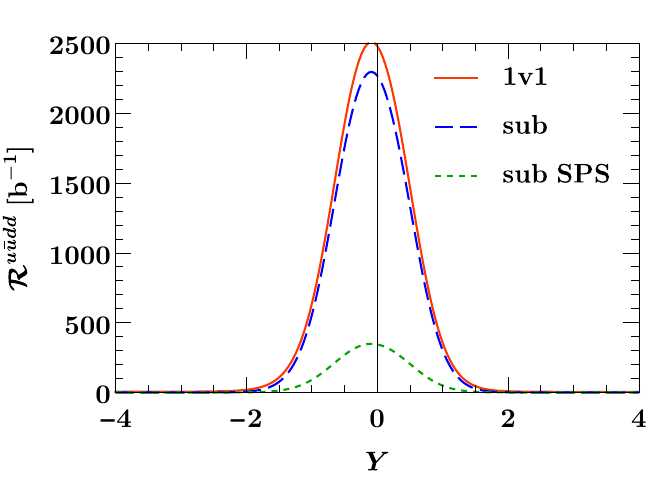}
}
\caption{\label{fig:lumi-sub-SPSvsDPS} Normalised double parton luminosities for
selected parton combination, with parton kinematics and scales set for $M_1 =
M_2 = 80 \gev$ as specified in \sect{\protect\ref{sec:parton-kin}}.  Shown are
the 1v1 component $\lum_{\text{1v1}}$ of the luminosity for DPS, the luminosity
$\lum_{\text{sub,SPS}}$ for the double counting subtraction term, and its
large-$y$ part $\lum_{\text{sub,SPS}}$ in the decomposition
\protect\eqref{sub-lumis}.  All luminosities are normalised as specified in
\protect\eqref{lumi-ratio-def}.}
\end{figure}

In \fig{\ref{fig:lumi-sub-SPSvsDPS}} we show this comparison for equal scales
$M_1 = M_2 = 80\gev$ and for parton combinations that allow splitting already at
LO . The total subtraction term is important in all cases; for the parton
combination $u \bar{u}, \bar{d} d$ at central rapidities it is nearly equal to
$\lum_{\text{1v1}}$.  We always find $| \lum_{\text{sub,SPS}} | \ll |
\lum_{\text{sub}} |$, i.e.\ $\lum_{\text{sub}} \approx \lum_{\text{sub,DPS}}$,
which reflects that the product of two splitting DPDs is most important at small
$y$.  In some cases $\lum_{\text{sub,SPS}}$ is comparable in size to the
difference $\lum_{\text{1v1}} - \lum_{\text{sub}}$ and in this sense has an
impact on the overall cross section.  This holds for $u \bar{u}, \bar{d} d$ and
also for the four-gluon case at NLO.  In the other cases shown in
\fig{\ref{fig:lumi-sub-SPSvsDPS}}, $\lum_{\text{sub,SPS}}$ is of minor
importance.

\begin{figure}
\centering
\subfloat[$u \bar{d}, \bar{d} u$, NLO, colour singlet]{
   \includegraphics[width=0.48\textwidth]{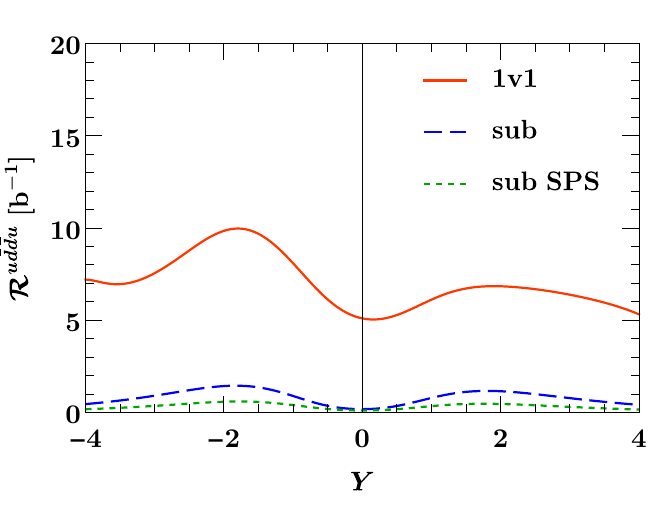}
}
\subfloat[$u \bar{d}, \bar{d} u$, NLO, colour octet]{
   \includegraphics[width=0.48\textwidth]{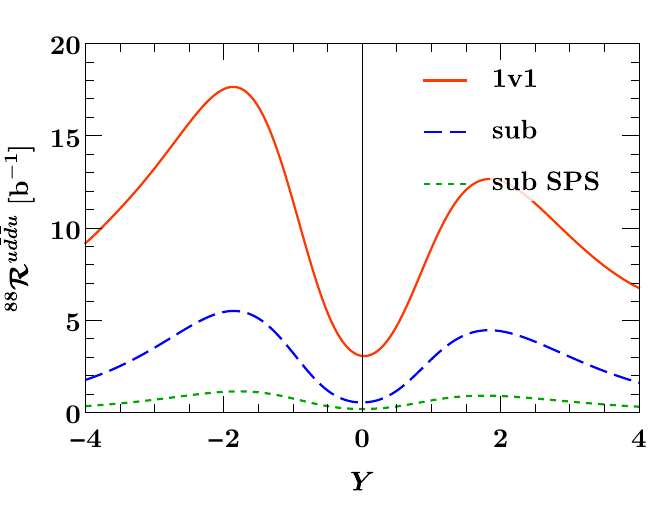}
}
\caption{\label{fig:lumi-sub-NLO-chan} As
\fig{\protect\ref{fig:lumi-sub-SPSvsDPS}}, but for a parton combination where
the DPDs splitting formula starts at NLO.}
\end{figure}

For parton combinations where splitting only starts at NLO, we find that the
subtraction term evaluated at NLO is generally of minor size, although not
necessarily negligible as seen for $u \bar{d}, \bar{d} u$ in
\fig{\ref{fig:lumi-sub-NLO-chan}}.  At LO (not shown here) the corresponding
subtraction terms are very close to zero, since the splitting DPDs vanish at the
initial scale and there is essentially no evolution in this case.

We now turn to the case of unequal scales with $M_1 = 80\gev$ and $M_2 =
10\gev$. The difference between the integrands for the DPS and SPS subtraction
now becomes important, and we need to specify the interpolating function in
\eqref{interpol-I}.  As a default we take $y_{\rho} = \ycut$ and
\begin{align}
   \label{default-r1-r2}
   r_1 &= 2
   \,,
   &
   r_2 &= 4
   \,,
\end{align}
which corresponds to a rather quick changeover from $I_{\text{sub,DPS}}(y)$ to
$I_{\text{sub,SPS}}(y)$ as a function of $y$.  We make this choice in order to
see how important $\lum_{\text{sub,SPS}}$ can plausibly be, since with larger
values of $r_2$ that term will decrease even further.  We also computed double
parton luminosities with the choices $r_1 = 2, r_2 = 6$ and $r_1 = 3, r_2 = 5$
and find that $\lum_{\text{sub}}$ and $\lum_{\text{1v1}} - \lum_{\text{sub}}$
barely change for the parton combinations considered in our study.

\begin{figure}
\centering
\subfloat[$g g, g g$, LO]{
   \includegraphics[width=0.48\textwidth]{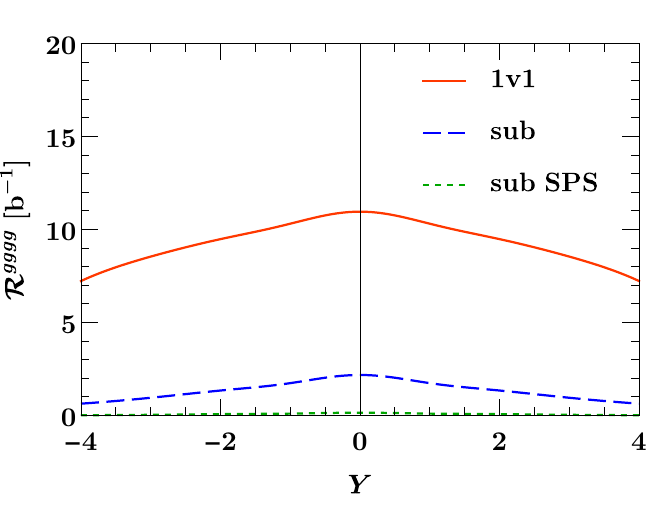}
}
\subfloat[$g g, g g$, NLO]{
   \includegraphics[width=0.48\textwidth]{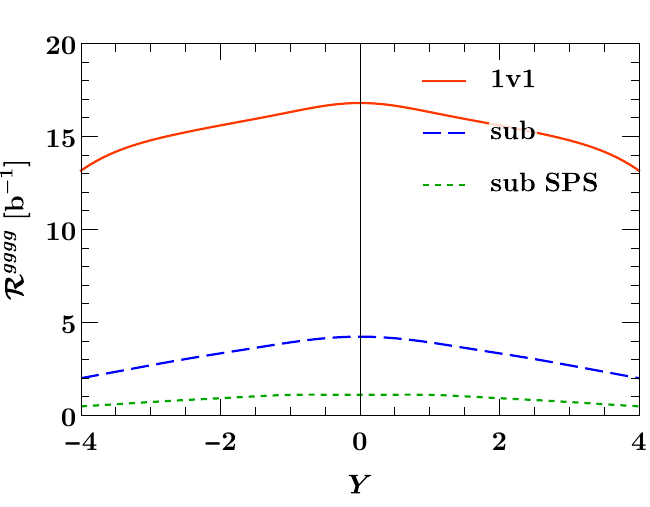}
}
\caption{\label{fig:lumi-sub-SPSvsDPS-80-10} As the first row of
\fig{\protect\ref{fig:lumi-sub-SPSvsDPS}}, but for kinematics with $M_1 = 80
\gev$ and $M_2 = 10 \gev$.}
\end{figure}

Comparing \fig{\ref{fig:lumi-sub-SPSvsDPS-80-10}} with the first row in
\fig{\ref{fig:lumi-sub-SPSvsDPS}}, we see that  in the four-gluon channel the
subtraction term with one small and one large scale is less important than with
two large scales.  We find a similar situation for the combination $u g, \bar{d}
g$ (not shown here).

%%%%%%%%%%%%%%%%%%%%%%%%%%%%%%%%%%%%%%%%

\begin{figure}
\centering
\subfloat[$u g, \bar{d} g$, LO]{
   \includegraphics[width=0.48\textwidth]{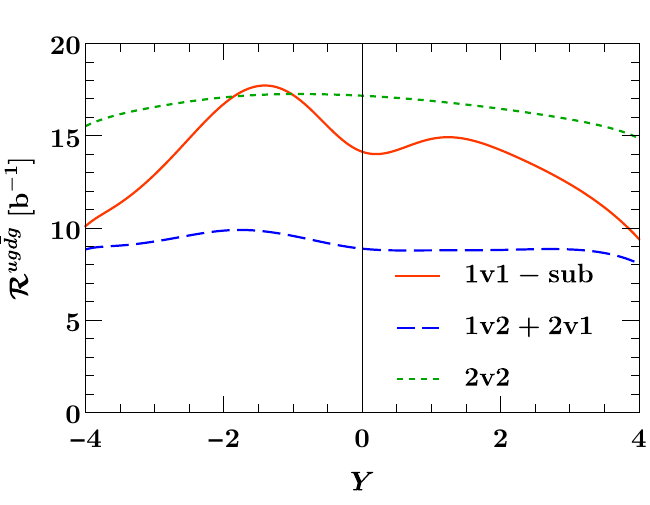}
}
\subfloat[\label{lumi-contrib-ug-dbarg} $u g, \bar{d} g$, NLO]{
   \includegraphics[width=0.48\textwidth]{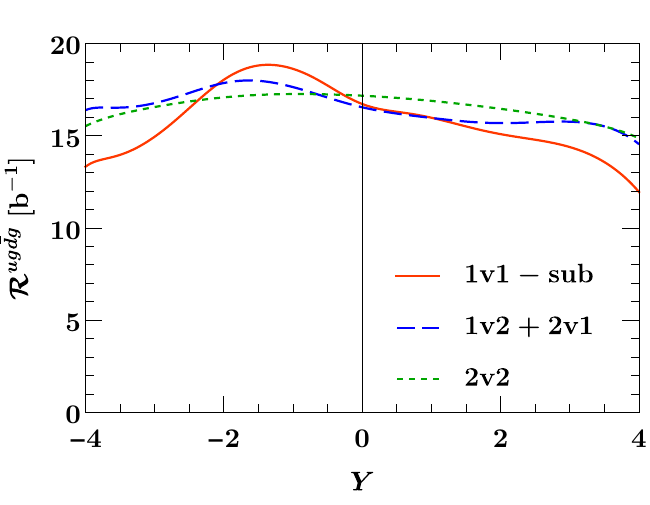}
}
\\[1.5em]
\subfloat[$c \bar{s}, \bar{b} c$, LO]{
   \includegraphics[width=0.48\textwidth]{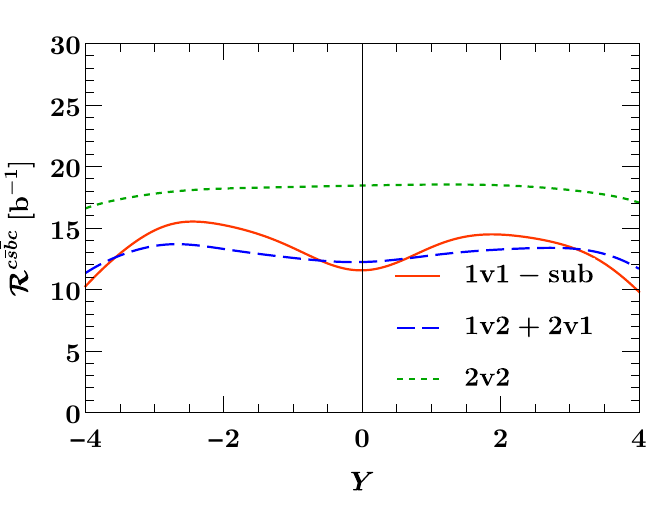}
}
\subfloat[$c \bar{s}, \bar{b} c$, NLO]{
   \includegraphics[width=0.48\textwidth]{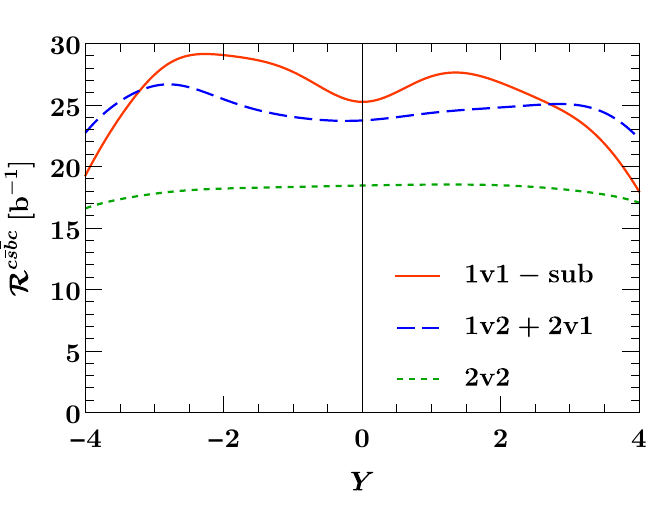}
}
\caption{\label{fig:lumi-contribs-LOvsNLO} Different contributions to normalised
double parton luminosities in kinematics with $M_1 = M_2 = 80 \gev$.  The sum of
the three contributions gives
$\mathcal{R}_{\text{DPS}} - \mathcal{R}_{\text{sub}}$.}
\end{figure}

\paragraph{Size of the subtracted 1v1 term.}
According to the master formula \eqref{full-Xsect}, the full cross section of a
process involves the subtracted luminosity $\lum_{\text{1v1}} -
\lum_{\text{sub}}$ plus the partial luminosities for 1v2, 2v1, and 2v2.  In
\figs{\ref{fig:lumi-contribs-LOvsNLO}} to \ref{fig:lumi-contribs-NLO-80-10} we
compare these three terms for different kinematics and parton combinations.
Figure~\ref{fig:lumi-contribs-LOvsNLO} shows that the relative importance
of the different contributions can change to some extent when going from LO to
NLO splitting.

\begin{figure}
\centering
\subfloat[\label{fig:lumi-contribs-qqbar} $u \bar{u}, \bar{d} d$, NLO]{
   \includegraphics[height=0.35\textwidth]{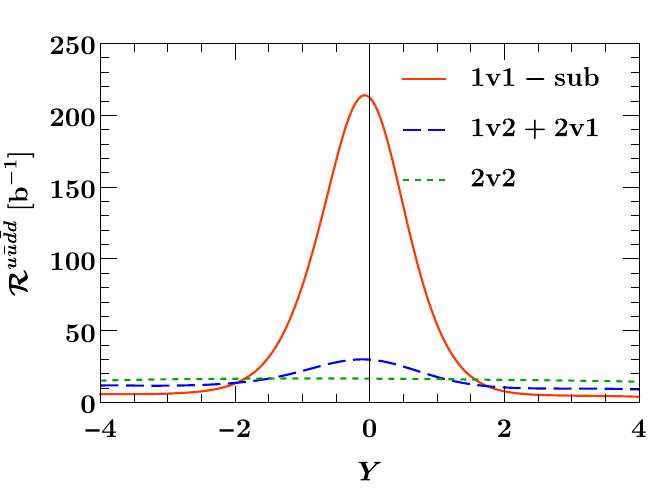}
}
\hfill
\subfloat[\label{lumi-contrib-gg-gg} $g g, g g$, NLO]{
   \includegraphics[height=0.35\textwidth]{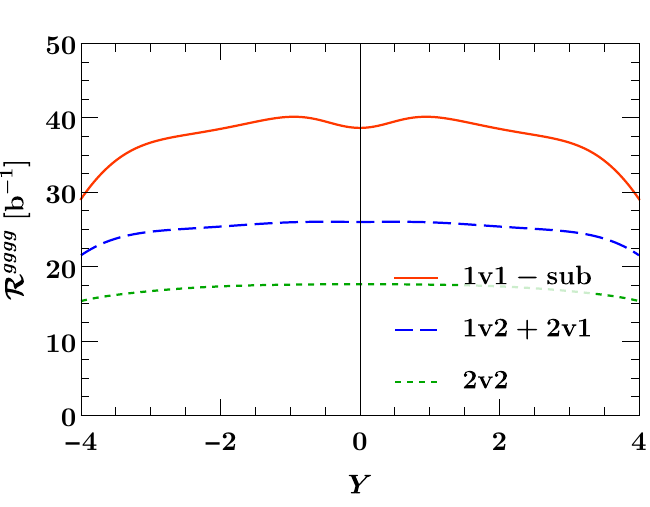}
}
\\[1.5em]
\subfloat[$u \bar{d}, \bar{d} u$, NLO]{
   \includegraphics[height=0.35\textwidth]{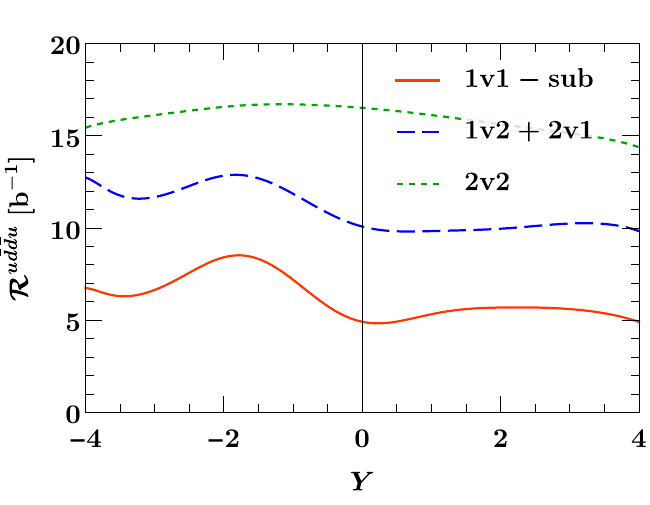}
}
\hfill
\subfloat[$u u, \bar{d} \bar{d}$, NLO]{
   \includegraphics[height=0.35\textwidth]{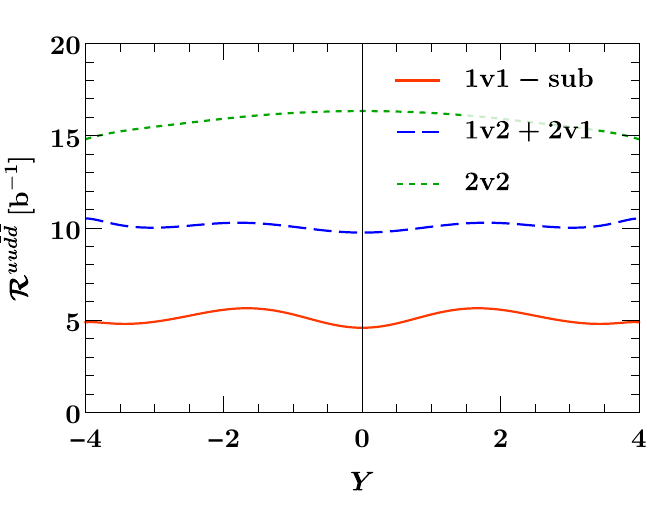}
}
\caption{\label{fig:lumi-contribs-NLO-80-80} As the right panels in
\fig{\protect\ref{fig:lumi-contribs-LOvsNLO}}, but for other parton
combinations.}
\end{figure}

\begin{figure}
\centering
\subfloat[$u g, \bar{d} g$, NLO]{
   \includegraphics[height=0.35\textwidth]{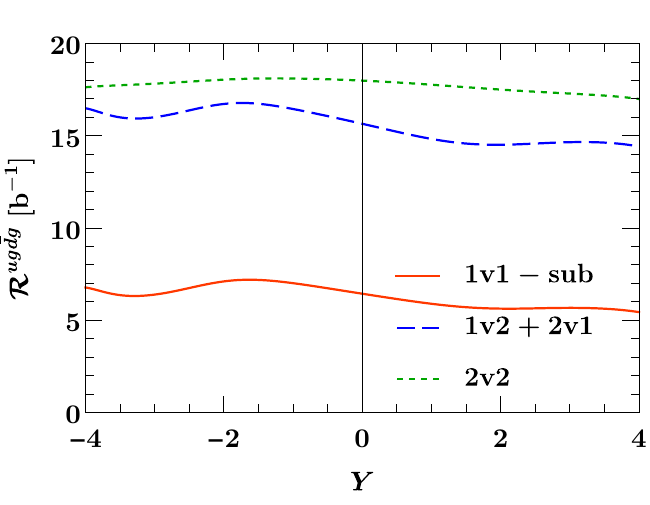}
}
\hfill
\subfloat[$g g, g g$, NLO]{
   \includegraphics[height=0.35\textwidth]{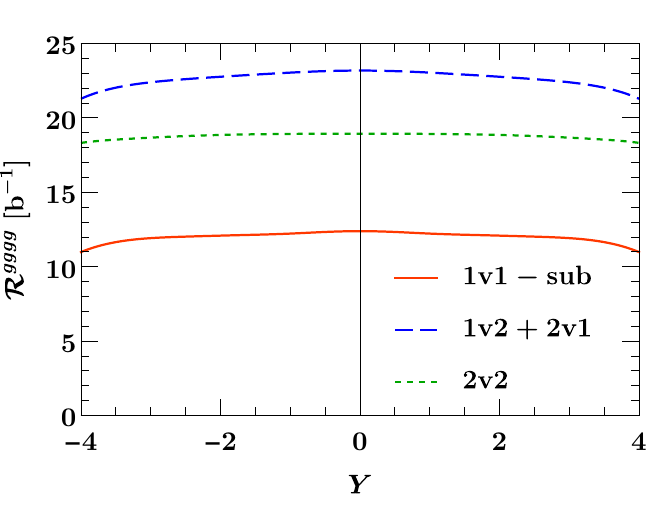}
}
\caption{\label{fig:lumi-contribs-NLO-80-10} Different contributions to
normalised double parton luminosities in kinematics with $M_1 = 80 \gev$ and
$M_1 = 10 \gev$.  The corresponding plots for equal scales are shown in
\figs{\protect\ref{lumi-contrib-ug-dbarg}} and
\protect\ref{lumi-contrib-gg-gg}.}
\end{figure}

In almost all considered channels we find that the contributions with or without
splitting DPDs are of comparable magnitude and thus all play a role in the final
result.

The only exception is the parton combination $u \bar{u}, \bar{d} d$ in
\fig{\ref{fig:lumi-contribs-qqbar}}, where for central rapidities $|Y| \lsim 2$
the 1v1 contribution is completely dominant, even after double counting
subtraction.  This is remarkable, because the subtracted 1v1 term in this case
is already the result of a substantial cancellation between $\lum_{\text{1v1}}$
and $\lum_{\text{sub}}$, as we saw in the bottom row of
\fig{\ref{fig:lumi-sub-SPSvsDPS}}.  The same is observed for other channels with
a $q \bar{q}$ pair in each DPD.
An important consequence is that at central rapidities these luminosities have
very little uncertainty due to modelling the intrinsic part of DPDs, which is
not the case for the other channels in \figs{\ref{fig:lumi-contribs-LOvsNLO}} to
\ref{fig:lumi-contribs-NLO-80-10}. Since the $q \bar{q}$ splitting DPDs are
concentrated at small $y$, there is also less dependence on how the perturbative
splitting form is extended to the non-perturbative $y$ region.

Comparing the plots for unequal scales $M_1 = 80\gev$ and $M_2 = 10\gev$ in
\fig{\ref{fig:lumi-contribs-NLO-80-10}} with their counterparts for $M_1 = M_2 =
80\gev$ in \figs{\ref{lumi-contrib-ug-dbarg}} and \ref{lumi-contrib-gg-gg}, we
observe that the subtracted 1v1 part is less important for unequal scales.

%%%%%%%%%%%%%%%%%%%%%%%%%%%%%%%%%%%%%%%%

\subsection{Subtracted double parton luminosities and their scale dependence}
\label{sec:lumi-scale-dep}

Let us finally study the dependence of the subtracted luminosities
$\lum_{\text{DPS}} - \lum_{\text{sub}}$ on the choice of the scales
$\mu_{\text{init}}(y)$ and $\nu$.  This dependence propagates into the overall
cross section, albeit with possible cancellations in the sum over different
parton and colour channels.  Apart from such cancellations, the uncertainty due
to choosing a scale directly translates into an uncertainty of cross section
predictions.

The panels on the left of the following figures show bands resulting from
varying $\mu_{\text{init}}(y)$ around the central choice $\muy(y)$ by a factor
2, with a lower bound $\mu_{\text{min}}$ as specified in
\eqref{mu-init-variation}. The value of $\nu$ is fixed to its default choice
$\min(\mu_1, \mu_2)$ in this case.

The panels on the right of the same figures show bands resulting from varying
the cutoff parameter $\nu$ by a factor 2 up and down while keeping the
interpolation function $\rho(y)$ fixed.  The initial scale for DPDs in
$\lum_{\text{DPS}}$ is kept at its default $\mu_{\text{init}}(y) = \muy(y)$ in
this case, but note that the initial scale of DPDs in the subtraction term
$\lum_{\text{sub,DPS}}$ is always $\mu_{\text{sub,init}} =
\mu_{\text{init}}(\ycut)$ and hence changes with $\ycut$.
If one prefers to keep $\mu_{\text{sub,init}}$ fixed, one can vary $\ycut$ and
$\mu_{\text{init}}$ simultaneously, or one can choose $\mu_{\text{sub,init}} =
\min(\mu_1, \mu_2)$ independent of $\ycut$.  We will not pursue these
possibilities here.

If follows from \eqn{\eqref{DPD-resid-scale-dep}} that the $\mu_{\text{init}}$
dependence of subtracted double parton luminosities decreases parametrically
with the order at which the splitting DPDs are computed.  We have shown in
\sect{\ref{sec:cutoff-dep}} that the dependence of subtracted luminosities on
$\nu$ is beyond the perturbative order of the calculation and hence also
decreases parametrically when that order is raised. Note that the same is
\emph{not} true (and not meant to be true) for the $\nu$ dependence of
$\lum_{\text{DPS}}$ without the double counting subtraction.  In the following
we investigate to which extent the parametric decrease in powers of $\alpha_s$
shows up at the quantitative level.

\begin{figure}
\centering
\subfloat[$\mu_{\text{init}}$ variation, $u \bar{u}, \bar{d} d$]{
   \includegraphics[width=0.48\textwidth]{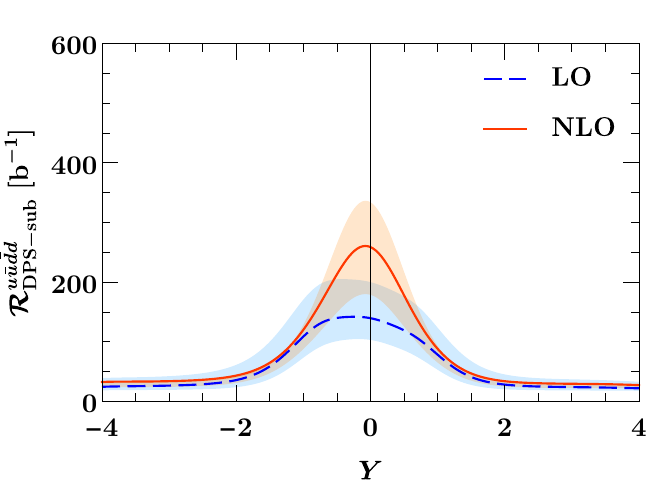}
}
\subfloat[$\nu$ variation, $u \bar{u}, \bar{d} d$]{
   \includegraphics[width=0.48\textwidth]{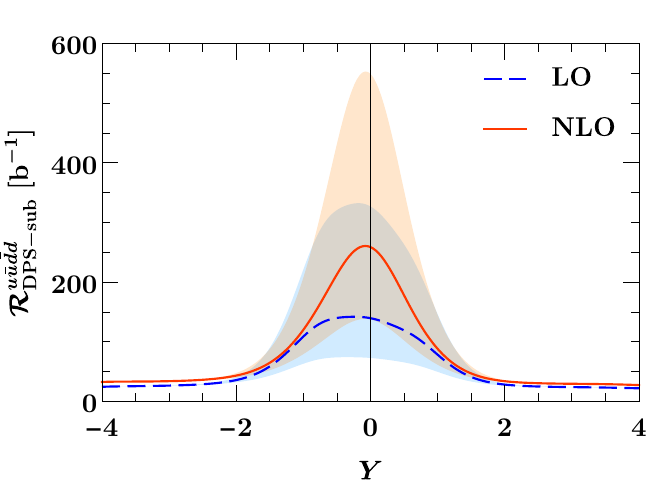}
}
\\[1.5em]
\subfloat[$\mu_{\text{init}}$ variation, $u \bar{d}, \bar{d} u$]{
   \includegraphics[width=0.48\textwidth]{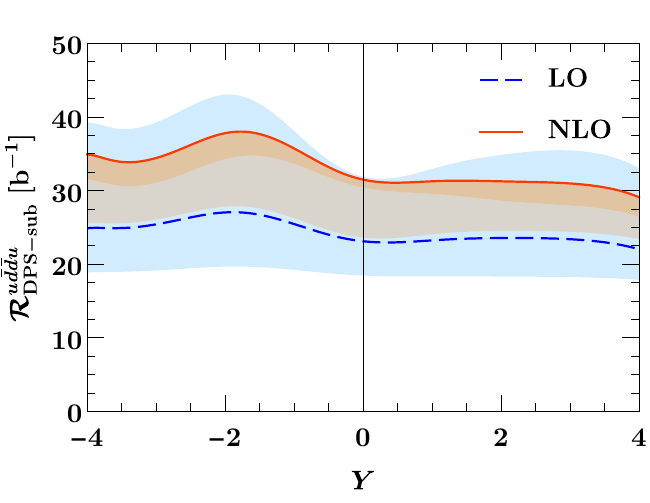}
}
\subfloat[$\nu$ variation, $u \bar{d}, \bar{d} u$]{
   \includegraphics[width=0.48\textwidth]{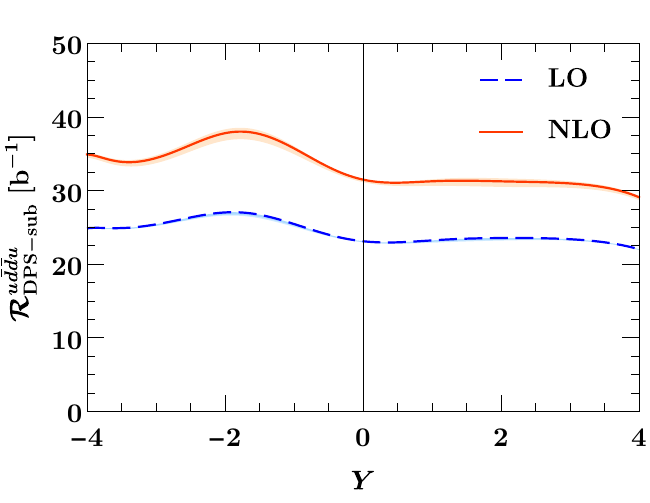}
}
\\[1.5em]
\subfloat[$\mu_{\text{init}}$ variation, $c \bar{s}, \bar{b} c$]{
   \includegraphics[width=0.48\textwidth]{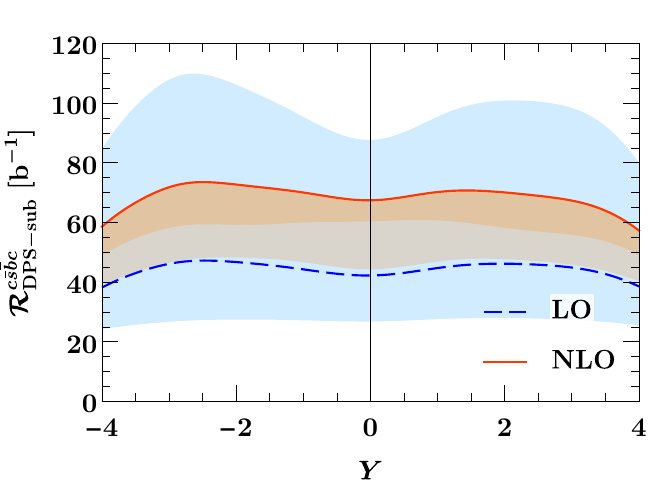}
}
\subfloat[$\nu$ variation, $c \bar{s}, \bar{b} c$]{
   \includegraphics[width=0.48\textwidth]{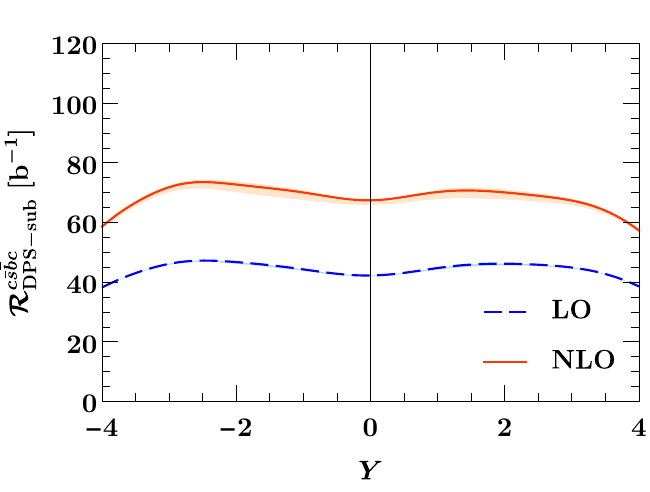}
}
\caption{\label{fig:scale-col-sing} Subtracted double parton luminosities and
their scale dependence in the colour singlet for $M_1 = M_2 = 80 \gev$. Further
parton combinations are shown in \fig{\protect\ref{fig:scale-col-sing-extra}}.}
\end{figure}

\begin{figure}
\centering
\subfloat[$\mu_{\text{init}}$ variation, $u \bar{u}, \bar{d} d$, colour octet]{
   \includegraphics[width=0.48\textwidth]{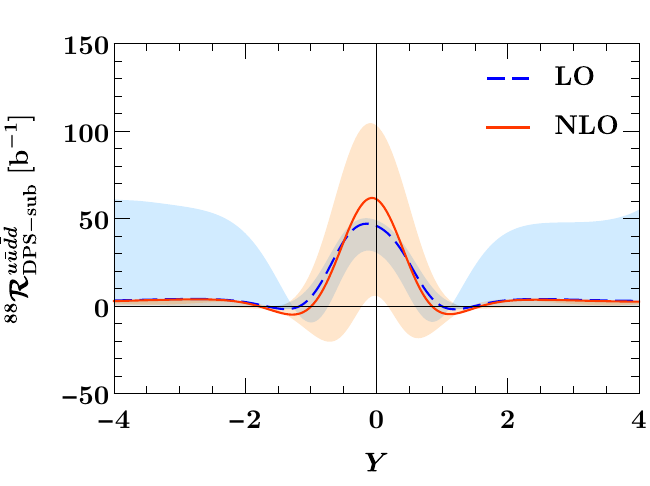}
}
\subfloat[$\nu$ variation, $u \bar{u}, \bar{d} d$, colour octet]{
   \includegraphics[width=0.48\textwidth]{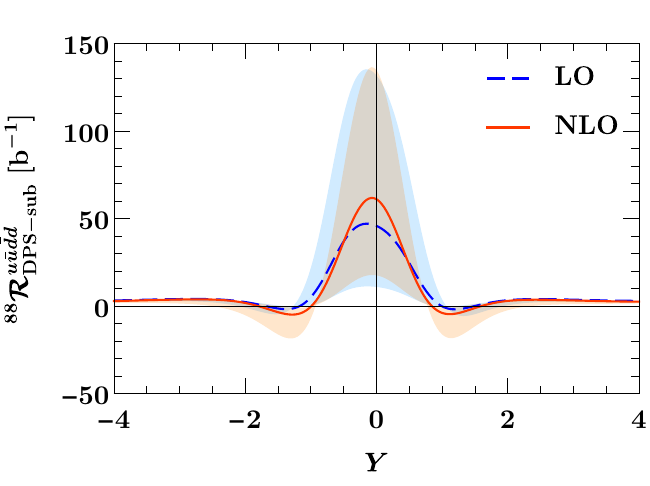}
}
\\[1.5em]
\subfloat[$\mu_{\text{init}}$ variation, $u \bar{d}, \bar{d} u$, colour octet]{
   \includegraphics[width=0.48\textwidth]{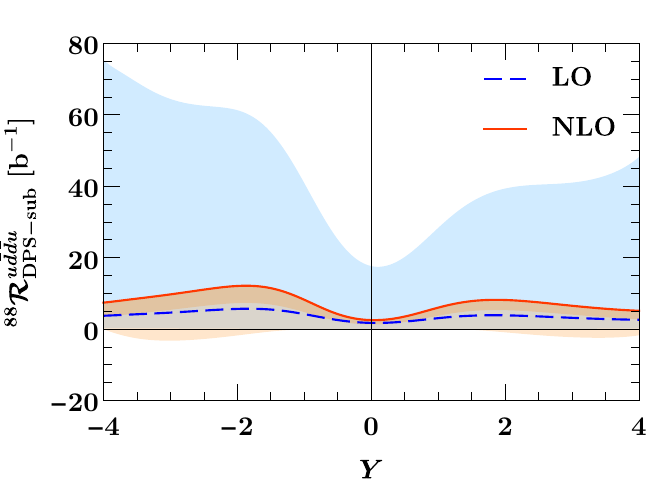}
}
\subfloat[$\nu$ variation, $u \bar{d}, \bar{d} u$, colour octet]{
   \includegraphics[width=0.48\textwidth]{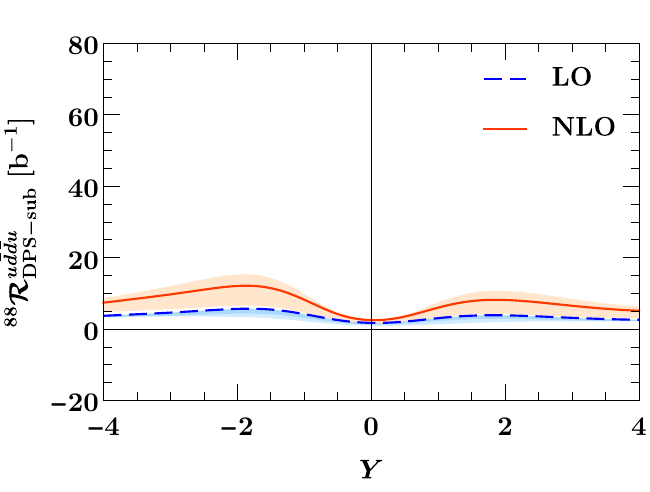}
}
\\[1.5em]
\subfloat[$\mu_{\text{init}}$ variation, $c \bar{s}, \bar{b} c$, colour octet]{
   \includegraphics[width=0.48\textwidth]{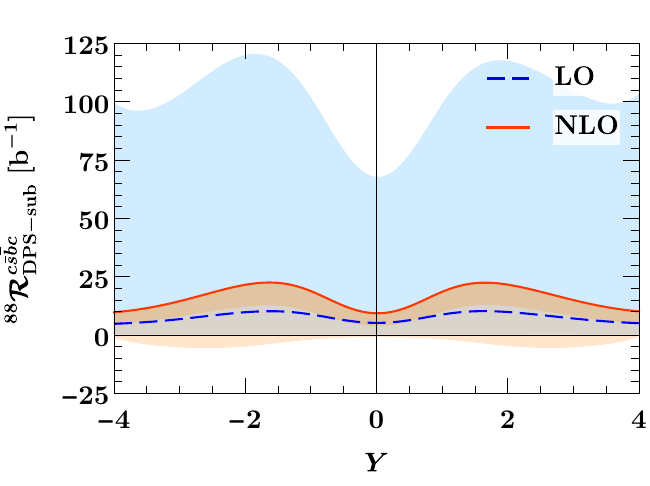}
}
\subfloat[$\nu$ variation, $c \bar{s}, \bar{b} c$, colour octet]{
   \includegraphics[width=0.48\textwidth]{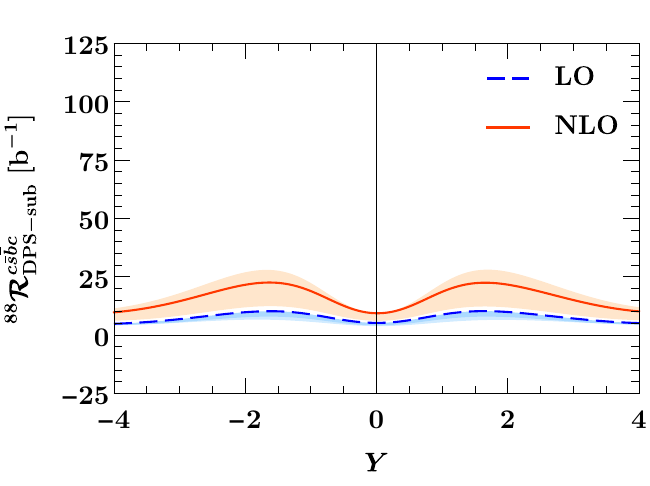}
}
\caption{\label{fig:scale-col-oct} As \fig{\protect\ref{fig:scale-col-sing}},
but for colour octet channels.  Further parton combinations are shown in
\fig{\protect\ref{fig:scale-col-oct-extra}}.}
\end{figure}

\begin{figure}
\centering
\subfloat[$\mu_{\text{init}}$ variation, $g g, g g$]{
   \includegraphics[width=0.48\textwidth]{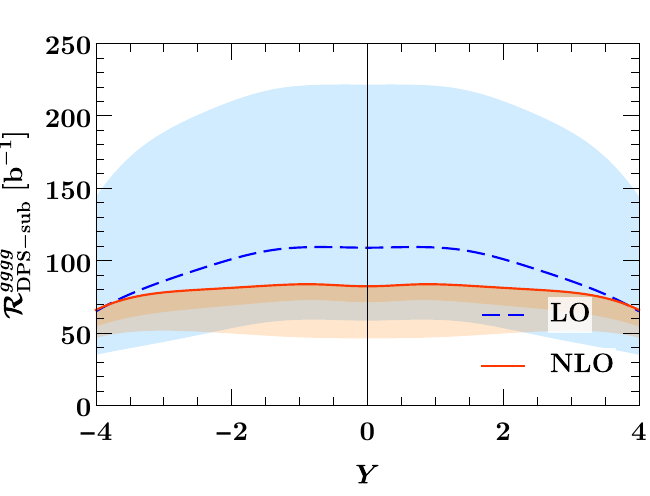}
}
\subfloat[$\nu$ variation, $g g, g g$]{
   \includegraphics[width=0.48\textwidth]{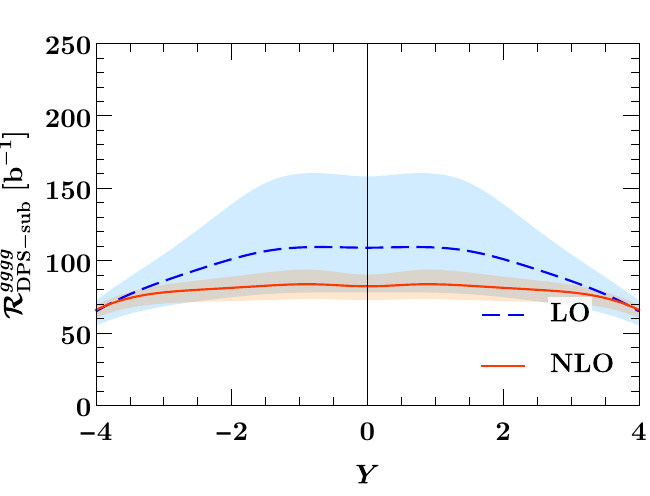}
}
\\[1.5em]
\subfloat[$\mu_{\text{init}}$ variation, $u g, \bar{d} g$]{
   \includegraphics[width=0.48\textwidth]{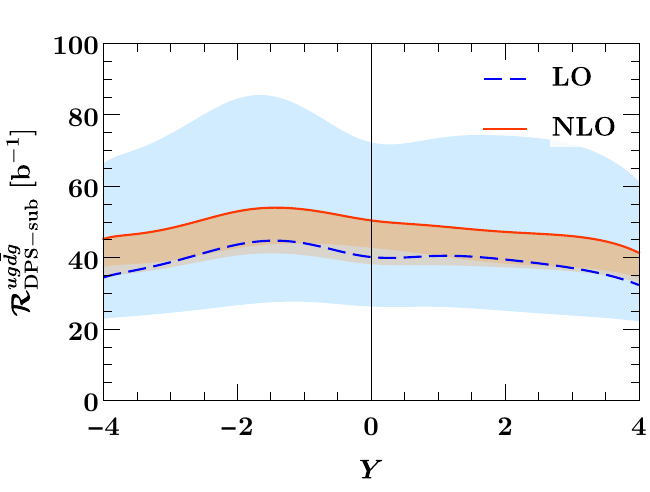}
}
\subfloat[$\nu$ variation, $u g, \bar{d} g$]{
   \includegraphics[width=0.48\textwidth]{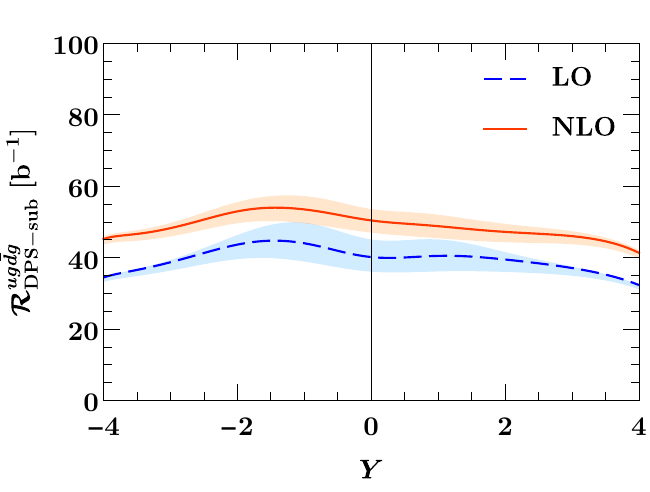}
}
\\[1.5em]
\subfloat[$\mu_{\text{init}}$ variation, $u u, \bar{d} \bar{d}$]{
   \includegraphics[width=0.48\textwidth]{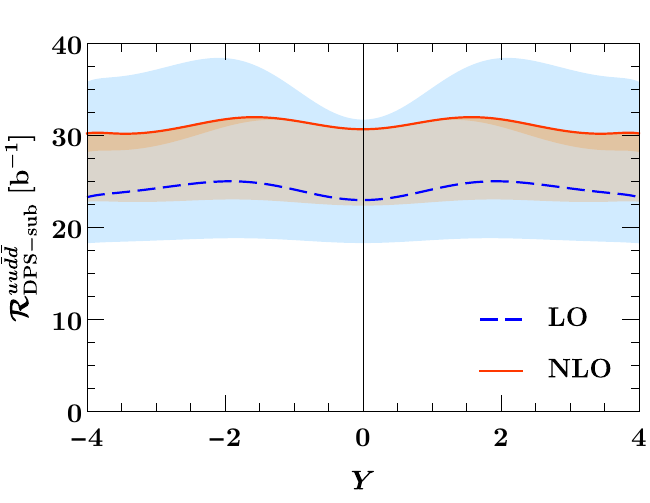}
}
\subfloat[$\nu$ variation, $u u, \bar{d} \bar{d}$]{
   \includegraphics[width=0.48\textwidth]{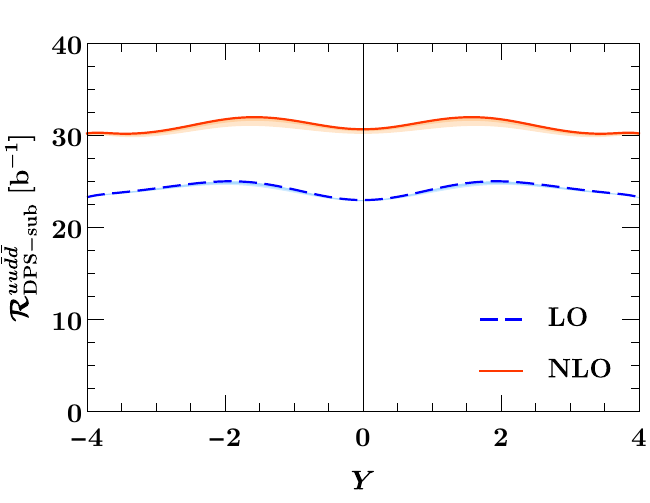}
}
\caption{\label{fig:scale-col-sing-extra} Further subtracted double parton
luminosities and their scale dependence in the colour singlet for $M_1 =
M_2 = 80 \gev$.}
\end{figure}

\begin{figure}
\centering
\subfloat[$\mu_{\text{init}}$ variation, $g g, g g$, $A A, A A$]{
   \includegraphics[width=0.48\textwidth]{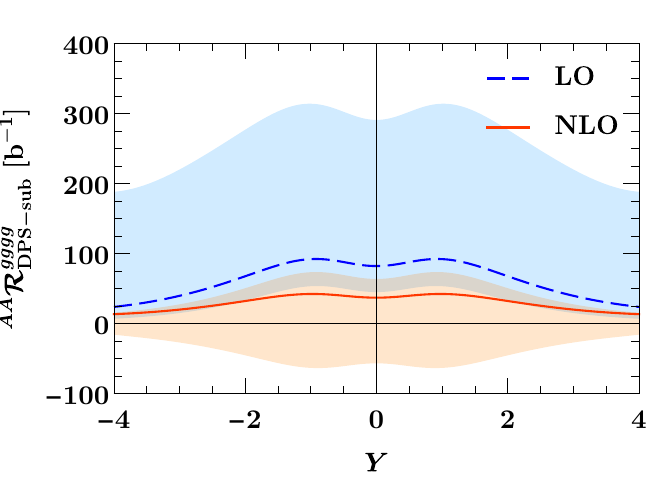}
}
\subfloat[$\nu$ variation, $g g, g g$, $A A, A A$]{
   \includegraphics[width=0.48\textwidth]{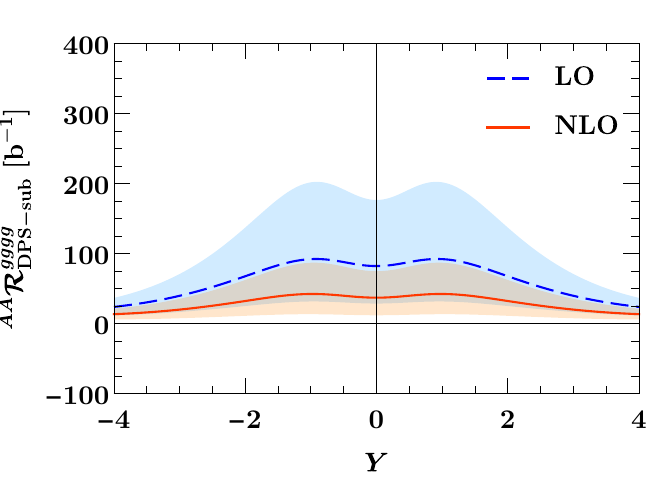}
}
\\[1.5em]
\subfloat[$\mu_{\text{init}}$ variation, $u g, \bar{d} g$, $8 A, 8 A$]{
   \includegraphics[width=0.48\textwidth]{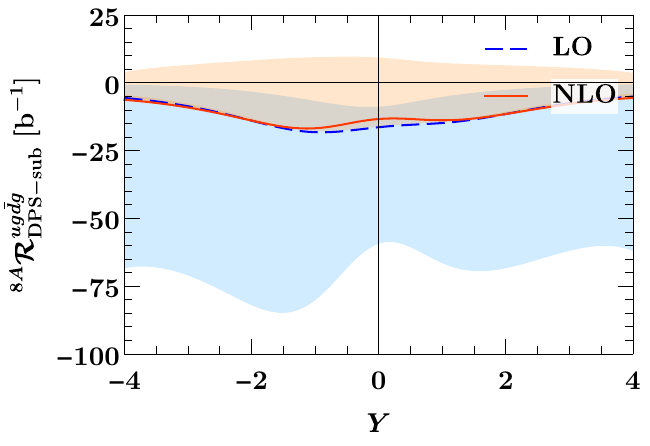}
}
\subfloat[$\nu$ variation, $u g, \bar{d} g$, $8 A, 8 A$]{
   \includegraphics[width=0.48\textwidth]{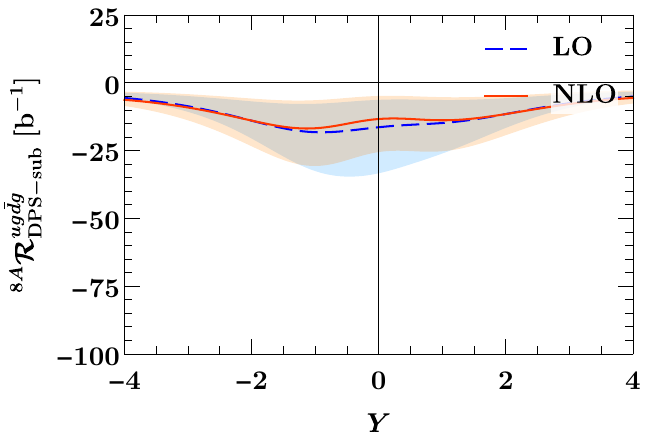}
}
\\[1.5em]
\subfloat[$\mu_{\text{init}}$ variation, $u u, \bar{d} \bar{d}$, colour octet]{
   \includegraphics[width=0.48\textwidth]{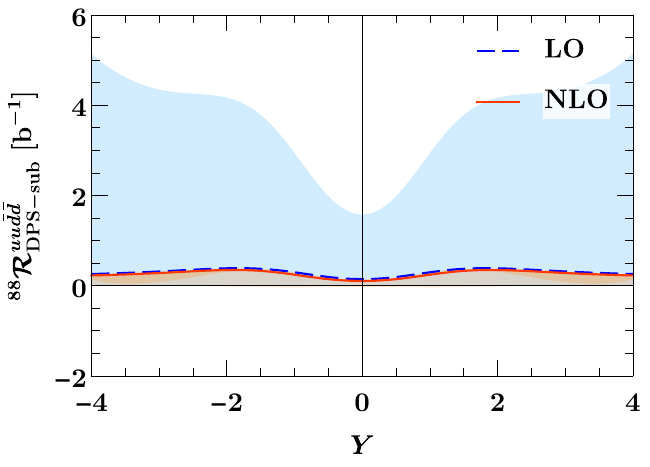}
}
\subfloat[$\nu$ variation, $u u, \bar{d} \bar{d}$, colour octet]{
   \includegraphics[width=0.48\textwidth]{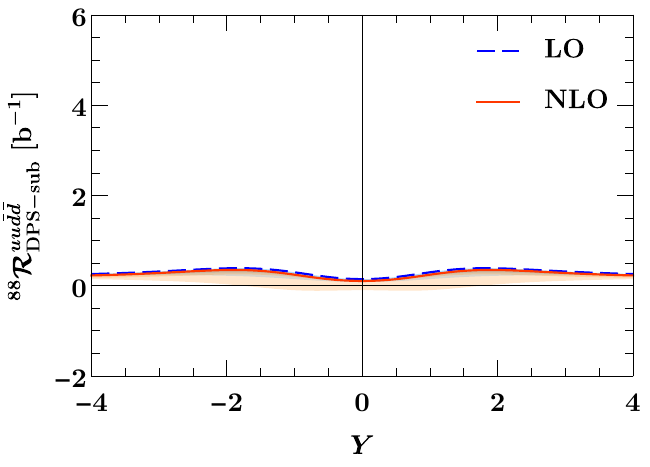}
}
\caption{\label{fig:scale-col-oct-extra} As
\fig{\protect\ref{fig:scale-col-sing-extra}}, but for colour octet channels.}
\end{figure}

\begin{figure}[t]
\centering
\subfloat[$\mu_{\text{init}}$ variation, $g g, g g$]{
   \includegraphics[width=0.48\textwidth]{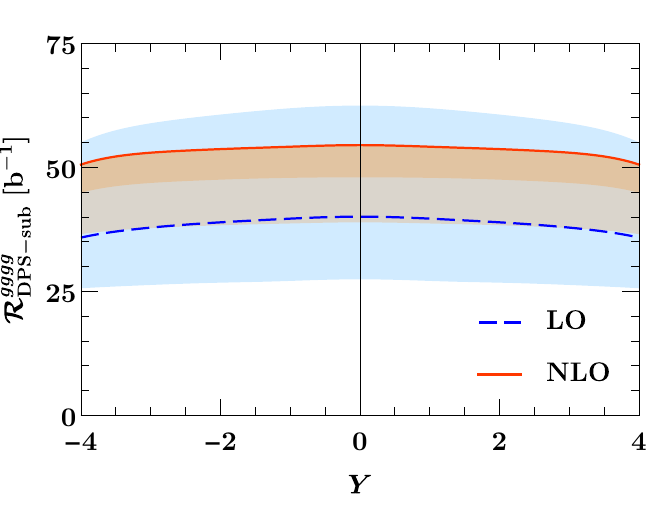}
}
\subfloat[$\nu$ variation, $g g, g g$]{
   \includegraphics[width=0.48\textwidth]{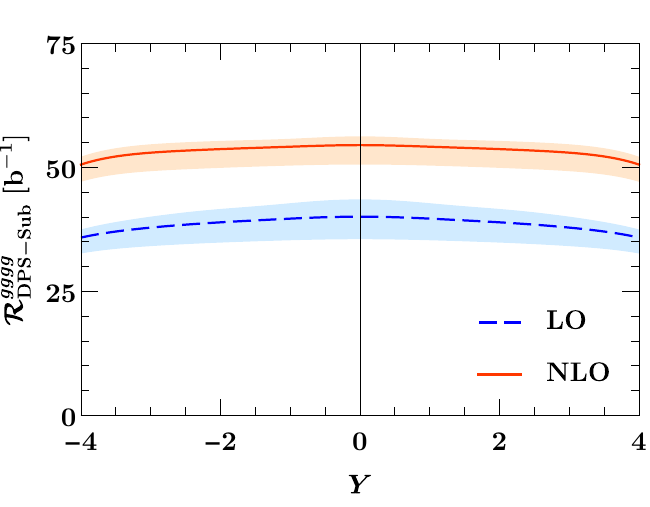}
}
\\[1.5em]
\subfloat[$\mu_{\text{init}}$ variation, $u g, \bar{d} g$]{
   \includegraphics[width=0.48\textwidth]{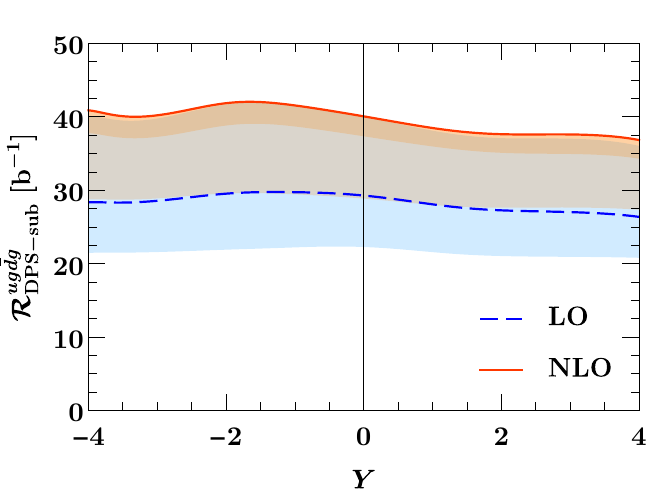}
}
\subfloat[$\nu$ variation, $u g, \bar{d} g$]{
   \includegraphics[width=0.48\textwidth]{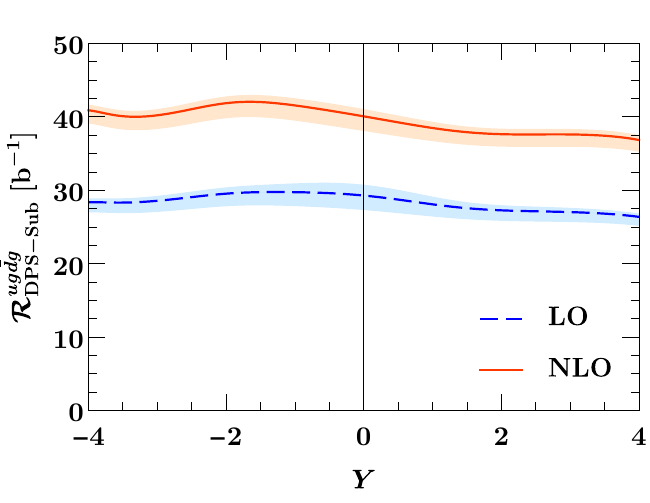}
}
\caption{\label{fig:scale-col-sing-80-10} As
\fig{\protect\ref{fig:scale-col-sing-extra}}, but for asymmetric scales $M_1 =
80 \gev$ and $M_2 = 10 \gev$.}
\end{figure}

\begin{figure}[t]
\centering
\subfloat[$\mu_{\text{init}}$ variation, $g g, g g$, $A A, A A$]{
   \includegraphics[width=0.48\textwidth]{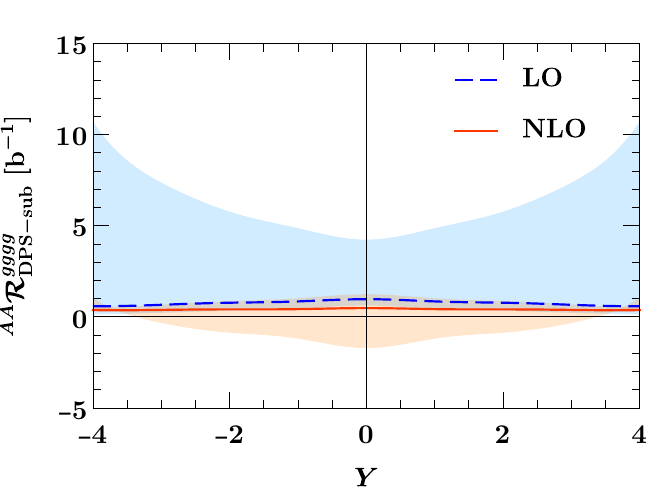}
}
\subfloat[$\nu$ variation, $g g, g g$, $A A, A A$]{
   \includegraphics[width=0.48\textwidth]{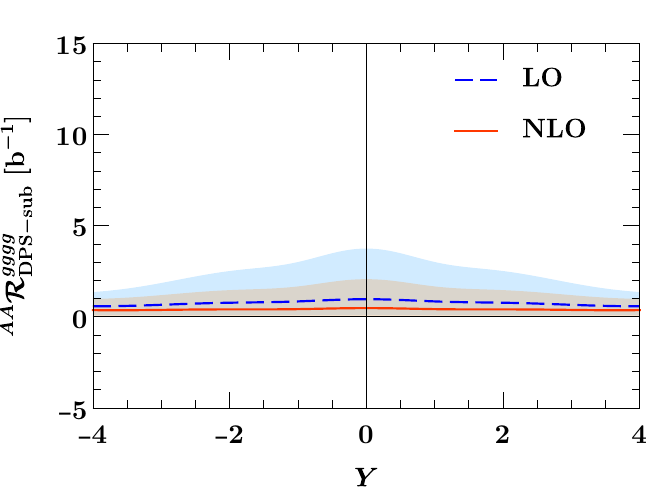}
}
\\[1.5em]
\subfloat[$\mu_{\text{init}}$ variation, $u g, \bar{d} g$, $8 A, 8 A$]{
   \includegraphics[width=0.48\textwidth]{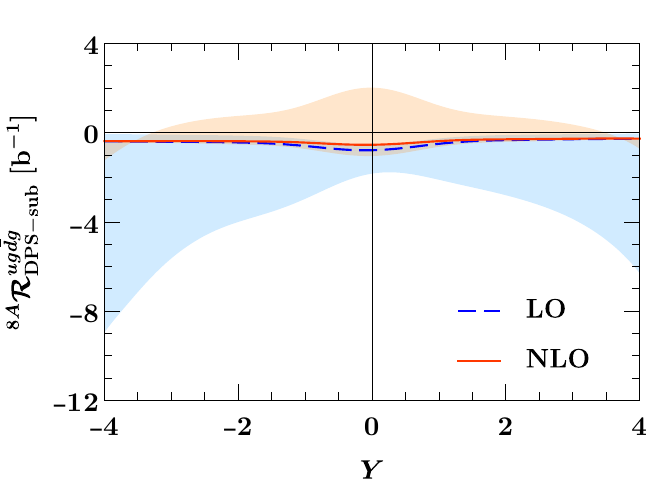}
}
\subfloat[$\nu$ variation, $u g, \bar{d} g$, $8 A, 8 A$]{
   \includegraphics[width=0.48\textwidth]{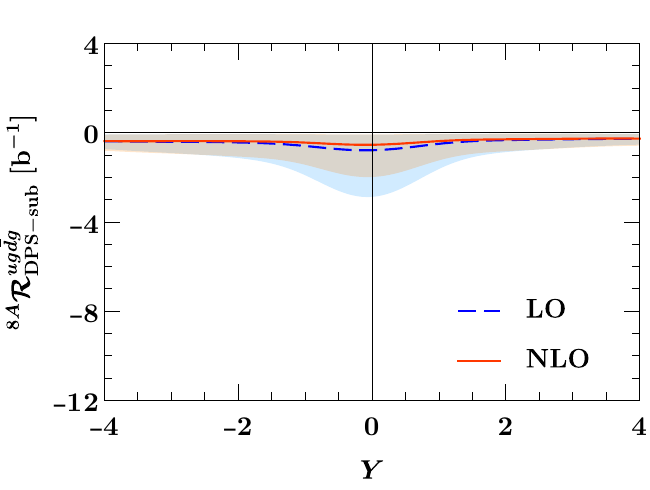}
}
\caption{\label{fig:scale-col-oct-80-10} As
\fig{\protect\ref{fig:scale-col-sing-80-10}}, but for colour octet channels.}
\end{figure}

Given the practical importance of subtracted luminosities and their
$\mu_{\text{init}}$ and $\nu$ dependence, the following plots will cover a
larger number of channels than the examples shown so far.
Figures~\ref{fig:scale-col-sing} to \ref{fig:scale-col-oct-extra} are for equal
scales $M_1 = M_2 = 80\gev$, and \figs{\ref{fig:scale-col-sing-80-10}} to
\ref{fig:scale-col-oct-80-10} are for unequal scales $M_1 = 80\gev$, $M_2 =
10\gev$.

% more parton combinations:

% unequal scales:

Subsequent pairs of figures show colour singlet and colour octet channels for
the same parton combinations.  We typically find that the central values for
octet channels are much smaller than for the colour singlet, although not
necessarily negligible.  Only in the four-gluon case at equal scales do we find
colour octet contributions of size similar to the singlet, as seen in the top
panels of \figs{\ref{fig:scale-col-sing-extra}} and
\ref{fig:scale-col-oct-extra}.

Not shown here are the colour combinations with symmetric instead of
antisymmetric octets.  We generally find that the $S S, S S$ luminosity for four
gluons is very similar in magnitude to $A A, A A$.  The same holds for $8 S, 8
S$ compared with $8 A, 8 A$ for the $u g, \bar{d} g$ combination.  The
mixed-symmetric $A S, A S$ luminosity for four gluons is tiny compared to $A A,
A A$.  At the starting scale $\muy(y)$, the smallness of $A S$ splitting DPDs
was already noticed in \cite{Diehl:2021wpp}.

For the central values of subtracted luminosities, we find in most cases that
the NLO corrections to DPD splitting are appreciable at the level of several
$10\%$.  In a few channels they reach a factor around 2, notably for $u \bar{u},
\bar{d} d$ at central rapidities in the colour singlet and for $g g, g g$ in the
colour octet case.

For scale variations we see a clear difference between $u \bar{u}, \bar{d} d$
and the other parton combinations shown here.  In the channels other than $u
\bar{u}, \bar{d} d$, we find the following:
\begin{itemize}
\item The variation with $\mu_{\text{init}}$ significantly decreases from LO to
NLO.  At NLO, the scale variation is highly asymmetric, with the central curve
being close to the edge of the band of variation.

Compared with the central value, the scale variation at NLO is moderate for the
colour singlet, whereas in the octet channels it is still large.
\item The variation with $\nu$ is typically much less than the one with
$\mu_{\text{init}}$.

In colour singlet channels, the $\nu$ dependence either decreases from LO to
NLO, or it is very small at both orders.  For colour octet channels there is no
uniform picture, with the variation decreasing from LO to NLO in some channels
and increasing in others.
\item With the exception of the four-gluon luminosity at equal scales, the
variation with $\mu_{\text{init}}$ and $\nu$ in the octet channel is moderate
compared with the size of the colour singlet result.
\end{itemize}
At high rapidity, the behaviour of the $u \bar{u}, \bar{d} d$ luminosity is
quite similar to the one just discussed.  By contrast, for central rapidities
we find that
\begin{itemize}
\item in the colour singlet channel, the variation with $\mu_{\text{init}}$ and
$\nu$ is very similar at LO and NLO.  At $Y=0$ we have a $\mu_{\text{init}}$
variation of $30$ to $40\%$ and a $\nu$ variation by about a factor of $2$, with
both variations being approximately symmetric around the central value.

In the colour octet channel, the $\nu$ variation is similar at both orders,
whilst the $\mu_{\text{init}}$ variation is much larger at NLO than at LO.
Compared with the central value of the colour singlet, the variation of the
colour octet luminosity remains moderate: at $Y=0$ it is at level of about
$25\%$.

We recall that for central rapidities the cancellation between the 1v1
luminosity and its subtraction term is by far the largest for $u \bar{u},
\bar{d} d$, see the bottom row in \fig{\ref{fig:lumi-sub-SPSvsDPS}}.  This is in
line with our earlier observation that the DPDs for $q \bar{q}$ are most
strongly concentrated at $y \sim \ycut$, where the difference between the DPDs
in the 1v1 term and its subtraction is small by design.  Given this strong
cancellation, a high sensitivity of the result to scale variations is perhaps
not too surprising.
\end{itemize}

Whilst the relative uncertainty due to scale choices is most pronounced for $q
\bar{q}$ channels at central rapidities, the uncertainty due to modelling the
intrinsic DPDs is smallest in this case, as we discussed in the context of
\fig{\ref{fig:lumi-contribs-qqbar}}.

%%%%%%%%%%%%%%%%%%%%%%%%%%%%%%%%%%%%%%%%%%%%%%%%%%%%%%%%%%%%%%%%%%%%%%%%%%%%%%%%

\newpage

\section{Splitting with massive quarks}
\label{sec:massive}

So far we have neglected the masses of the heavy quarks $c, b, t$ in the
computation of DPD splitting kernels.  For a given flavour $Q$ this is adequate
if the characteristic scale $1/y$ of the splitting is much larger than the quark
mass $m_Q$. Conversely, a heavy quark decouples from the splitting process if
its mass is much larger than $1/y$.   In the region where $1/y \sim m_Q$, the
quark mass should be taken into account when computing the splitting process in
the DPD. For colour singlet DPDs we gave a detailed analysis of this in
\cite{Diehl:2022dia}.  In the present section we generalise the results from
\sects{3}, 5.5, and 6.4 in \cite{Diehl:2022dia} to colour non-singlet channels,
which is straightforward.  We then proceed to a numerical analysis of evolved
splitting DPDs that include quark mass effects.

For each heavy flavour $Q$, we keep the quark mass $m_Q$ finite in the region
$\alpha \ms m_Q < \mu_y < \beta \ms m_Q$, where we write again $\mu_y = b_0 /
y$. In that region we replace the massless splitting
formula~\eqref{split-master} by
\begin{align}
   \label{split-master-heavy}
   \pr{R_1 R_2}{F}_{a_1 a_2}^{n_F + 1}(x_1, x_2, y; \mu, \mu,\zeta)
   &
   =
   \frac{1}{\pi\ms y^2}\, \sum_{a_0}\,
   \prn{R_1 R_2}{V}^{Q, n_F}_{a_1 a_2, a_0} \conv{12} f^{n_F}_{a_0}
   \,,
\end{align}
where ${F}_{a_1 a_2}^{n_F + 1}$ is a DPD with $n_f + 1$ active flavours
(including the heavy one) and $f^{n_F}_{a_0}$ is a PDF with $n_f$ active
flavours.  We omit the superscript ``spl'' on $F$ for brevity.
The massive splitting kernels $V^{Q, n_F}$ are the analogues of the massless
kernels in \eqref{split-master} and depend on the additional dimensionless
variable $y\ms m_Q$.  One should take $1/\alpha$ and $\beta$ sufficiently large,
such that
\begin{enumerate}
\item at scales $\mu_y < \alpha \ms m_Q$ the heavy quark approximately decouples
in the splitting process.  DPDs including the flavour $Q$ are generated by the
usual matching procedure, as discussed around \eqn{\eqref{gg-matching}}.
\item at scales $\mu_y > \beta \ms m_Q$ the quark $Q$ is well approximated as
massless.
\end{enumerate}
At the same time, $1/\alpha$ and $\beta$ should not be taken too large: the
massive splitting kernels contain terms with $\ln (\mu_y / m_Q)$ raised to a
power that increases with the order in $\alpha_s$, and when these logarithms
become too large, the fixed-order truncation of the DPD splitting kernels
becomes a poor approximation.

At LO the only channel in which heavy partons are produced is $g \to Q
\overline{Q}$.  The corresponding kernel for unpolarized quarks reads
\begin{align}
   \label{split-LO-heavy}
   &
   \prn{R_1 R_2}{V}^{Q (1)}_{Q \overline{Q}, g}(z_1, z_2, y \ms m_Q)
   \notag \\
   &\quad
   =
   c_{q \bar{q}, g}(R_1 R_2) \;
   \delta(1 - z_1 - z_2) \, T_F \, (y \ms m_Q)^2 \,
   \Bigl[ (z_1^2 + z_2^2) \ms K_1^2(y \ms m_Q) + K_0^2(y \ms m_Q) \Bigr]
   \,,
\end{align}
where $K_1$ and $K_0$ are modified Bessel functions of the second kind.  Its
polarised counterparts are given in \sect{3.1} of \cite{Diehl:2022dia}.
Leading-order kernels with one observed heavy parton are zero, and kernels
$\prn{R_1 R_2}{V}^{Q (1)}_{a_1 a_2, a_0}$ for light partons $a_1$ and $a_2$ are
equal to their massless counterpart $\prn{R_1 R_2}{V}^{(1)}_{a_1 a_2, a_0}$ in
\eqref{split-LO}.

The NLO kernels including mass effects have not been computed so far.  Their
behaviour for $\mu_y \ll m_Q$ and for $\mu_y \gg m_Q$ is however known, as is
their scale dependence.  Using these constraints, we can write
\begin{align}
   \label{Vm-ansatz}
   \prn{R_1 R_2}{V}^{Q, n_F (2)}_{a_1 a_2, a_0}
   &
   =
   \prn{R_1 R_2}{V}^{n_F [2,0]}_{a_1 a_2, a_0}
   + \prn{R_1 R_2}{V}^{n_F [2,1]}_{a_1 a_2, a_0} \ln \frac{y^2 \ms m_Q^2}{b_0^2}
   + \prn{R_1 R_2}{V}^{I}_{a_1 a_2, a_0}(z_1, z_2, y \ms m_Q)
   \notag \\
   & \quad
   + g_{0}(y \ms m_Q) \,
      \biggl(
         \prn{R_1 R_2}{V}^{n_F + 1 [2,0]}_{a_1 a_2, a_0}
         - \prn{R_1 R_2}{V}^{n_F [2,0]}_{a_1 a_2, a_0}
      \biggr)
   \notag \\
   & \quad
   + g_{1}(y \ms m_Q) \,
      \biggl(
         \prn{R_1 R_2}{V}^{n_F + 1 [2,1]}_{a_1 a_2, a_0}
         - \prn{R_1 R_2}{V}^{n_F [2,1]}_{a_1 a_2, a_0}
      \biggr)
   \nonumber\\
   & \quad
   + \ln \frac{\mu^2}{m_Q^2} \;
      \prn{R_1 R_2}{v}^{n_F, \text{RGE}}_{a_1 a_2, a_0}(z_1, z_2)
   + \frac{\prn{R_1}{\gamma}_J^{(0)}}{2} \,
     \biggl( L_y L_\zeta - \frac{L_y^2}{2} - \frac{\pi^2}{12} \biggr) \,
     \prn{R_1 R_2}{V}^{Q (1)}_{a_1 a_2, a_0}
   \,,
\end{align}
where $V^{n_f [2,0]}(z_1, z_2)$ and $V^{n_f [2,1]}(z_1, z_2)$ are the massless
splitting kernels in the decomposition \eqref{split-NLO-logs}, evaluated for
$n_f$ active flavours.
The $\mu$ dependent part of \eqref{Vm-ansatz} involves the function
\begin{align}
   \label{vRGE}
   \prn{R_1 R_2}{v}^{n_F,\, \text{RGE}}_{a_1 a_2, a_0}
      &= \sum_{b_1^{}, \Rp{1}} \prn{R_1^{} \Rpbar{1}}{P}_{a_1 b_1}^{n_F + 1 (0)}
            \conv{1} \prn{\Rp{1} R_2^{}}{V}^{Q (1)}_{b_1 a_2, a_0}
         + \sum_{b_2^{}, \Rp{2}} \prn{R_2^{} \Rpbar{2}}{P}_{a_2 b_2}^{n_F + 1 (0)}
            \conv{2} \prn{R_1^{} \Rp{2}}{V}^{Q (1)}_{a_1 b_2, a_0}
   \nonumber\\
   & \quad {}
      - \sum_{b_0} \prn{R_1 R_2}{V}^{Q (1)}_{a_1 a_2, b_0}
         \conv{12} \prn{1 1}{P}_{b_0 a_0}^{n_F (0)}
      + \frac{\beta_0^{n_F + 1}}{2} \;
         \prn{R_1 R_2}{V}^{Q (1)}_{a_1 a_2, a_0}
\end{align}
with $\beta_0^{n_F} = 11 C_A / 3 - 2 \ms n_F / 3$.
In addition to the convolution $\otimes_{12}$ introduced in
\eqref{split-master}, the expression \eqref{vRGE} involves convolution products
\begin{align}
   P \conv{1} V
   &=
   \int_{x_1}^{1} \frac{d z}{z} \, P(z) \,
      V\biggl( \frac{x_1}{z}, x_2, \ldots \biggr)
\end{align}
for the first parton and its analogue $\otimes_{2}$ for the second parton.
In the second and third lines of \eqref{Vm-ansatz} we have the functions
\begin{align}
   \label{gi-default}
   g_0(w)
   &=
   w^2 K_1(w)^2
   \,,
   &
   g_1(w)
   &=
   - w^2 K_0(w) K_2(w)
   \,,
\end{align}
which have the limiting behaviour
\begin{align}
   \label{gi-limits}
   g_{i}(w)
   &
   \overset{w \to 0}{\longrightarrow}
      \biggl( \ln \frac{w^2}{b_0^2} \biggr)^{i}
   \,,
   &
   g_{i}(w)
   &
   \overset{w \to \infty}{\longrightarrow}
   0
   &
   \text{ for } i = 0, 1
   \,.
\end{align}

Finally, the functions $V^{I}_{a_1 a_2, a_0}$ satisfy
\begin{align}
   \label{VI-limits}
      \prn{R_1 R_2}{V}^{I}_{a_1 a_2, a_0}(z_1, z_2, y \ms m_Q)
      \overset{y \to 0}{\longrightarrow} 0 \,,
   && \prn{R_1 R_2}{V}^{I}_{a_1 a_2, a_0}(z_1, z_2, y \ms m_Q)
      \overset{y \to \infty}{\longrightarrow} 0
\end{align}
and specify the transition from small to large $y$.  For specific parton
combinations we have the additional constraints $\prn{R_1 R_2}{V}^{I}_{g g, g}
\propto \delta(1 - z_1 - z_2)$ and $\prn{R_1 R_2}{V}^{I}_{q \bar{q}, g} = 0$.

Without a massive two loop calculation, the functions ${V}^{I}$ remain unknown.
In
our numerical studies we set them to zero for all parton and colour channels.
This does not yield a correct NLO prediction in all regions of $y$, but it
provides a smooth interpolation between the correct limiting behaviour for $y
\ll 1/m_Q$ and $y \gg 1/m_Q$, and it has the correct dependence on the scales
$\mu$ and $\zeta$.  In this sense we regard it as an improvement over using the
massless scheme at NLO, in which the region $y \sim 1/m_Q$ is treated in a less
realistic way, with often large unphysical discontinuities of the DPDs at the
transition point $\mu_y = \gamma m_Q$.

%%%%%%%%%%%%%%%%%%%%%%%%%%%%%%%%%%%%%%%%

\paragraph{Several heavy flavours}

The masses of charm and bottom quarks are rather close to each other, so that a
region of scales $\mu$ with $m_c \ll \mu \ll m_b$ is hard to identify.   It is
therefore useful to consider a scheme in which $u, d, s$ are treated as massless
and both $c$ and $b$ as heavy.  The matching of PDFs from $n_f = 3$ to $n_f = 5$
active flavours in one step has been worked out in \cite{Blumlein:2018jfm} at
two loops, and three-loop calculations for the relevant parton transitions are
ongoing \cite{Ablinger:2025nnq}.

Using the two-mass scheme for DPDs in the relevant region of $y$, we evaluate
initial conditions with mass effects in the splitting as follows:
\begin{align}
   \label{massive-cb}
   n_f &= 4,
   &
   m_c &\neq 0
   &&
   \text{ for } \mu_y < \alpha \ms m_b ,
   \notag \\
   n_f &= 5,
   &
   m_c &\neq 0,
   \quad
   m_b \neq 0
   &&
   \text{ for } \alpha \ms m_b < \mu_y < \beta \ms m_c ,
   \notag \\
   n_f &= 5,
   &
   m_c &= 0,
   \quad
   m_b \neq 0
   &&
   \text{ for } \beta \ms m_c < \mu_y < \beta \ms m_b
\intertext{and}
   \label{massive-t}
   n_f &= 6,
   &
   m_c &= m_b = 0,
   \quad
   m_t \neq 0
   &&
   \text{ for } \alpha \ms m_t < \mu_y < \beta \ms m_t ,
   \notag \\
   n_f &= 6,
   &
   m_c &= m_b = 0,
   \quad
   m_t = 0
   &&
   \text{ for } \beta \ms m_t < \mu_y
   \,.
\end{align}
Here we have assumed that $\alpha \ms m_b < \beta \ms m_c$ and $\alpha \ms m_t >
\beta \ms m_b$, which is the case for our parameter choices in the following.

The DPD splitting kernels for the three regimes in \eqref{massive-cb} are $V^{c,
3}$, $V^{c b}$, and  $V^{b, 4}$, respectively, where $V^{c b}$ denotes
the splitting kernels for 3 light flavours plus massive charm and bottom.

The analogue of the general form \eqref{Vm-ansatz} for $V^{c b}$ at NLO is given
for colour singlet DPDs in \eqs{(6.34)} to (6.38) in \cite{Diehl:2022dia}.  To
generalise those expressions to all colour channels, one should add labels $R_1,
R_2$ to $V^{c b}$ in the same way as in \eqref{Vm-ansatz} and \eqref{vRGE} here,
and add a double logarithmic term
\begin{align}
   \frac{\prn{R_1}{\gamma}_J^{(0)}}{2} \,
   \biggl( L_y L_\zeta - \frac{L_y^2}{2} - \frac{\pi^2}{12} \biggr) \,
   \prn{R_1 R_2}{V}^{c b (1)}_{a_1 a_2, a_0}
\end{align}
with the massive LO kernels ${V}^{c b (1)}_{a_1 a_2, a_0}$ for the colour
singlet being specified in and below \eqn{(3.11)} of \cite{Diehl:2022dia}.

%%%%%%%%%%%%%%%%%%%%%%%%%%%%%%%%%%%%%%%%

\subsection{Parton kinematics and scales}

A comprehensive numerical study of mass effects in DPD splitting at LO was
carried out in \cite{Diehl:2022dia}, where several shortcomings of the LO
approximation were identified.  In the present work we will see how the
situation changes when NLO effects are included in the approximate manner
described below \eqref{VI-limits}.

We consider the same kinematics as in \cite{Diehl:2022dia} and evaluate DPDs and
double parton luminosities for
\begin{align}
   \label{dijet-kin}
   n_f &= 5
   \,,
   &
   M_1 &= M_2 = 25 \gev
   \,,
   &
   \sqrt{s} &= 14 \tev
   \,,
   \\
   \label{top-kin}
   n_f &= 6
   \,,
   &
   M_1 &= M_2 = 1 \tev
   \,,
   &
   \sqrt{s} &= 100 \tev
   \,.
\end{align}
As in \sect{\ref{sec:massless}} we take $Y = Y_1 = - Y_2$ in double parton
luminosities.

The scale of $25 \gev$, which may for instance appear in dijet production, is a
compromise between being high enough to compute parton-level cross sections with
$n_f = 5$ massless quarks and low enough to be sensitive to quark mass effects
in the DPD splitting process.

Parton momentum fractions for the settings \eqref{dijet-kin} and \eqref{top-kin}
are
\begin{align}
   x_1
   =
   \bar{x}_1
   &\approx
   \begin{cases}
      1.8 \times 10^{-3}  & \text{ if $M_1 = 25 \gev$ and $\sqrt{s} = 14\tev$}
      \\
      1.0 \times 10^{-2}  & \text{ if $M_1 = 1 \tev$ and $\sqrt{s} = 100\tev$}
   \end{cases}
\end{align}
for $Y_1 = 0$ and
\begin{align}
   x_1
   &\approx
   \begin{cases}
      9.7 \times 10^{-2} \\
      5.5 \times 10^{-1}
   \end{cases}
   \qquad
   \bar{x}_1
   \approx
   \begin{cases}
      3.3 \times 10^{-5}  & \text{ if $M_1 = 25 \gev$ and $\sqrt{s} = 14\tev$}
      \\
      1.8 \times 10^{-4}  & \text{ if $M_1 = 1 \tev$ and $\sqrt{s} = 100\tev$}
   \end{cases}
\end{align}
for $Y_1 = 4$.  The corresponding fractions $x_2$ and $\bar{x}_2$ are obtained
by symmetry considerations.

In the following studies, initial conditions of the DPDs are evaluated for the
$y$ dependent number of active flavours in the chosen flavour scheme, given by
\eqref{nf-choice} or by \eqref{massive-cb} and \eqref{massive-t}.  The DPDs
are then evolved to the final scales $\mu_1 = \mu_2 = M_1$ and the rapidity
parameter $\zeta = M_1 M_2 / (x_1 x_2)$, with appropriate flavour number
matching at scales $\mu = m_Q$.

%%%%%%%%%%%%%%%%%%%%%%%%%%%%%%%%%%%%%%%%

\subsection{Splitting DPDs with or without mass effects}
\label{sec:massive-dpds}

Both in the massless flavour scheme laid out in \sect{\ref{sec:massless}} and in
the massive scheme just presented, the number of active flavours in the initial
conditions of DPDs is changed at specific values of $y$.  This leads to
discontinuities in the $y$ dependence of the resulting DPDs, even after they
have been evolved to final scales and matched to final $n_f$ values.  Such
discontinuities can be considered as artefacts of the approximations in a given
scheme, so that the size of these discontinuities can be regarded as an
indicator for the quality of the scheme.
Specifically, we will chose the parameters $\alpha$ and $\beta$ in
\eqref{massive-cb} and \eqref{massive-t} so as to minimise discontinuities of
evolved DPDs in $y$.  Of course, small unphysical discontinuities are necessary
but not sufficient for having realistic results.  We will shortly encounter
examples for this statement.

In the following we speak of the ``massless scheme'' in the case where splitting
DPDs are initialised with massless splitting kernels for $n_f$ given by
\eqref{nf-choice}, and of the ``massive scheme'' in the case where they are
initialised with massless or massive kernels as specified in \eqref{massive-cb}
and \eqref{massive-t}.  In the following discussion and plots, we use the scale
$\mu_y = b_0 / y$ that determines the initial conditions of the splitting DPDs,
rather than the distance $y$.

Not surprisingly, we find that in general the largest discontinuities occur for
DPDs with one or two heavy partons.  In the following plots we therefore
concentrate on such cases.  For DPDs with only light partons we find that
discontinuities are often most pronounced in the $g g$ channel.

\begin{figure}[t]
\centering
\subfloat[$c \bar{c}$, $\gamma = 1$]{
   \includegraphics[height=0.35\textwidth]{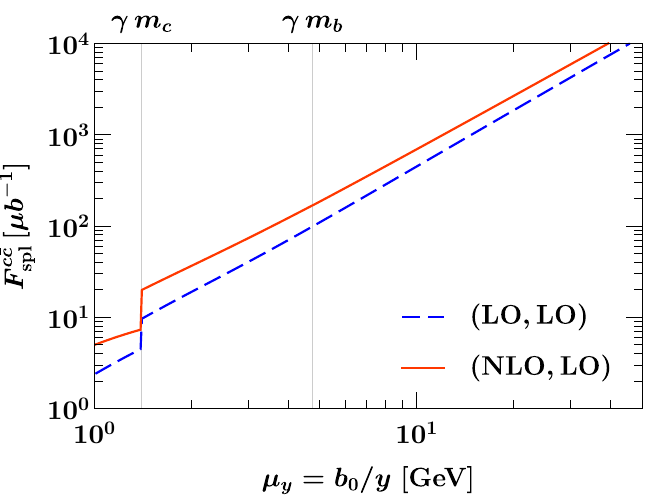}
}
\subfloat[\label{fig:massless-bbar} $b \bar{b}$, $\gamma = 1$]{
   \includegraphics[height=0.35\textwidth]{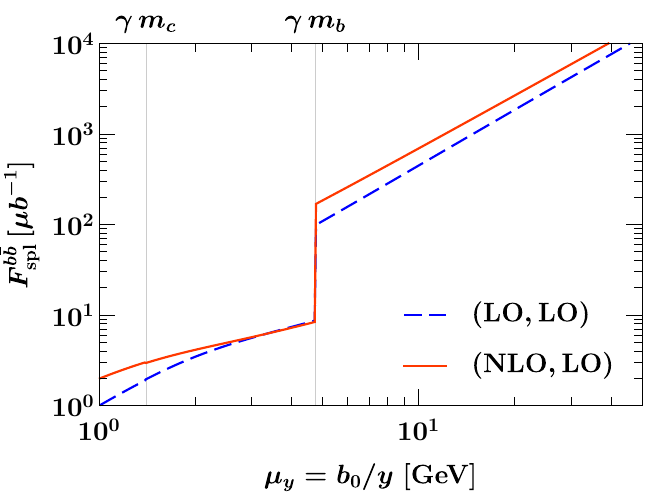}
}
\\[1.5em]
\subfloat[$c b$, $\gamma = 1$]{
   \includegraphics[height=0.35\textwidth]{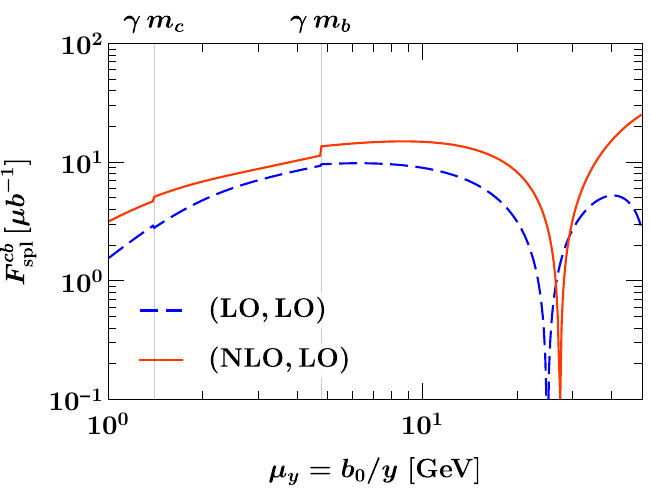}
}
\subfloat[\label{fig:massless-gb} $g b$, $\gamma = 1$]{
   \includegraphics[height=0.35\textwidth]{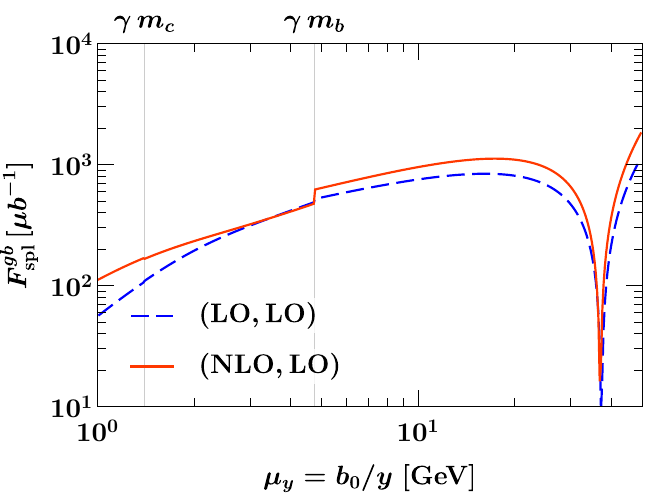}
}
\caption{\label{fig:dpds-dijet-massless} Splitting DPDs for $x_1 = x_2 \approx
1.8 \times 10^{-3}$.  Distributions are initialised in the massless scheme
\protect\eqref{nf-choice} with $\gamma = 1$, followed by evolution and flavour
matching to $\mu_1 = \mu_2 = 25\gev$ and $n_f = 5$.  The specification of
perturbative orders is the same as in \fig{\ref{fig:dpds-g}}.}
\end{figure}

In \fig{\ref{fig:dpds-dijet-massless}} we show colour singlet DPDs in the
massless scheme.  Very large discontinuities are seen for $Q \overline{Q}$
distributions at $\mu_y = m_Q$, which reflects that the direct splitting $g \to
Q \overline{Q}$ does not contribute for $\mu_y < m_Q$.  The same holds in the
colour octet channel, as seen in \fig{\ref{fig:octet-massless}}.
For all channels shown here, we find that discontinuities at NLO are somewhat
larger than at LO.  This is an example in which smaller discontinuities do not
indicate a better approximation.

\begin{figure}
\centering
\subfloat[$b \bar{b}$, $1/\alpha = \beta = 3$]{
   \includegraphics[width=0.48\textwidth]{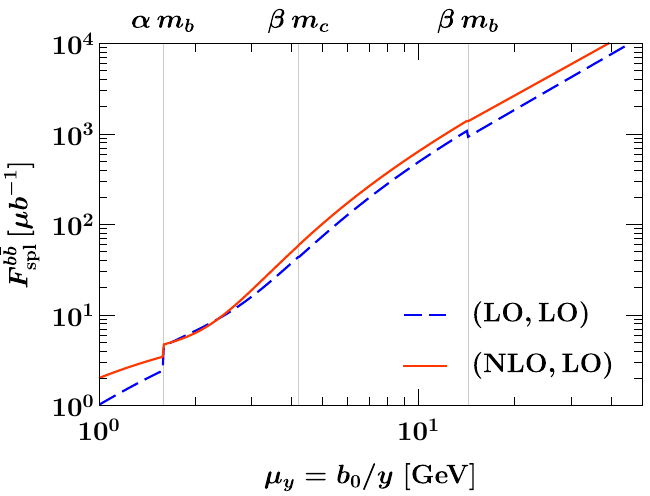}
}
\subfloat[$b \bar{b}$, $1/\alpha = \beta = 4$]{
   \includegraphics[width=0.48\textwidth]{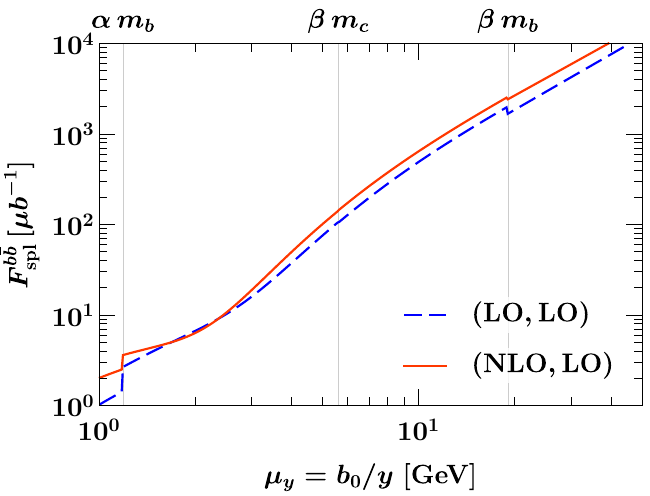}
}
\\[1.5em]
\subfloat[$c b$, $1/\alpha = \beta = 3$]{
   \includegraphics[width=0.48\textwidth]{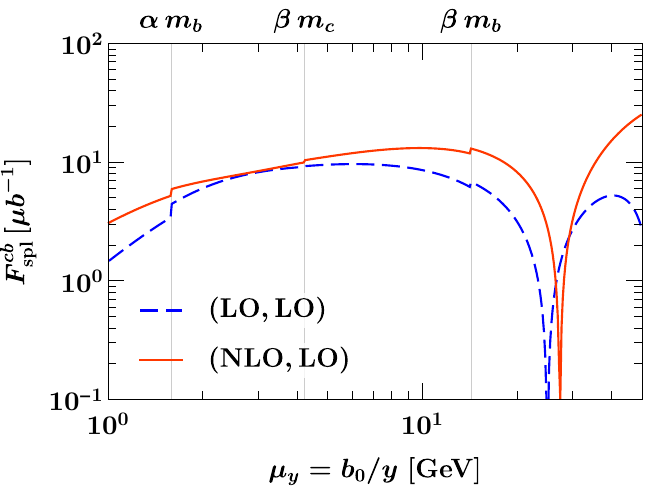}
}
\subfloat[$c b$, $1/\alpha = \beta = 4$]{
   \includegraphics[width=0.48\textwidth]{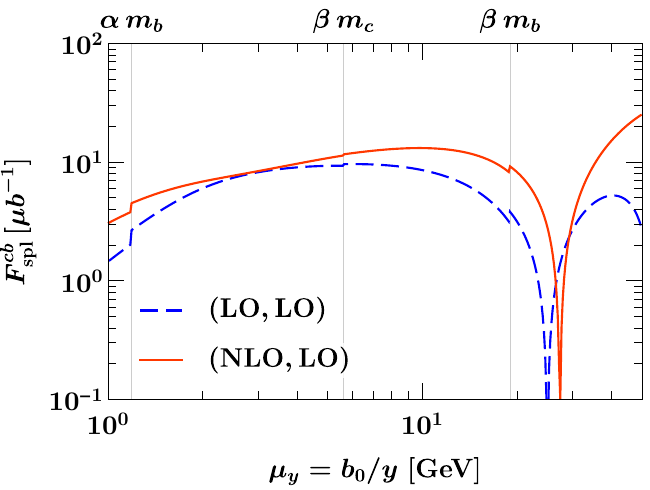}
}
\\[1.5em]
\subfloat[$g b$, $1/\alpha = \beta = 3$]{
   \includegraphics[width=0.48\textwidth]{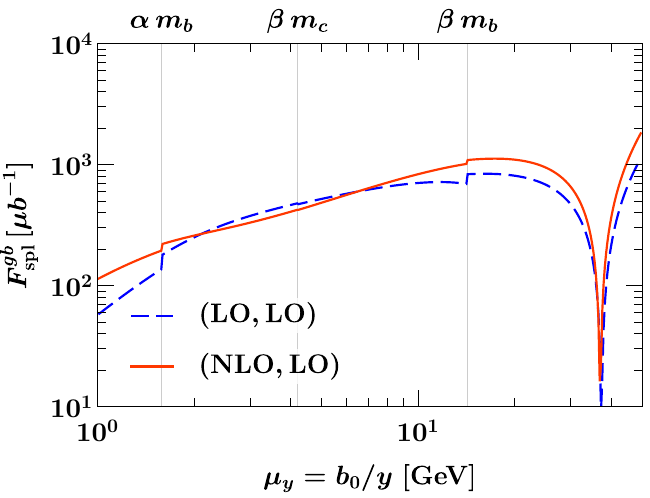}
}
\subfloat[$g b$, $1/\alpha = \beta = 4$]{
   \includegraphics[width=0.48\textwidth]{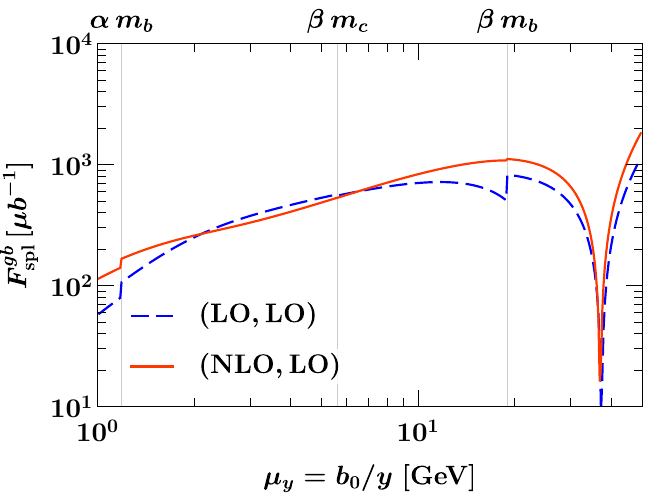}
}
\caption{\label{fig:dpds-dijet-massive} The same DPDs as in
\fig{\protect\ref{fig:massless-bbar}} to \protect\ref{fig:massless-gb}, but
initialised in the massive scheme \protect\eqref{massive-cb} with two different
choices of $1/\alpha = \beta$.}
\end{figure}

\begin{figure}
\centering
\subfloat[$c \bar{c}$, $1/\alpha = \beta = 3$]{
   \includegraphics[width=0.48\textwidth]{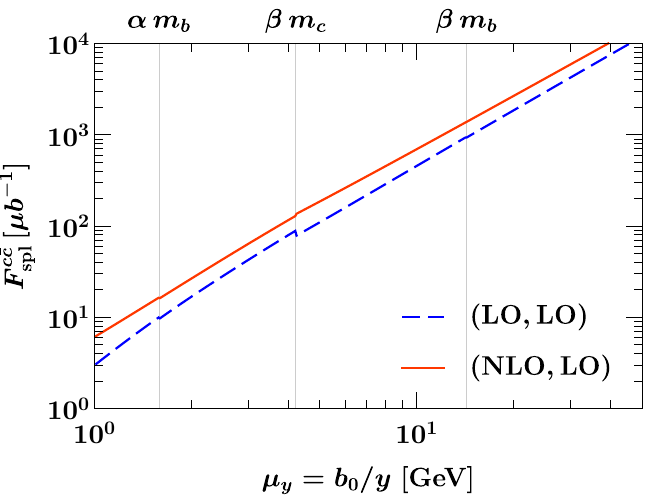}
}
\subfloat[$g g$, $1/\alpha = \beta = 3$]{
   \includegraphics[width=0.48\textwidth]{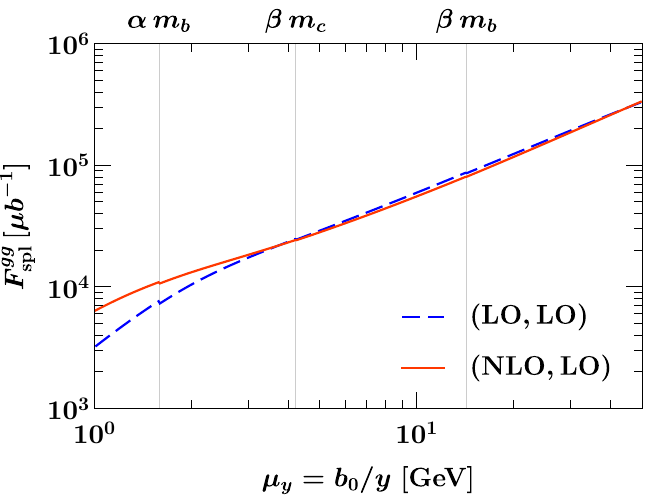}
}
\caption{\label{fig:dpds-dijet-massive-extra} As the left panels of
\fig{\protect\ref{fig:dpds-dijet-massive}}, but for further parton
combinations.}
% \end{figure}
%
\vspace{4em}
%
% \begin{figure}
\centering
\subfloat[\label{fig:octet-massless} $b \bar{b}$, colour octet,
   massless $\gamma = 1$]{
   \includegraphics[width=0.48\textwidth]{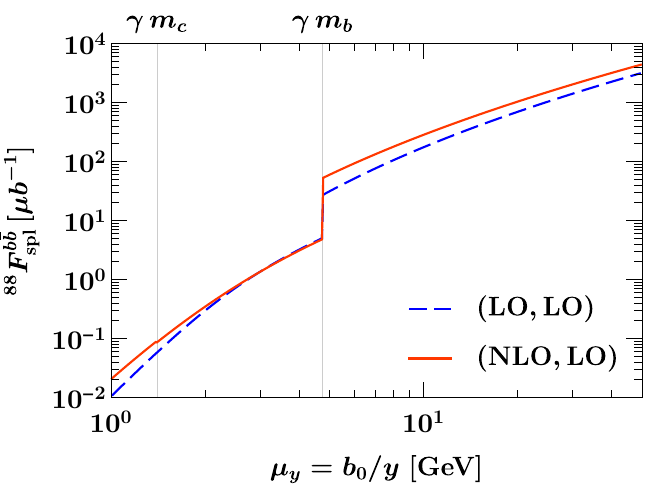}
}
\subfloat[\label{fig:octet-massive} $b \bar{b}$, colour octet,
   massive $1/\alpha = \beta = 3$]{
   \includegraphics[width=0.48\textwidth]{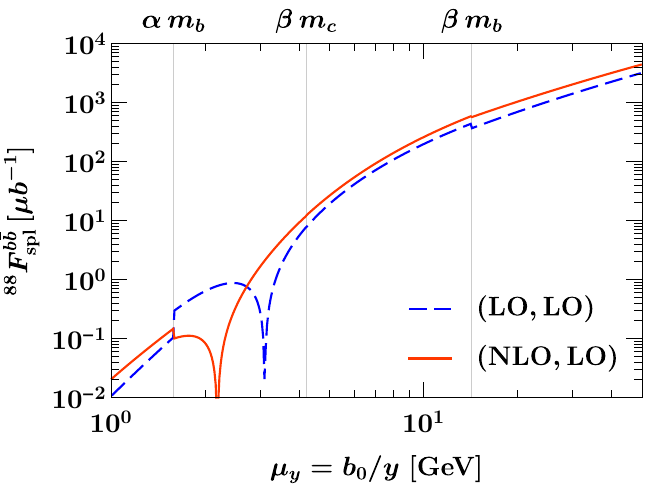}
}
\caption{\label{fig:dpds-dijet-bbar-oct} Colour octet DPDs for $b \bar{b}$ with
the same settings as in \fig{\protect\ref{fig:dpds-dijet-massless}} (left) or
\protect\ref{fig:dpds-dijet-massive-extra} (right).}
\end{figure}

The situation is very different in the massive scheme, as shown in
\figs{\ref{fig:dpds-dijet-massive}} and \ref{fig:dpds-dijet-massive-extra}.  The
discontinuities for $b \bar{b}$ and $c \bar{c}$ distributions are significantly
reduced compared with the massless scheme.  Moreover, discontinuities tend to be
smaller at NLO than at LO.  This holds in particular for the discontinuity of
the $g b$ distribution at $\mu_y = \beta \ms m_b$, which is rather pronounced at
LO. The reason for this is discussed in detail in \sect{4.1.2} of
\cite{Diehl:2022dia}: if one uses the massive scheme at LO, then for $\mu_y <
\beta m_b$ the sequential splitting processes $g \to b \bar{b} \to g b \bar{b}$
and $g \to g g \to g b \bar{b}$ require DGLAP evolution of the DPD, which is
strongly suppressed for $\mu_y$ close to the final scale $25\gev$.  This leads
to an underestimation of the DPD at $\mu_y < \beta \ms m_b$.  To minimise this
effect we took the rather small value $\beta = 2$ in \cite{Diehl:2022dia}. At
NLO, both splitting processes are taken into account in the DPD splitting
kernels, and the problem just described is absent. Consequently, there is no
problem in setting $\beta$ to 3 or 4, as is seen in the bottom row of
\fig{\ref{fig:dpds-dijet-massive}}.

We remark in passing that the colour octet $b \bar{b}$ distribution has a change
of sign in the massive scheme, which is absent in the massless one (see
\fig{\ref{fig:octet-massive}}).  Since in the corresponding $y$ region the
DPD is many orders of magnitude smaller than its maximal value, we do not
investigate this issue further here.

% six flavours

\begin{figure}
\centering
\subfloat[$t \bar{t}$, $\gamma = 1$]{
   \includegraphics[height=0.35\textwidth]{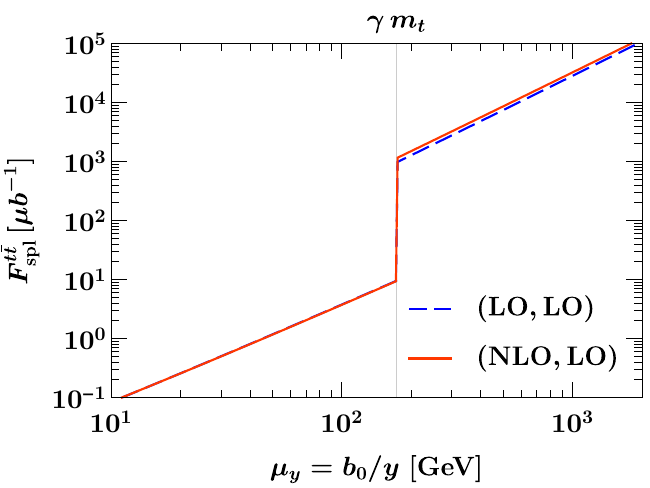}
}
\subfloat[$g t$, $\gamma = 1$]{
   \includegraphics[height=0.35\textwidth]{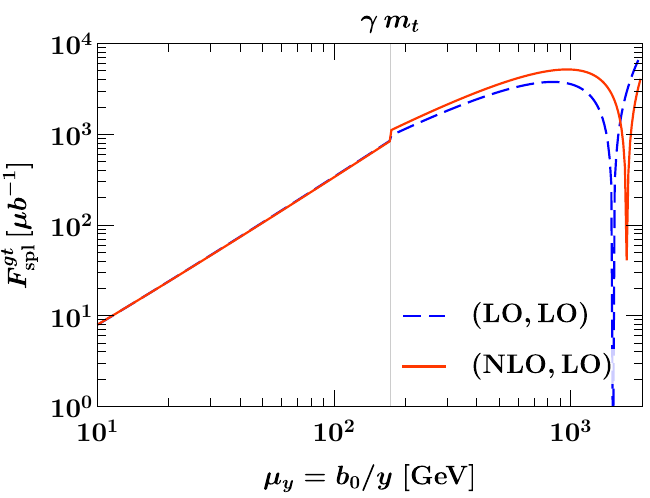}
}
\caption{\label{fig:dpds-ttbar-massless} Splitting DPDs with top quarks for $x_1
= x_2 = 0.01$. Distributions are initialised in the massless scheme
\protect\eqref{nf-choice}  with $\gamma = 1$, followed by evolution and flavour
matching to $\mu_1 = \mu_2 = 1\tev$ and $n_f = 6$.}
% \end{figure}
%
\vspace{2em}
%
% \begin{figure}
\centering
\subfloat[$t \bar{t}$, $1/\alpha = \beta = 3$]{
   \includegraphics[width=0.48\textwidth]{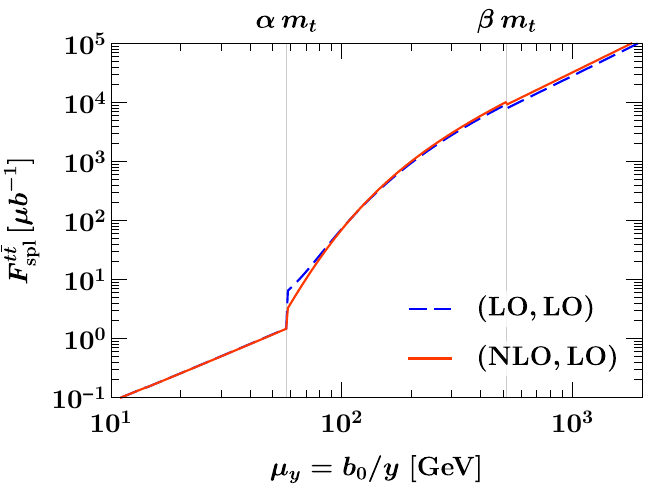}
}
\subfloat[$t \bar{t}$, $1/\alpha = \beta = 4$]{
   \includegraphics[width=0.48\textwidth]{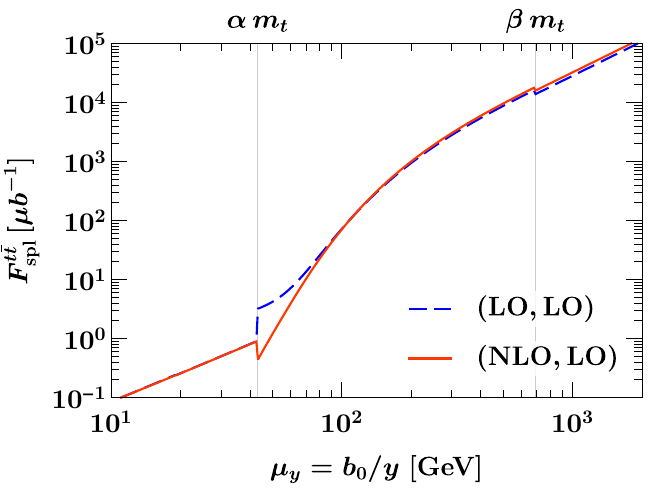}
}
\\[1.5em]
\subfloat[$g t$, $1/\alpha = \beta = 3$]{
   \includegraphics[width=0.48\textwidth]{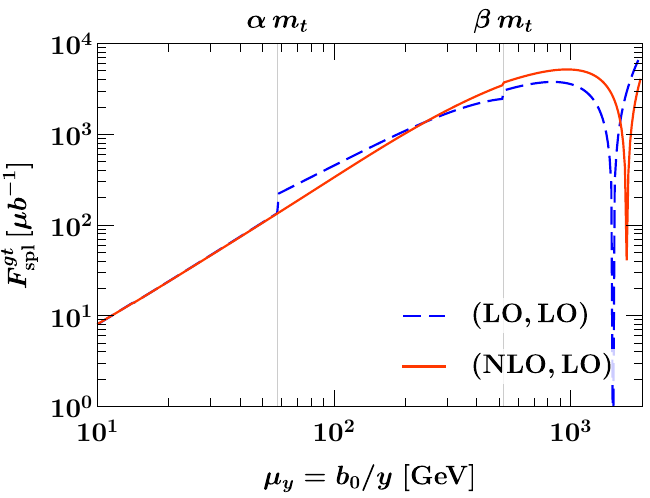}
}
\subfloat[$g t$, $1/\alpha = \beta = 4$]{
   \includegraphics[width=0.48\textwidth]{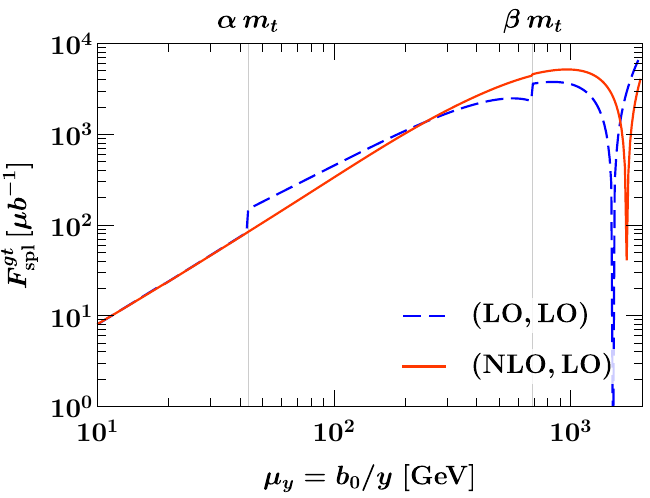}
}
\caption{\label{fig:dpds-ttbar-massive} The same DPDs as in
\fig{\protect\ref{fig:dpds-ttbar-massless}}, but initialised in the massive
scheme \protect\eqref{massive-cb}, \protect\eqref{massive-t} with two different
choices of $1 / \alpha = \beta$.}
\end{figure}

Examples for distributions involving top quarks are shown in
\figs{\ref{fig:dpds-ttbar-massless}} and \ref{fig:dpds-ttbar-massive}.  The
differences between the massless and massive schemes, and between LO and NLO
splitting are qualitatively similar to what we observed for charm and bottom
quarks.  We note that in the massless scheme, the discontinuity of the $t
\bar{t}$ distribution at $\mu_y = \gamma \ms m_t$ spans two orders of magnitude
both at LO and at NLO.

For the channels shown here, both $1 / \alpha = 3$ and $1/\alpha = \beta = 4$
yield reasonably small discontinuities in $y$ for NLO splitting DPDs.  In the
following we take the smaller value $3$ as our default.

%%%%%%%%%%%%%%%%%%%%%%%%%%%%%%%%%%%%%%%%

\subsection{Double parton luminosities: scheme parameter dependence}
\label{sec:massive-lumis}

We find that in the massive scheme the inclusion of NLO effects has
qualitatively similar effects on double parton luminosities as in the massless
case, both for the central values and for the size of scale variations.  We
will therefore not show corresponding plots here.

\begin{figure}
\centering
\subfloat[$c \bar{c}, \bar{b} b$, 1v1, LO]{
   \includegraphics[width=0.48\textwidth]{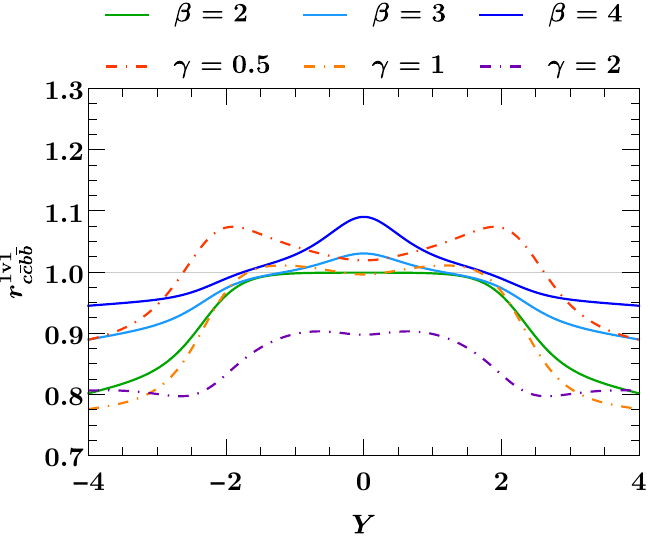}
}
\subfloat[$c \bar{c}, \bar{b} b$, 1v1, NLO]{
   \includegraphics[width=0.48\textwidth]{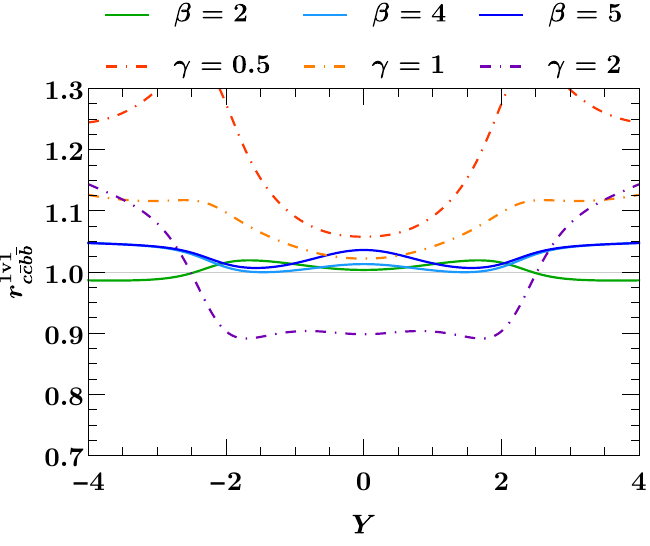}
}
\\[1.5em]
\subfloat[$c \bar{c}, \bar{b} b$, 1v2+2v1, LO]{
   \includegraphics[width=0.48\textwidth]{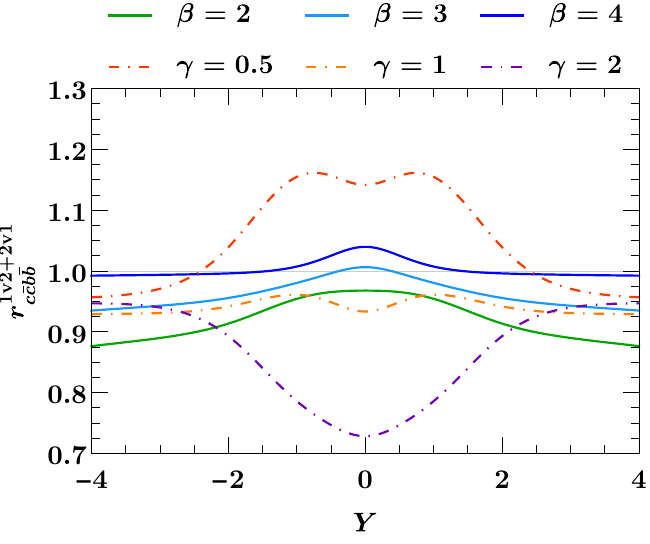}
}
\subfloat[$c \bar{c}, \bar{b} b$, 1v2+2v1, NLO]{
   \includegraphics[width=0.48\textwidth]{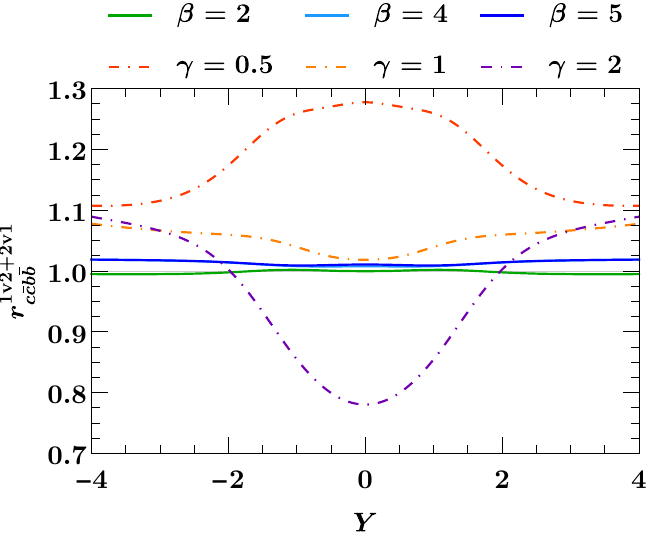}
}
\\[1.5em]
\subfloat[$c b, g g$, 1v1, LO]{
   \includegraphics[width=0.48\textwidth]{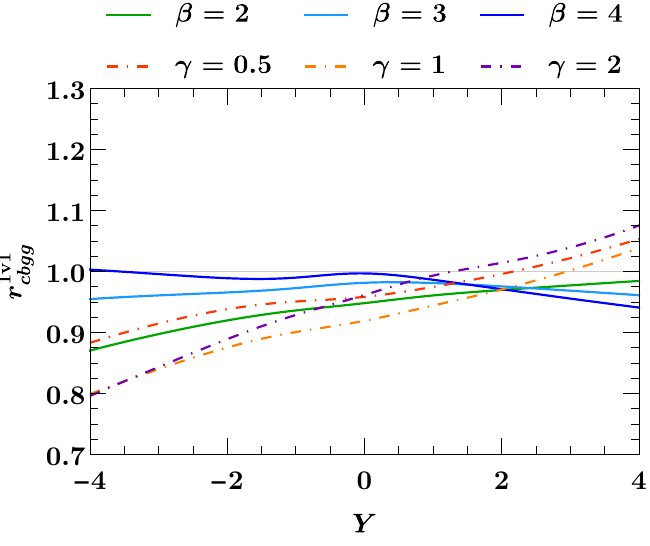}
}
\subfloat[$c b, g g$, 1v1, NLO]{
   \includegraphics[width=0.48\textwidth]{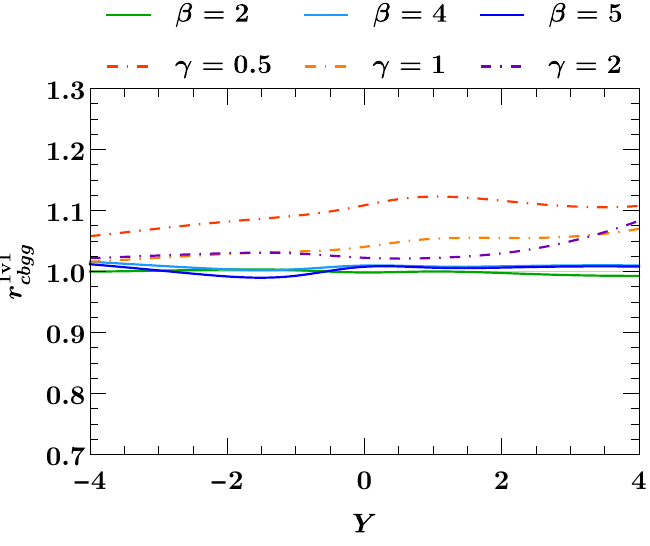}
}
\caption{\label{fig:massive-lumis-dijets} Double parton luminosity ratios
\protect\eqref{massive-lumi-r} in the kinematic setting
\protect\eqref{dijet-kin}.  All plots refer to the colour singlet channel.  Note
that at NLO the ratio $r(\beta)$ with $\beta=3$ is equal to $1$ per definition
and therefore not shown.  We include $\beta=5$ in the comparison instead.}
\end{figure}

\begin{figure}
\centering
\subfloat[$t \bar{t}, \bar{t} t$, 1v1, LO]{
   \includegraphics[width=0.48\textwidth]{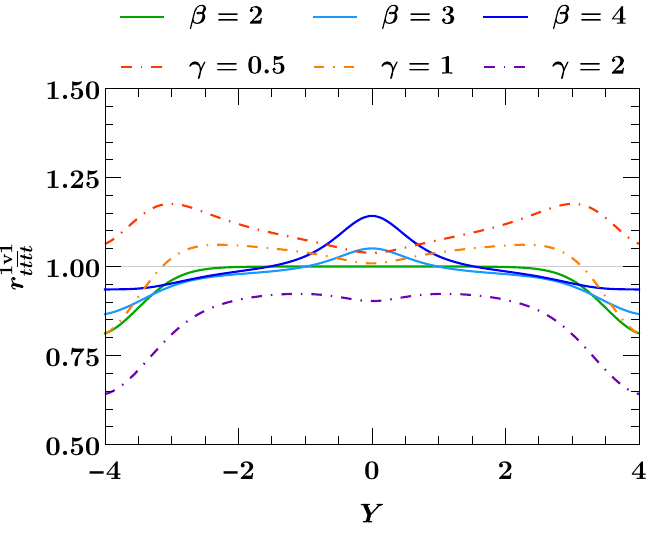}
}
\subfloat[$t \bar{t}, \bar{t} t$, 1v1, NLO]{
   \includegraphics[width=0.48\textwidth]{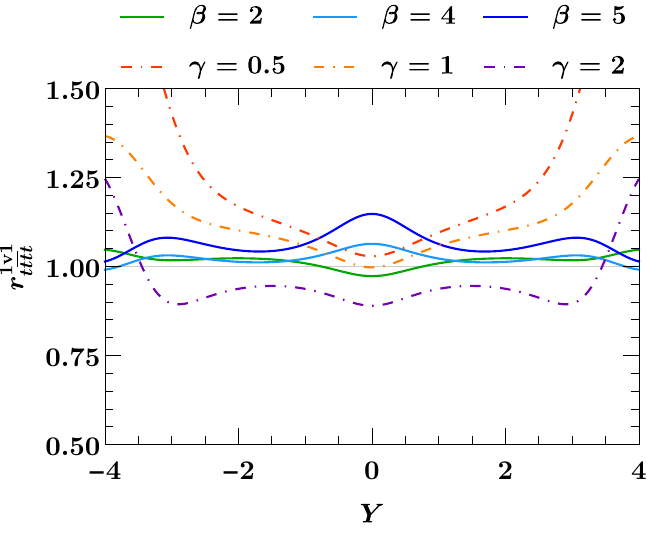}
}
\\[1.5em]
\subfloat[$t \bar{t}, \bar{t} t$, 1v2+2v1, LO]{
   \includegraphics[width=0.48\textwidth]{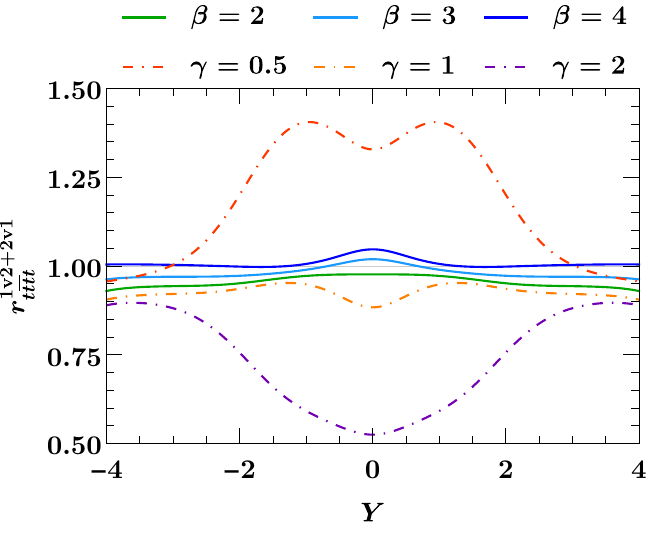}
}
\subfloat[$t \bar{t}, \bar{t} t$, 1v2+2v1, NLO]{
   \includegraphics[width=0.48\textwidth]{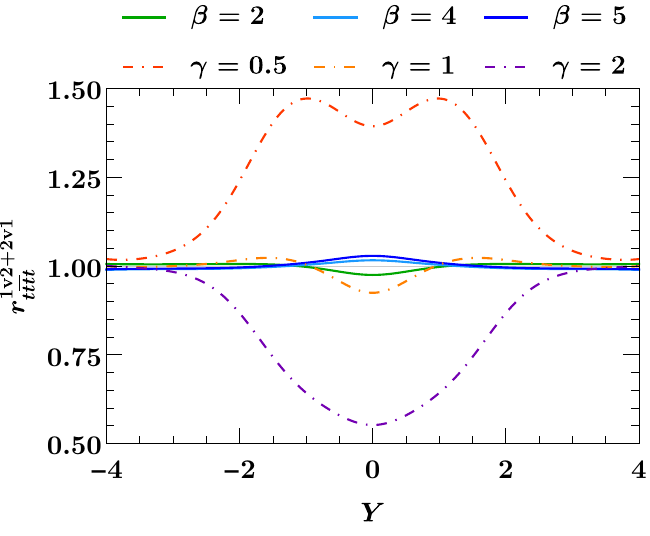}
}
\\[1.5em]
\subfloat[$g t, \bar{t} g$, 1v1, LO]{
   \includegraphics[width=0.48\textwidth]{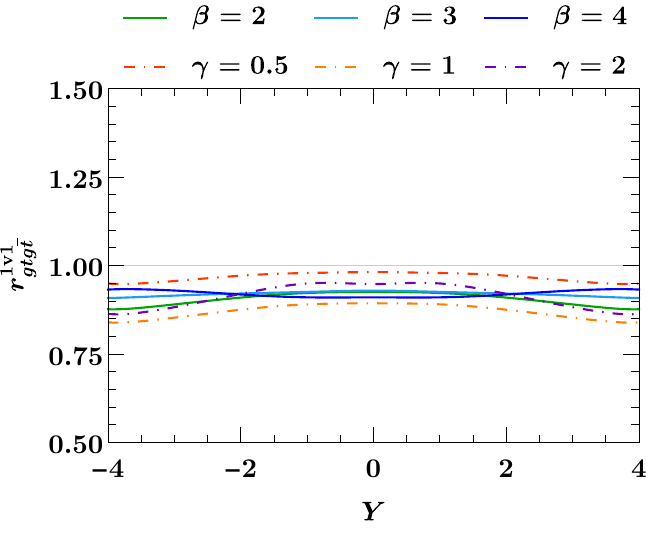}
}
\subfloat[$g t, \bar{t} g$, 1v1, NLO]{
   \includegraphics[width=0.48\textwidth]{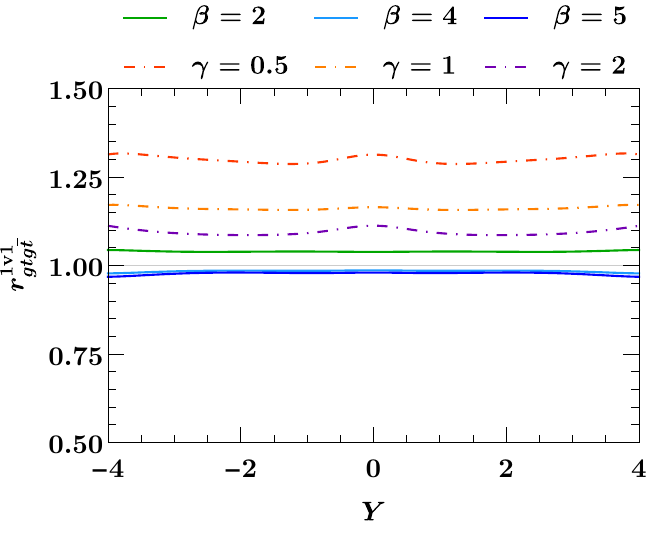}
}
\caption{\label{fig:massive-lumis-ttbar} As
\fig{\protect\ref{fig:massive-lumis-dijets}}, but for top quarks in the
kinematic setting \protect\eqref{top-kin}.}
\end{figure}

Instead we now revisit the dependence of double parton luminosities on the
parameters $\alpha, \beta$ or $\gamma$ in the massive or the massless scheme.
We take
\begin{align}
   \label{alpha-beta-choices}
   \alpha_0 &= 1/3
   \,,
   &
   \beta_0 &= 3
   &&
   \text{at NLO,}
   \notag \\
   \alpha_0 &= 1/4
   \,,
   &
   \beta_0 &= 2
   &&
   \text{at LO}
\end{align}
as baseline parameters in the massive scheme, where the choice for LO
corresponds to our analysis in \cite{Diehl:2022dia}.  In
\figs{\ref{fig:massive-lumis-dijets}} to \ref{fig:massive-lumis-ttbar} we show
the ratios
\begin{align}
   \label{massive-lumi-r}
   r(\beta)
   &=
   \frac{\lum_{\text{massive}}(1/\alpha = \beta)}{
         \lum_{\text{massive}}(\alpha = \alpha_0, \beta = \beta_0)}
   \,,
   &
   r(\gamma)
   &=
   \frac{\lum_{\text{massless}}(\gamma)}{
         \lum_{\text{massive}}(\alpha = \alpha_0, \beta = \beta_0)}
\end{align}
for selected parton combinations, either for the pure splitting contributions
(1v1) or for the mixed contributions (1v2 + 2v1). The intrinsic part of the DPDs
is modelled as described in \sect{\ref{sec:massless-lumis}}.

\begin{figure}
\centering
\subfloat[$c \bar{c}, \bar{b} b$, colour octet, 1v1, LO]{
   \includegraphics[width=0.48\textwidth]{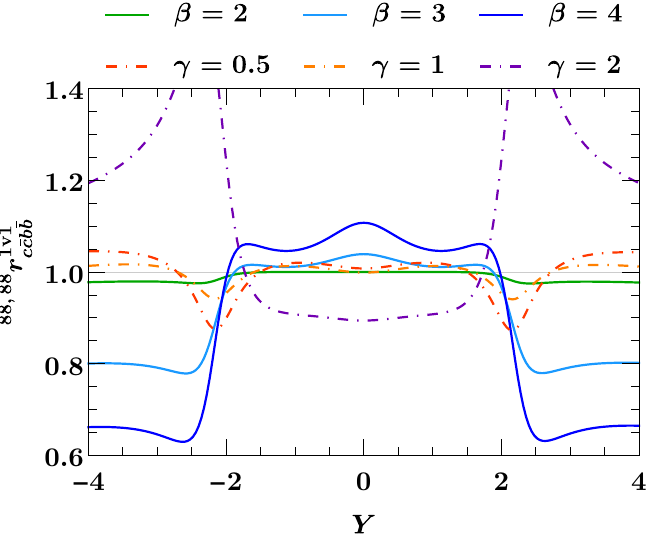}
}
\subfloat[$c \bar{c}, \bar{b} b$, colour octet, 1v1, NLO]{
   \includegraphics[width=0.48\textwidth]{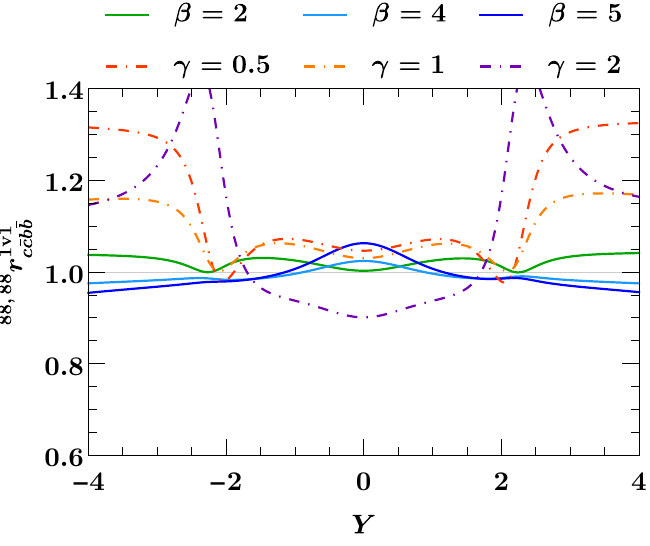}
}
\\[1.5em]
\subfloat[$c b, g g$, 1v1, colour octet, LO]{
   \includegraphics[width=0.48\textwidth]{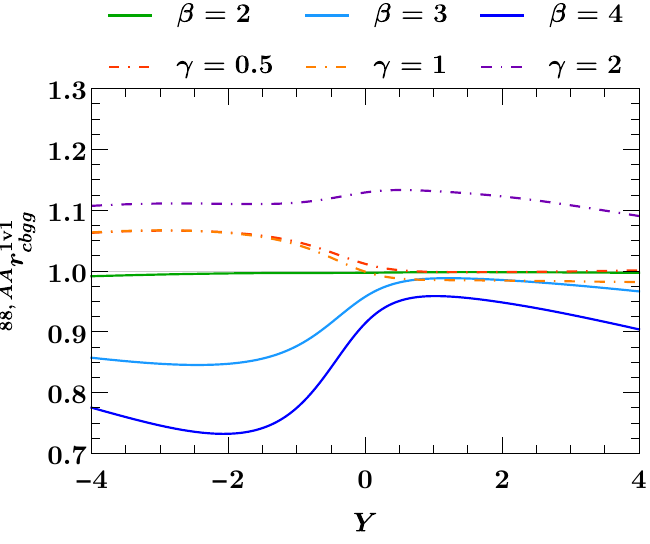}
}
\subfloat[$c b, g g$, 1v1, colour octet, NLO]{
   \includegraphics[width=0.48\textwidth]{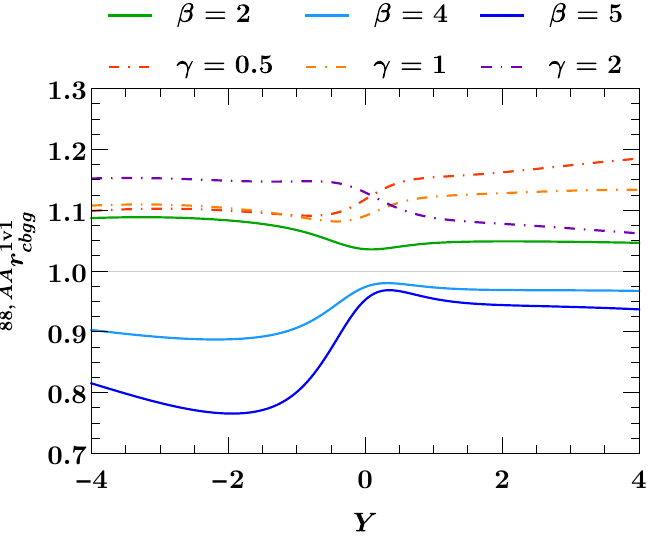}
}
\caption{\label{fig:massive-lumis-dijets-octet} As
\fig{\protect\ref{fig:massive-lumis-dijets}}, but for the colour octet channel.}
\end{figure}

The variation of the ratio $r(\beta)$, shown as solid lines in the figures,
quantifies how luminosities change within the massive scheme when the parameters
$\alpha$ and $\beta$ are varied.  We find that in general this variation is
smaller at NLO than at LO (except for the combination $c b, g g$ in the colour
octet channel).  With the inclusion of NLO effects, the influence of the scheme
parameters on the luminosities is at the level of 5 to 15\% in the colour
singlet and slightly larger in the octet case.

The ratio $r(\gamma)$, shown as dashed lines in the figures, quantifies to which
extent the massless scheme is able to reproduce the (more realistic)
massive scheme at the level of double parton luminosities.  For
$c$ and $b$ quarks (\figs{\ref{fig:massive-lumis-dijets}} and
\ref{fig:massive-lumis-dijets-octet}) an agreement at the level of 15\% can be
achieved with $\gamma = 1/2$ at LO.  At NLO the agreement is better with $\gamma
= 1$ than with $\gamma = 1/2$ and again reaches about 15\% in the cases
considered here.

For $t$ quarks (\fig{\ref{fig:massive-lumis-ttbar}}) the choice $\gamma = 1$ is
preferred at both LO and NLO, with deviations from the massive results reaching
20\% at LO and 40\% at NLO.  In this case, the possibility to approximately
reproduce the massive scheme with the massless one degrades rather than improves
when going from LO to NLO.  This does, of course, not imply that results are
more accurate at LO than at NLO.

\section{Summary}
\label{sec:summary}

We have performed a detailed quantitative study of parton splitting in DPDs,
i.e.\ of the mechanism in which the two observed partons originate from the
splitting of a single one.  Our study proceeds at the level of DPDs and of
double parton luminosities, which are products of DPDs integrated over the
transverse distance $y$ between the partons.  We considered several combinations
of momentum fractions and scales that are relevant for double parton scattering
at the LHC, in particular for the production of combinations of $W$ bosons,
$\jpsi$, and dijets.  In addition we considered kinematics in which quark mass
effects in the splitting process are important for DPDs with bottom or top
quarks.

The ``1v1'' contributions to the DPS cross section, which involve parton
splitting in both DPDs, have an overlap with higher-loop corrections to single
parton scattering.  In the formalism of \cite{Diehl:2017kgu} this is addressed
by a double counting subtraction term in the overall cross section, along with a
lower cutoff $\ycut$ on the integral over $y$.  We have presented a new
construction of this subtraction term, which is both flexible and simple to
implement in practical calculations.

Calculations within our scheme depend on the initial scale $\mu_{\text{init}}$
of DPD evolution, at which the DPD splitting process is evaluated.  We have
shown that the variation of the overall cross section with $\mu_{\text{init}}$
and $\ycut$ within appropriate limits can be appreciable, although it is
formally beyond the accuracy of the computation. This can for instance happen
when contributions with an additional power of $\alpha_s$ are initiated by
partons with a higher density than the partons initiating the lower-order
contribution (e.g.\ by gluons instead of quarks at small $x$, or by quarks
instead of gluons at large $x$).

These are the main findings of our study for double parton luminosities that
include the double counting subtraction term:
\begin{enumerate}
\item \label{state} In many cases, the scale dependence is huge when the
splitting in DPDs is evaluated at LO, whereas at NLO it is considerably reduced.
The dependence on $\ycut$ is typically smaller than the one on
$\mu_{\text{init}}$, and it often decreases further when going from LO to NLO.
\item The preceding statement holds in particular for parton combinations that
contribute to $W^+ W^+$ or $W^- W^-$ pair production.  We find comparable
contributions from parton splitting and from ``intrinsic'' parton pairs in the
proton (which must be modelled).  This is remarkable because the parton
splitting part only starts at NLO for this process.
\item A notable exception from statement \ref{state} are parton combinations
with a $q \bar{q}$ pair in each proton, which contribute in particular to $W^+
W^-$ and $Z Z$ production.  At central rapidities of the produced bosons, the
$\mu_{\text{init}}$ variation is moderate and the  $\ycut$ variation is large,
both at LO and at NLO.
The fact that the double parton luminosity is strongly dominated by $y$ close to
$\ycut$ indicates that SPS contributions having an overlap with DPS should be
much more important than the subtracted DPS cross section $\sigma_{\text{DPS}} -
\sigma_{\text{sub}}$ (see the discussion in \cite{Diehl:2017kgu}).

We therefore expect that, although the computation of $\sigma_{\text{DPS}} -
\sigma_{\text{sub}}$ is affected by large uncertainties at the presently
available perturbative order, the same is not true for the overall cross
section.  We note that SPS contributions overlapping with DPS start at order
$\alpha_s^2$ (NNLO) for $W^+ W^-$ and $Z Z$ production and are known, as well as
their corrections with an additional power of $\alpha_s$ \cite{Glover:1988rg,
Campbell:2016ivq, Caola:2015psa, Caola:2016trd, Grazzini:2018owa, Caola:2015rqy,
Grazzini:2020stb, Agarwal:2024pod}.
\item Colour non-singlet luminosities are most often smaller than their
colour-singlet counterparts, with the exception of the four-gluon channel.  They
are dominated by values of $y$ in the perturbative region, because larger values
of $y$ are strongly suppressed by rapidity evolution.
\end{enumerate}

The mass $m_Q$ of a heavy quark should be taken into account when computing
splitting DPDs at $y \sim 1/m_Q$.  The corresponding kernels are known at LO,
whilst at NLO we currently have only ``approximate'' kernels that have ($i$) the
correct scale behaviour required by the evolution equations and ($ii$) the
correct limiting behaviour for $y\ll 1/m_Q$ and $y\gg 1/m_Q$, while
interpolating smoothly between these two regimes in a model dependent way.  Our
main findings in this ``massive scheme'' are the following:
\begin{enumerate}
\item Unphysical discontinuities occur at $y$ values where one switches between
treating a quark flavour as massless, massive, or absent in the splitting
kernels.  These are greatly decreased when going from LO to approximate NLO.
\item In line with this, the dependence of double parton luminosities on the $y$
values where the above transitions occur is reduced when going from LO to
approximate NLO.
\item Double parton luminosities in the massive scheme differ from those in the
scheme where only massless DPD splitting kernels are used.  By adjusting the $y$
values at which a new flavour is included in the massless splitting DPDs, one
can obtain an \emph{approximate} global agreement between the two schemes, at
the level of several 10\% in the cases we considered.  For top quarks, this
agreement happens to be worse at approximate NLO than at LO.
\end{enumerate}

Overall, we find that the inclusion of NLO effects in splitting DPDs
significantly improves the stability of predictions for double parton
scattering.  It would be good to have NLO splitting kernels also for polarised
partons, and to include heavy quark masses in a more accurate way than the one
described here.  We leave this to future work.

\section*{Acknowledgements}

It is a pleasure to thank Frank Tackmann for useful discussions.
The numerical results in this work have been obtained with the
\chilipdf\ library \cite{Diehl:2021gvs, Diehl:2023cth}, which is under
development.  We gratefully acknowledge the contributions of our collaborators
Florian Fabry, Oskar Grocholski, Riccardo Nagar, and Frank Tackmann to that
project.
The Feynman graphs in this manuscript were produced with JaxoDraw
\cite{Binosi:2003yf, Binosi:2008ig}.

This work has been supported by the Deutsche Forschungsgemeinschaft (DFG,
German Research Foundation) -- grant number 409651613 (Research Unit FOR 2926).
This project has received funding from the European Research Council (ERC)
under the European Union's Horizon 2020 research and innovation programme
(Grant agreement No. 101002090 COLORFREE).

\appendix

%%%%%%u%%%%%%%%%%%%%%%%%%%%%%%%%%%%%%%%%%%%%%%%%%%%%%%%%%%%%%%%%%%%%%%%%%%%%%%%%

% the following lines create an entry in the table of contents
\phantomsection
\addcontentsline{toc}{section}{References}

\bibliography{num-split.bib}

\end{document}